\documentclass[fleqn,usenatbib]{mnras}

\usepackage{newtxtext,newtxmath}

\usepackage[T1]{fontenc}

\DeclareRobustCommand{\VAN}[3]{#2}
\let\VANthebibliography\thebibliography
\def\thebibliography{\DeclareRobustCommand{\VAN}[3]{##3}\VANthebibliography}

\usepackage{graphicx}	
\usepackage{amsmath}	
\usepackage{threeparttable}
\usepackage{wasysym}
\usepackage{comment}
\usepackage{multirow}
\usepackage{hhline}
\usepackage{hyperref}
\usepackage{cancel}
\usepackage{soul}

\usepackage{tikz}
\usetikzlibrary{decorations.pathreplacing}
\usetikzlibrary{arrows}

\newcommand{\kms}{km\,s$^{-1}$} 

\title[Spectroscopic analysis of RGB stars in nine open clusters]{Spectroscopic analysis of RGB stars in nine open clusters\thanks{Based on observations obtained with Brazilian time at the Canada-France-Hawaii Telescope (CFHT) under the program 09BB02 and on observations collected at the European Southern Observatory under ESO programmes 089.D-0716(A) and 089.D-0716(B).}}

\author[S. O. Cantanhêde et al.]{Saulo de Oliveira Cantanhêde$^{1,2}$\thanks{E-mail: saulo.cantanhede@outlook.com} \href{https://orcid.org/0000-0002-9695-7119}{\includegraphics[scale=0.04]{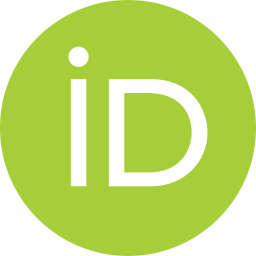}},
Alan Alves-Brito$^{1}$ \href{https://orcid.org/0000-0001-5579-2138}{\includegraphics[scale=0.04]{ORCID_iD.png}},
Rodolfo Smiljanic$^{3}$ \href{https://orcid.org/0000-0003-0942-7855}{\includegraphics[scale=0.04]{ORCID_iD.png}},
Beatriz Barbuy$^{4}$ 
\href{https://orcid.org/0000-0001-9264-4417}{\includegraphics[scale=0.04]{ORCID_iD.png}},\newauthor
Nadège Lagarde$^{5}$
\href{https://orcid.org/0000-0003-0108-3859}{\includegraphics[scale=0.04]{ORCID_iD.png}},
Corinne Charbonnel$^{6,7}$
\href{https://orcid.org/0000-0002-6449-6194}{\includegraphics[scale=0.04]{ORCID_iD.png}},
Pierre North$^{8}$
\href{https://orcid.org/0000-0002-9098-4896}{\includegraphics[scale=0.04]{ORCID_iD.png}}
\\
$^{1}$Departamento de Astronomia, Universidade Federal do Rio Grande do Sul, Av. Bento Gonçalves, 9500, Agronomia, 91501-970, Porto Alegre, RS, Brazil\\
$^{2}$Current Affiliation: Department of Physics, Andrews University, 8975 Old US 31, Berrien Springs, MI, 49104, USA\\
$^{3}$Nicolaus Copernicus Astronomical Center, Polish Academy of Sciences, ul. Bartycka 18, 00-716, Warsaw, Poland\\
$^{4}$Universidade de São Paulo, IAG, R. do Matão, 1226, Cidade Universitária, 05508-900, São Paulo, SP, Brazil\\
$^{5}$Institut UTINAM, CNRS UMR6213, University Bourgogne Franche-Comté, OSU THETA Franche-Comté-Bourgogne, Observatoire de Besançon,\\BP 1615, 25010 Besançon Cedex, France\\
$^{6}$Department of Astronomy, University of Geneva, Chemin Pegasi 51, 1290 Versoix, Switzerland\\
$^{7}$IRAP, CNRS UMR 5277 and Université de Toulouse, 14 Avenue Edouard Belin, 31400 Toulouse, France\\
$^{8}$Institute of Physics, Laboratory of Astrophysics, École Polytechnique Fédérale de Lausanne (EPFL), Observatoire de Sauverny, 1290 Versoix, Switzerland
}

\date{Accepted 2026 June 5. Received 2026 May 27; in original form 2025 November 13}

\pubyear{2026}

\begin{document}
\label{firstpage}
\pagerange{\pageref{firstpage}--\pageref{lastpage}}
\maketitle

\begin{abstract}
Stellar clusters are crucial tools for studying the age, spatial distribution, dynamics, kinematics, and chemical composition of different Galactic stellar populations. In this work, we used red giant stars from open clusters to better understand the extra-mixing process through the CNO abundances and $^{12}$C/$^{13}$C, $^{16}$O/$^{17}$O and $^{16}$O/$^{18}$O isotopic ratios determined using high-quality spectra in the visible and near-infrared regions. We analysed the radial velocities and chemical composition of 22 K-type giant stars from nine open clusters (NGC188, NGC2682, NGC3680, NGC5822, IC4756, NGC6633, NGC3532, NGC6281, and NGC5460). High-resolution and high signal-to-noise spectra of stars in the NGC188 cluster were obtained with the ESPaDOnS spectrograph at the CFHT in the visible region. The stars in the other clusters were observed with the CRIRES spectrograph at the VLT. We used \textsc{iraf} to compute radial velocities and \textsc{Turbospectrum} and \textsc{moog} for the chemical analysis. The values obtained for the radial velocities and abundances of the sample are similar to those found in the literature. The results in the visible and infrared support the occurrence and predicted mass dependence of thermohaline mixing on the red giant branch and of rotation-induced mixing on the main sequence. Variations of the initial abundances of $^{17}$O and $^{18}$O may be needed to explain the dispersion of the oxygen isotopic ratios in red giant stars.
\end{abstract}




\section{Introduction}

Stars in open clusters share the same age, distance, motion, and initial chemical composition, making them ideal objects to understand the chemical and dynamical evolution of the Milky Way. In particular, studying these stellar clusters, we may obtain essential information about the Galactic disk, such as star formation history, chemical enrichment, structure, kinematics, and evolution of the disk, as well as stellar evolution itself \citep[e.g.,][and references therein]{sestito1,sestito3,sestito2,joshi,cantat-gaudin,cantat2020, GAIA-comparacao, soubiran,penhasuarez, donor2020, Loaiza-Tacuri2023, Reyes2024}.

However, from a stellar evolution perspective, there are many open issues that prevent a complete understanding of all the details of how red-giant stars evolve. Standard models, for example, consider spherical symmetry and hydrostatic equilibrium, neglecting effects due to diffusion, rotation, mass loss, overshoot, and magnetic fields \citep{mathis,atomic-diffusion,salaris-cassisi2017,aerts}. In these models, convection is the only mechanism that brings the elements produced in the stellar interior to the surface during the first dredge-up between the main sequence and the red-giant branch (RGB) \citep{iben1,maciel,maciel2,salaris, JoyceTayar2023}.

In the RGB phase, the surface abundances of Li, C, N, and Na, as well as the surface $^{12}$C/$^{13}$C isotopic ratio are changed, as clearly described for the first time by \citet{iben67}. In this case, the $^{12}$C/$^{13}$C isotopic ratio is a measurement of the processing of $^{12}$C to $^{13}$C through the CNO cycle during the main sequence. In low ($0.8\leqslant M \leqslant 2.2\,M_{\astrosun}$) and intermediate ($2.2\leqslant M \leqslant 8.0\,M_{\astrosun}$) mass stars, standard models of stellar evolution show that the surface abundances of $^{13}$C, $^{14}$N, and $^{17}$O increase while the abundances of Li, $^{12}$C, and $^{18}$O decrease as a function of initial stellar mass and metallicity \citep{charbonnel-mixing,lagarde}. However, observations of low-mass RGB stars have shown that there is an additional increase of nitrogen and decrease of carbon and $^{12}$C/$^{13}$C in evolutionary phases after the first dredge-up, contrary to expectations of the standard models. An extra mixing process is needed to explain the differences between the predictions of the standard models and the observations \citep{1998A&A...332..204C, 2000A&A...354..169G, chaname, rodolfo2009, grazina2016, Fraser2022, McCormick2023, Lagarde2024}. This behaviour is observed in field stars and in stars belonging to open and globular clusters, with an indication of a universal process, which is independent of the star environment and common to $\sim$ 95\% of low-mass stars \citep{mixing,charbonnel,lagarde2019}.

\citet{eggleton} proposed that the extra-mixing process is probably caused by the  molecular-weight inversion created by the $^3$He($^3$He,2p)$^4$He reaction in the outer part of the hydrogen-burning shell in RGB stars. In this nuclear reaction, two nuclei are transformed into three, and the local mean molecular weight decreases when moving from the stellar surface towards the centre \citep{1972ApJ...172..165U}. \citet{charbonnel-zahn} identified the mixing mechanism as the double-diffusive instability called thermohaline diffusion \citep{kippenhahn}. 

Thermohaline diffusion is a double-diffusive instability that occurs when there is a combination of gradients in chemical composition and temperature. Heat acts as a stabilising agent that diffuses more rapidly than the destabilising agent, the local mean molecular weight \citep{denissenkov,karakas-lattanzio,salaris-cassisi2017}. \citet{charbonnel-mixing} showed that the thermohaline mixing explains simultaneously the behaviour of $^{12}$C/$^{13}$C, [N/C] and Li in low mass RGB stars \citep[see also][]{rodolfo2009,grazina2013,grazina2015,grazina2016,bagdonas,drazultimo} while simultaneously providing an explanation for the evolution of the $^3$He abundance in the Galaxy \citep{eggleton,lagarde2}. However, there are doubts about whether the efficiency of thermohaline convection is enough to explain the observations \citep{2010A&A...521A...9C,Brown2013,Henkel2017,TayarJoyce22, McCormick2023}. The possible interaction between thermohaline convection and other magneto-hydrodynamical processes is still under investigation \citep{Maeder2013,Garaud2019,2019ApJ...870L...5H, Fraser2024}. 

Carbon isotopic ratios and carbon and nitrogen abundances are sensitive to stellar evolution processes. However, the abundance of oxygen (dominated by $^{16}$O) remains approximately constant since the formation of the star. Therefore, oxygen abundances are useful to study and trace Galactic chemical evolution \citep{grazina2016,magrini,franchini}. On the other hand, the minor isotopes of oxygen ($^{17}$O and $^{18}$O) can still be useful diagnostics of stellar evolution \citep{Dearborn1992}.

Standard stellar evolution models indicate that the isotopic ratios of oxygen change only in deep regions of stars. During hydrogen burning, $^{17}$O is formed from $^{16}$O through the ON cycle (reactions $^{16}$O(p,$\gamma$)$^{17}$F($e^+,\nu$)$^{17}$O), and thus the $^{16}$O/$^{17}$O isotopic ratio is a measurement of how much processing happened to $^{16}$O \citep{1997A&AS..123..241F,referencia-esquema,lebzelter2015}. As discussed in \citet{charbonnel-mixing}, rotation-induced mixing during the main sequence is expected to lower the post dredge-up values of $^{16}$O/$^{17}$O and $^{16}$O/$^{18}$O isotopic ratios, compared to standard models. Otherwise, it is expected that the thermohaline mixing should weakly affect $^{16}$O/$^{17}$O surface ratio and produce a small increase of the $^{16}$O/$^{18}$O ratio when the stars cross the luminosity bump on the RGB. The carbon and oxygen isotopic ratios are expected to change from approximately solar values\footnote{Solar values of $^{12}$C/$^{13}$C, $^{16}$O/$^{17}$O and $^{16}$O/$^{18}$O isotopic ratios are 89, 2700 and 489, respectively \citep{lodders}.} to values $^{12}$C/$^{13}$C $\lesssim20$, $^{16}$O/$^{17}$O $\sim500$ and $^{16}$O/$^{18}$O $\gtrsim600$ while the star advances in the RGB phase \citep{charbonnel-mixing,Halabi2015}. Therefore, and in addition to the $^{12}$C/$^{13}$C isotopic ratio and the CNO abundances, the oxygen isotopic ratios $^{16}$O/$^{17}$O and $^{16}$O/$^{18}$O can be used as tracers of the mixing phenomenon and of its dependence on stellar mass.

To achieve a better understanding of the extra-mixing phenomenon, it is still necessary to improve observations, both in terms of quantity and quality, to include stars in a substantial range of mass and metallicity \citep{grazina2015,Fraser2022, Lagarde2024}. As open clusters have a wide range of ages, Galactocentric distances, and masses, they are ideal targets to assemble a sample of red giants to study the extra-mixing process by analysing stellar parameters, metallicities, isotopic ratio, and [X/Fe] abundance ratios of different elements \citep[e.g.,][]{grazina2015,penhasuarez,Magrini2021,Boesgaard2022}.

As few open clusters have heavily populated RGBs for a robust study of stellar evolutionary properties, studying the extra-mixing phenomenon in stars with near-solar metallicity requires the combination of observations from objects in different clusters. Stars in open clusters share the same age and initial chemical composition. Moreover, precise masses can be determined by isochrone fitting. In addition, it is possible to establish their evolutionary stage using the cluster's colour-magnitude diagram (CMD). Thus, the main goal of this paper is to perform a spectroscopic characterisation of 22 RGB stars from nine open clusters. We investigated the extra mixing phenomenon through CNO abundances and $^{12}$C/$^{13}$C, $^{16}$O/$^{17}$O and $^{16}$O/$^{18}$O isotopic ratios using high-resolution spectra in visible and near-infrared regions. We determine stellar atmospheric parameters, chemical abundances of several species and radial velocities, which all provided constraints to ascertain the membership of the stars to the analysed clusters.

In Section 2, we present the observations and the data reduction process. In Section 3, the methods for the determination of the radial velocity, atmospheric parameters, and chemical abundances are presented. In Section 4, the results of the analysis and their implications are presented and discussed. Finally, in Section 5 we draw our conclusions.

\section{Observations and Data Reduction}\label{sec:observacao}

Our sample consists of high-resolution spectra of 22 red giant stars in 9 open clusters (NGC188, NGC2682, NGC3680, NGC5822, IC4756, NGC6633, NGC3532, NGC6281, and NGC5460). The observations are described in Table \ref{observacao}, which includes the exposure time, signal-to-noise ratio (S/N), and the date of the spectrum observation. The table also includes the evolutionary stage estimated in this work (see Section \ref{sec:analysis}) and binary star information taken from the literature. We numbered the NGC188 stars analysed here, starting from the upper RGB toward the lower RGB, passing through first the clump and then the bump (Fig.~\ref{cmd}).

Spectra of 10 stars in the field of the NGC188 cluster were obtained at visible wavelengths ($3690-10480$ \AA). The rest of the sample (12 stars) was observed in the $M$ band of the near-infrared ($45615-46784$ \AA). Observations of NGC188 stars were conducted using the Echelle SpectroPolarimetric Device for the Observation of Stars \citep[ESPaDOnS,][]{espadons} on the 3.6 m Canada-France-Hawaii Telescope (CFHT) at Mauna Kea, Hawaii, USA. The spectra have $R=68\,000$, S/N $\sim15-42$, and were obtained in `star+sky' mode during a few nights in September and December 2009. The detector was a CCD with 2048x4500 pixels with a size of 13.5 $\mu$m. NGC188 data reduction was carried out using the Upena\footnote{ \url{http://www.cfht.hawaii.edu/Instruments/Upena/}} pipeline (version 1.0), through the reduction package Libre-ESpRIT \citep[version 2.12, 2006 April 20;][]{donati-upena}. The pipeline gives reduced and normalised spectra, and we remove the telluric lines using the \textsc{iraf}\footnote{\url{http://www.iraf.noao.edu/}} task \texttt{telluric}.

Infrared (IR) observations were made with the CRyogenic high-resolution InfraRed Echelle Spectrograph \citep[CRIRES,][]{crires} at the 8m Very Large Telescope (VLT), at Cerro Paranal, Chile. The spectra have $R=100\,000$, S/N $\sim83-152$, and were obtained during nights of April (run A) and May (run B) 2012. CRIRES was equipped with an array of four Aladin III InSb detectors that provided 4096x512 pixels of 27 $\mu$m. The near-IR sample was reduced through the EsoRex\footnote{\url{https://www.eso.org/sci/software/cpl/esorex.html}} pipeline (version 3.13.2) made available by the European Southern Observatory (ESO). Continuum normalisation was done using the \textsc{iraf} task \texttt{continuum}. As there are many telluric lines of molecules of CO, N$_2$O, H$_2$O and N$_2$ in the observed region, ESO recommends using the Molecfit\footnote{\url{ https://www.eso.org/sci/software/pipelines/skytools/molecfit}} software \citep[version 1.5.9;][]{smette2015,kaush2015} to remove telluric contamination, which was thus the procedure we used in this case.

\begin{table*}
    \caption{Observational data.}
    \begin{threeparttable}
    \begin{tabular}{lcccccccc}\hline\hline
    Star & 1 & 2 & 3 & 4 & 5 & 6 & 7 & Ref. \\ \hline
    \multicolumn{9}{c}{ ESPaDOnS@CFHT } \\ \hline
    NGC188-3018 & 1 & 11.368 & 600 & 36 & 2009 Nov 24 & upper RGB or early-AGB & no & a \\
    NGC188-1001 & 2 & 11.922 & 600 & 25 & 2009 Nov 30 & upper RGB or early-AGB & no & a \\
    NGC188-2072 & 3 & 12.416 & 900 & 30 & 2009 Sept 09 & upper RGB & yes & a, b \\
    NGC188-2026 & 4 & 13.618 & 2700 & 23 & 2009 Dec 01 & clump & no & a \\
    NGC188-1116 & 5 & 13.933 & 2700 & 26 & 2009 Sept 09 & clump & yes & a, b, c \\
    NGC188-3140 & 6 & 13.927 & 3600 & 20 & 2009 Nov 30 & before bump & no & a \\
    NGC188-1061 & 7 & 14.158 & 3600 & 29 & 2009 Dec 01 & --- & no & a \\
    NGC188-2187 & 8 & 14.074 & 3600 & 42 & 2009 Dec 02 & before bump & no & a \\
    NGC188-1006 & 9 & 14.886 & 3525 & 15 & 2009 Sept 30, Oct 04, and Nov 29 & lower RGB & no & a \\
    NGC188-2194 & 10 & 14.950 & 3525 & 18 & 2009 Sept 09 & lower RGB & yes & a, b \\ \hline
    \multicolumn{8}{c}{ CRIRES@VLT } \\ \hline
    NGC2682-MMU6495 & 11 & 5.933 & 300 & 157 & 2012 Apr 08 and May 01 & upper RGB or early-AGB & no & d \\
    NGC3680-44 & 12 & 7.165 & 400 & 122 & 2012 May 02 & upper RGB or early-AGB & no & e \\
    NGC5822-1 & 13 & 5.966 & 400 & 137 & 2012 Apr 09 & upper RGB or early-AGB & no & f \\
    NGC5822-240 & 14 & 6.254 & 400 & 169 & 2012 May 02 & upper RGB or early-AGB & no & f \\
    IC4756-69 & 15 & 6.537 & 400 & 112 & 2012 May 02 & clump & yes & f \\
    NGC6633-78 & 16 & 4.087 & 60 & 160 & 2012 May 02 & upper RGB or early-AGB & no & e \\
    NGC6633-100 & 17 & 5.674 & 400 & 185 & 2012 May 02 & lower RGB or clump & no & g \\
    NGC3532-MMU19 & 18 & 5.463 & 300 & 106 & 2012 Apr 09 & clump & no & a\\
    NGC3532-MMU649 & 19 & 5.550 & 300 & 133 & 2012 Apr 09 & --- & no & a \\
    NGC6281-3 & 20 & 5.357 & 300 & 107 & 2012 Apr 09 & lower RGB or clump & no & a \\
    NGC6281-4 & 21 & 5.455 & 300 & 193 & 2012 May 02 & lower RGB or clump & no & a \\
    NGC5460-MMU17 & 22 & 4.930 & 100 & 114 & 2012 May 02 & clump & no & a \\ \hline
    \end{tabular}
    Labels: (1) Star identification (ID) in this work, (2) Magnitude in $V$-band for objects 1 to 10 \citep{ngc188}, and $K_s$-band for objects 11 to 22 \citep{2MASS}, (3) Exposure time (s), (4) S/N measured at $\sim6073$ \AA\,for objects 1 to 10 (except object 7, at 6075 \AA) and at $\sim46720$ \AA\,for all objects 11 to 22 , (5) Observation dates, (6) Evolutionary stage visually estimated from their position in the CMD, except for objects 7 (NGC188-1061) and 19 (NGC3532-MMU64) that are non-members of their clusters (see Section \ref{sec:analysis}) and (7) Binary.\\
    References regarding the binary nature: (a) \citet{geller2008}, (b) \citet{geller2009}, (c) \citet{jacobson2011}, (d) \citet{mermilliod2007}, (e) \citet{mermilliod95}, (f) \citet{mermilliod90} and (g) \citet{mermilliod89}.
    \label{observacao}
    \end{threeparttable}
\end{table*}

\section{Analysis}

\subsection{Radial velocities}

Radial velocities were determined by measuring the wavelength of selected spectral lines in the observations and comparing the measured value to their rest-frame position from theory and experiments. These reference line lists were created with the help of atomic and molecular databases, i.e. the National Institute of Standards and Technology (NIST) Atomic Spectra Database \citep{NIST_ASD}\footnote{\url{https://www.nist.gov/pml/atomic-spectra-database}} and the HIgh-resolution TRANsmission molecular absorption database \citep[HITRAN,][]{hitran}\footnote{\url{https://www.hitran.org}}, literature \citep{goorvitch}, the Arcturus spectral atlas in the visible \citep{hinkle-visivel}, and the Arcturus and 10 Leo atlases in the IR \citep[][]{hinkle-infra,10Leo}. Arcturus and 10 Leo are K-type stars well studied in the literature. We selected well-defined clean absorption lines with easy identification (strong lines, doublets, and triplets) and without telluric contamination in the wavelength regions considered here (3690 to 10480 \AA, in the visible and 45615 to 46784 \AA, in the IR). For the NGC188 subsample, we created a list with 297 lines. For the IR subsample, the list has 61 lines. We computed the radial velocity using the \textsc{iraf} \texttt{rvidlines} task. The heliocentric correction for the IR subsample was obtained through the Astropy\footnote{\url{http://www.astropy.org/}} package \citep[][]{astropy}. The final value of the radial velocity of a cluster is the mean of the values of each star belonging to that cluster.

\subsection{Cluster parameters and evolutionary stages}\label{sec:analysis}

\begin{figure}
    \includegraphics[angle=-90,width=\columnwidth]{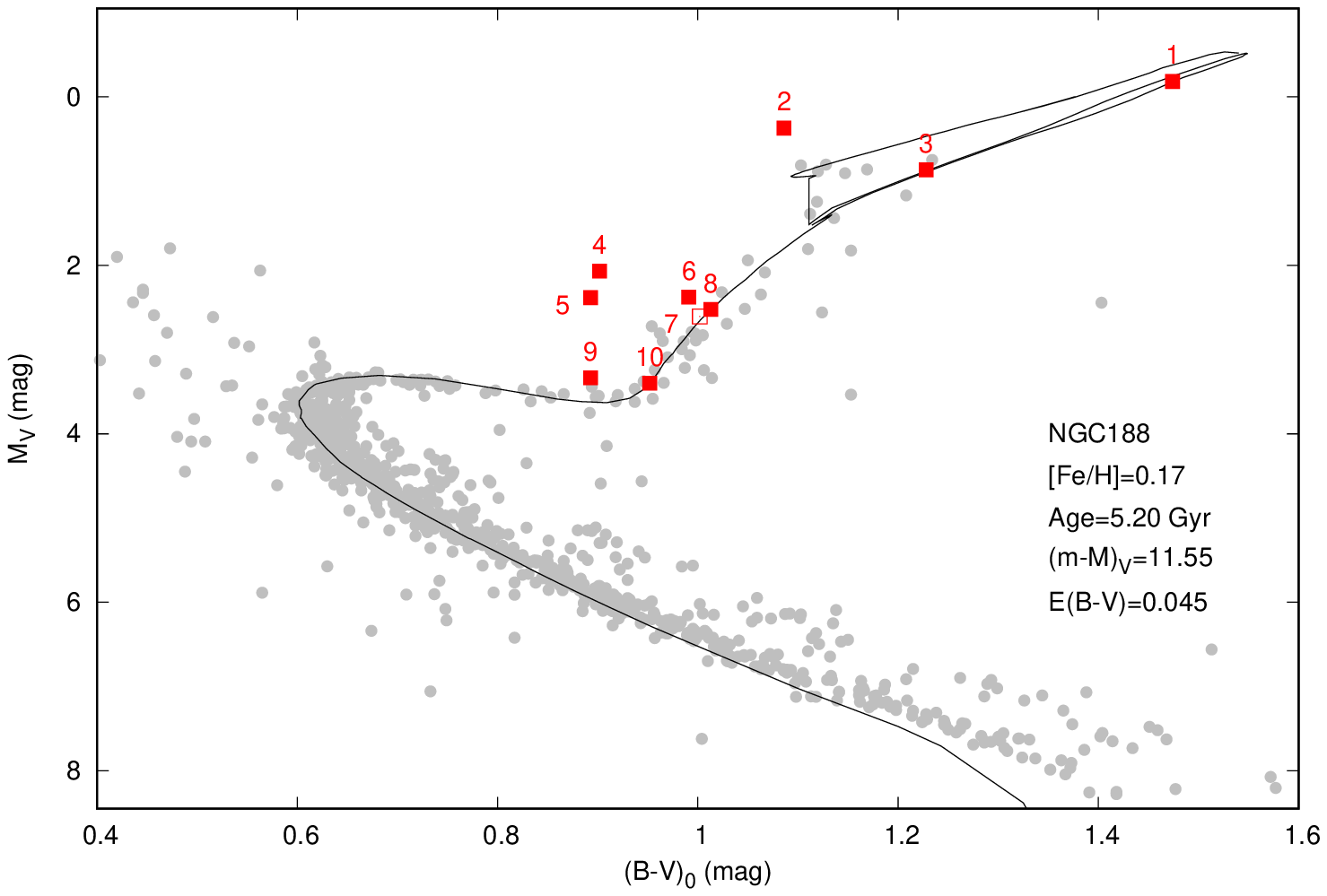}
    \caption{Isochrone fitting using PARSEC \citep{parsec} isochrone to the CMD of the NGC188 cluster, with data from \citep{ngc188}, used to determine colour excess $E(B-V)$, distance modulus $(m-M)_V$, age, [Fe/H], turn-off mass $M_{\textrm{TO}}$, and the evolutionary stage of the stars. Filled squares indicate cluster members, while open squares indicate non-members.}
    \label{cmd}
\end{figure}

We use the isochrone fitting method to obtain the cluster parameters. These parameters were later used as input to calculate the stellar atmospheric parameters from photometry. We use the \textsc{parsec} isochrones \citep{parsec} in the colour-magnitude diagram (CMD) in a plot of the colour $(B-V)_0$ and the absolute magnitude in the $V$-band, $\mathcal{M}_V$. By visual inspection, we estimate the colour excess $E(B-V)$, distance modulus ($m-\mathcal{M}$)$_V$, age, metallicity\footnote{In this work, we consider iron abundance [Fe/H] as indicator of metallicity (see Section \ref{sec:metallicity}).} [Fe/H], turn-off mass $M_{\textrm{TO}}$ and the evolutionary stage of the stars. The cluster parameters that we obtained can be found in Table \ref{fotometria-aglomerados}, and Figure \ref{cmd} shows an example of fit for the NGC188 cluster.

We estimate the mass of the stars in the cluster using the turn-off mass. As a first approximation, we assumed all stars have the same mass and that little or no mass loss happened between the main-sequence turn-off and the current evolutionary stage of the stars. That is a good approximation, as no strong mass loss is expected during the RGB phase \citep{gilroy1991,reddy,stello}.

\begin{table}
    \caption{Clusters parameters obtained through isochrone fitting.}
    \begin{tabular}{cccccc}\hline\hline
    Open & [Fe/H] & Age & $(m-\mathcal{M})_V$ & $E(B-V)$ & $M_{\textrm{TO}}$ \\
    Cluster & (dex) & (10$^9$ yr) & (mag) & (mag) & ($M_{\astrosun}$) \\ \hline
    NGC188 &  +0.17 & 5.20 & 11.55 &  0.045 &  1.45 \\
    NGC2682 &  +0.03 & 2.56 & 9.85 &  0.077 &  1.30 \\
    NGC3680 &  +0.04 & 1.79 & 10.15 &  0.065 &  1.50 \\
    NGC5822 &  +0.02 & 0.96 & 9.75 &  0.095 &  1.53 \\
    IC4756 &  +0.07 & 0.70 & 8.95 &  0.220 &  2.16 \\
    NGC6633 &  +0.10 & 0.43 & 8.60 &  0.260 &  2.55 \\
    NGC3532 &  +0.00 & 0.31 & 8.59 &  0.025 &  2.89 \\
    NGC6281 &  +0.00 & 0.21 & 9.15 &  0.195 &  3.21 \\
    NGC5460 &  $-$0.15 & 0.20 & 9.35 &  0.135 &  3.30 \\ \hline
    \end{tabular}
    $M_{\textrm{TO}}$: Turn-off mass; $(m-\mathcal{M})_V$: Distance modulus.
    \label{fotometria-aglomerados}
\end{table}

We estimated the evolutionary stage of the red giants in each cluster considering their relative positions in the CMD. The results are shown in Table \ref{observacao}. According to Figure \ref{cmd}, most of the NGC188 stars analysed here (objects 1 to 10) were identified as belonging to the RGB (either in its lower or upper parts or at the bump position). Although stars 2 (NGC188-1001), 4 (NGC188-2026), and 5 (NGC188-1116) lie far from the cluster's isochrone, the analysis of radial velocities and proper motion performed by \citet[][]{geller2008} indicate that these objects are members of the cluster with a probability of 95\%. The authors also commented that, in previous photometric studies, the red giant branch of NGC188 is identified as broad and not well defined. However, the reasons behind this are uncertain. The \emph{Gaia} DR2 data for the NGC188 star sample analysed in this work indicated that all objects are cluster members, except object 7 (NGC188-1061) \citep{cantat-gaudin2020}.

\citet{harris-mcclure} analysed the radial velocity and spectroscopy of object 2 (NGC188-1001) and argued that the object is a fast rotating star with X-ray emission, identified as a FK Comae star (FK Com), presenting strong and variable chromospheric emission in X-ray \citep{belloni}. For object 5 (NGC188-1116), \citet{geller2009} argued that it is a photometric variable star, probably an eclipsing binary system. We further discuss the binary nature of that object in Section \ref{sec:binaries}. Also, \emph{Gaia} DR3 classifies both objects as RS Canum Venaticorum (RS CVn) type variable \citep[][]{gaiadr3}. RS CVn objects are close binary systems; however, object 2 (NGC188-1001) is identified as a single star in the literature and in the recent analysis of the \textit{Gaia} DR3 astrometric parameters (see the normalised unit weight error (RUWE) parameter in Table \ref{resultado1} in Appendix \ref{sec:ew2}). Therefore, since these objects are variable stars, this could explain their location in the CMD.

For objects 4 (NGC188-2026) and 5 (NGC188-1116), the \emph{Gaia} DR3 gives parallaxes $\pi=0.509\pm 0.011$~mas and $\pi=0.501\pm 0.013$~mas, respectively \citep{gaiadr3}. These values correspond to the distances $d=1965\pm 52$~pc and $d=1996\pm 52$~pc, respectively. This is consistent within $2\,\sigma$ with the mean parallax of the cluster obtained by \citet{cantat-gaudin2020}, and with the distance modulus given in Table~\ref{fotometria-aglomerados}.

The stars in the remaining clusters (objects 11 to 22) are mostly in the upper RGB or at the red clump (the core He burning phase). These different evolutionary stages are essential for the characterisation of the extra-mixing process along with the giant evolution stages.

\subsection{Binaries}\label{sec:binaries}

Table \ref{observacao} indicates that some stars in the sample were identified in the literature as binary stars. NGC188 has a large population of binary stars \citep[$\gtrsim 100$ objects;][]{geller2009,geller2012}. It is important to characterise the stellar sample analysed in this work, as the evolution and analysis of possible interacting binaries is different from the case of single stars \citep{karakas-lattanzio}.

Part of our NGC188 sample is composed of binary stars (see Table \ref{observacao}). Non-interacting binaries evolve as single stars \citep{karakas-lattanzio}, and \citet{penhasuarez} indicate that we do not expect a significant difference between abundances of non-interacting binary and single stars of the same open cluster, and we can analyse the sample ignoring possible differences related to the sample binarity. Therefore, we kept objects 3 (NGC188-2072) and 10 (NGC188-2194) in our analysis.

\citet{geller2009} calculated the orbital period of several spectroscopic binaries in NGC188. Object 3 (NGC188-2072) is a single-lined spectroscopic binary (SB1) with a period of $344.01\pm0.05$ days and masses for the primary and secondary stars of $M_1<1.14\,M_{\astrosun}$ and $M_2<1.10\,M_{\astrosun}$, respectively. Object 5 (NGC188-1116) is a double-lined spectroscopic binary (SB2) with a period of $35.178\pm0.005$ days and masses for the primary and secondary stars of $M_1=1.14\,M_{\astrosun}$ and $M_2=1.09\,M_{\astrosun}$, respectively. \cite{vats2018} argue that the location of that object in the CMD (see Figure \ref{cmd}) could be explained by the combination of its binary components (an unevolved red giant and a less-evolved member).

\citet{gondoin2005} detected that object 5 (NGC188-1116) presents variability in X-ray emission, indicating a possible chromospheric activity related to fast stellar rotation with turbulent motions. These motions generate interactions of induced magnetic fields inside the star and emission in X-rays, classifying the star as a RS CVn system \citep{berdyugina,gaiadr3, Song2023}. The X-ray emission could be explained by the faster rotation of one or both stars in that binary compared to single stars in the cluster \citep{vats2018}. During our spectroscopic analysis, we did not detect the spectrum of the companion star in object 5 (NGC188-1116), probably because of its rapid rotation.

Object 10 (NGC188-2194) is an SB1 with a period of $1029\pm7$ days and masses for the primary and secondary stars of $M_1=1.13\,M_{\astrosun}$ and $M_2<0.64\,M_{\astrosun}$, respectively. The NGC188 cluster contains a rich binary population and a variety of candidates for post-interaction objects \citep{geller2009,geller2012}.

\citet{mermilliod90,mermilliod2007} identified object 15 (IC4756-69) as a spectroscopic binary with a large orbital period (1994 days) and masses for the primary and secondary stars of $M_1=2\,M_{\astrosun}$ and $M_2=0.54\,M_{\astrosun}$, respectively. The secondary star is probably a white dwarf \citep{rodolfo2009,vanderswaelmen}. Recently, \emph{Gaia} DR3 also classified this object as a non-single star from the analysis of its proper motion \citep{gaiadr3}.

\subsection{Atmospheric parameters}

\subsubsection{Photometry}\label{sec:photometry}

We first estimated the stellar atmospheric parameters using photometric calibrations. The photometric temperature was obtained using calibrations from \citet[][see also the erratum \citeauthor{alonso-erratum} \citeyear{alonso-erratum}]{alonso} for the colours $V-K$ and $J-K$. These colours are good indicators of temperature in giants of type G and K \citep{alonso,sales-silva,Huang}. These calibrations are weakly dependent on metallicity and luminosity. For example, considering the calibration of $V-K$, an error of 0.5 dex in [Fe/H] implies an error of about 0.7\% in temperature. Also, an error of 0.05 in magnitude implies a mean error of 0.7 to 1.0\%. Using the $J-K$ calibration, an error of 0.03 mag implies mean errors of 1.7 to 2.5 per cent in temperature, and this calibration does not depend on [Fe/H] \citep[][]{alonso}.

In addition, to reduce the uncertainty in the photometric temperature, we also adopted the calibrations of \citet{van-Belle} and \citet{Huang} for the $V-K$ colour and of \citet{alonso} for the $B-V$ and $J-H$ colours. We note here that star 16 (NGC6633-78) presents a very large uncertainty in its magnitude ($\sim0.294$ mag) due to the low quality of its photometry \citep[Photometric quality flag Qflg=DDE;][]{2MASS}. As the final estimate of photometric effective temperature ($T_{\textrm{eff}}$), we consider the mean of the values obtained with each calibration.

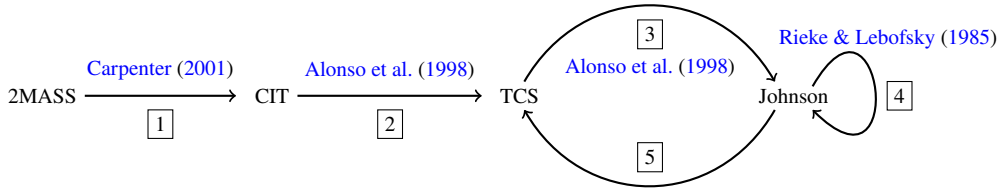
\begin{figure*}
\begin{tikzpicture}[scale=1.25]
\centering
\draw[thick,->] (0,0) node[anchor=east] {2MASS} -- (1.6,0) node [midway,label=above:{\small \cite{CIT}}] {};
\draw[thick,->] (2.25,0) node[anchor=east] {\small CIT} -- (4.2,0) node [midway,label=above:{\small \cite{TCS}}] {};
\node (a) at (4.6,0) {\small TCS};
\node (c) at (5.98,0.3) {\small \cite{TCS}};
\draw[thick,->] (4.65,0.15)  to [out=60,in=120,looseness=1.2] (7.3,0.15);
\draw[thick,<-] (4.65,-0.15)  to [out=-60,in=-120,looseness=1.2] (7.3,-0.15);
\node (a) at (7.5,0) {\small Johnson};
\draw[thick,->] (7.7,0.1)  to [out=60,in=-45, looseness=15] (7.7,-0.15);
\node (c) at (8.5,0.6) {\small \cite{rieke}};
\draw node[draw] at (0.8,-0.3) (1) {1};
\draw node[draw] at (3.23,-0.3) (1) {2};
\draw node[draw] at (5.98,-0.65) (1) {5};
\draw node[draw] at (5.98,0.65) (1) {3};
\draw node[draw] at (8.62,0) (1) {4};
\end{tikzpicture}
\caption{Diagram with steps evolved in the transformations between photometric systems.}
\label{diagrama2}
\end{figure*}

As the calibrations of \citet{alonso} use magnitudes in the system of the Telescopio Carlos S\'anchez (TCS), we need to first transform the magnitudes from the 2MASS system to the TCS system, and then de-redden them, before calculating the temperature. A diagram with the steps involved in the transformations between the photometric systems is shown in Figure \ref{diagrama2}. Transformations were made from the 2MASS system to the Caltech (CIT) system \citep{CIT} and then to the TCS system \citep{TCS}. As extinction expressions are given in the Johnson system by \citet{rieke}, we still need to transform from the TCS system to the Johnson system \citep[][]{TCS} and then return the corrected magnitudes to the TCS system. The mean difference between colours in the 2MASS and TCS systems is about $\sim$ 0.26 mag in $V-K$, $\sim$ 0.08 mag in $J-K$ and $\sim$ 0.05 mag in $J-H$. These differences show that proper transformation is necessary between photometric systems for greater precision in the photometric analysis.

Using the mean photometric effective temperature derived before, we calculated the photometric surface gravity $\log g$ considering the relation with the canonical parameters of the Sun: $T_{\textrm{eff},\astrosun}=5772$ K, $\log g_{\astrosun}=4.44$ dex, $\mathcal{M}_{\textrm{bol},\astrosun}=4.74$ mag \citep[see][]{resolucao-IAU1,resolucao-IAU2}:
\begin{align}
\log g & = 4.44+\log\left(\frac{M}{M_{\astrosun}}\right)+4\log\left(\frac{T_{\textrm{eff}}}{5772}\right)+\nonumber\\
&+\frac{2}{5}[m_V-3.09\,E(B-V)-(m-\mathcal{M})_V+BC_V-4.74],
\end{align}
\noindent where $M$ is the stellar mass in units of solar mass ($M_{\astrosun}$), $m_V$ is the observed star magnitude in the $V$ band, $E(B-V)$ is the colour excess, $(m-\mathcal{M})_V$ is the distance modulus, and $BC_V$ is the bolometric correction in the $V$ band. The distance modulus $(m-\mathcal{M})_V$, the colour excess $E(B-V)$ and the stellar mass were found through isochrone fitting of the cluster CMD. The bolometric correction $BC_V$ was obtained using the relations of \citet[][]{alonso}, considering the range of $T_{\textrm{eff}}$ values and metallicity ([Fe/H]) of the stars.

As photometric microturbulent velocity ($\xi$), we adopt the mean of the values estimated using the relation of \citet[][]{gratton} and the relation obtained in the \emph{Gaia}-ESO Survey \citep{Worley2024}. In these calculations, we used the values of photometric $T_{\textrm{eff}}$ and $\log g$ obtained as described above.

\subsubsection{Spectroscopy}

Spectroscopic atmospheric parameters were calculated for the sample of NGC188 stars using manual measurements of equivalent width (EW) of \ion{Fe}{i} and \ion{Fe}{ii} spectral lines. The analysis was done using local thermodynamic equilibrium (LTE) one-dimensional (1D) plane-parallel model atmospheres calculated with the \textsc{atlas9} code \citep{kurucz} and with the spectral analysis code \textsc{Turbospectrum} \citep{turbospectrum}. We used spectral lines from the line list used in \citet{rodolfo2009}, expanded with lines from \citet{arthur}. For the Fe lines, we used the $\log~gf$ values recommended in the compilation done by the Gaia-ESO Survey \citep{Heiter2021}. In this analysis, we considered lines with excitation potential $\chi$ < 5 eV and 30 < EW < 120 m\AA. The constraint for lines with low excitation potential was applied in order to minimise the dependence with the temperature, considering the LTE approximation used in our models for red giant stars \citep[see][]{Gray2005}. The upper limit in EW is enforced to guarantee that the selected spectral lines present a Gaussian profile (i.e., are not saturated). Furthermore, as the NGC188 stars have spectra with relatively low S/N ($\sim$ 15$-$42, see Table \ref{observacao}), precise EW measurements for weak lines are challenging due to the noise in the spectrum. At this S/N level, the continuum placement becomes uncertain, and fluctuations can mimic weak spectral lines \citep[see ][]{arthur}. Consequently, we excluded spectral lines with EW < 30 m\AA\, to ensure that EW measurements are restricted to lines that can be clearly distinguishable from continuum noise.

\begin{figure}
    \includegraphics[height=\columnwidth, angle=-90]{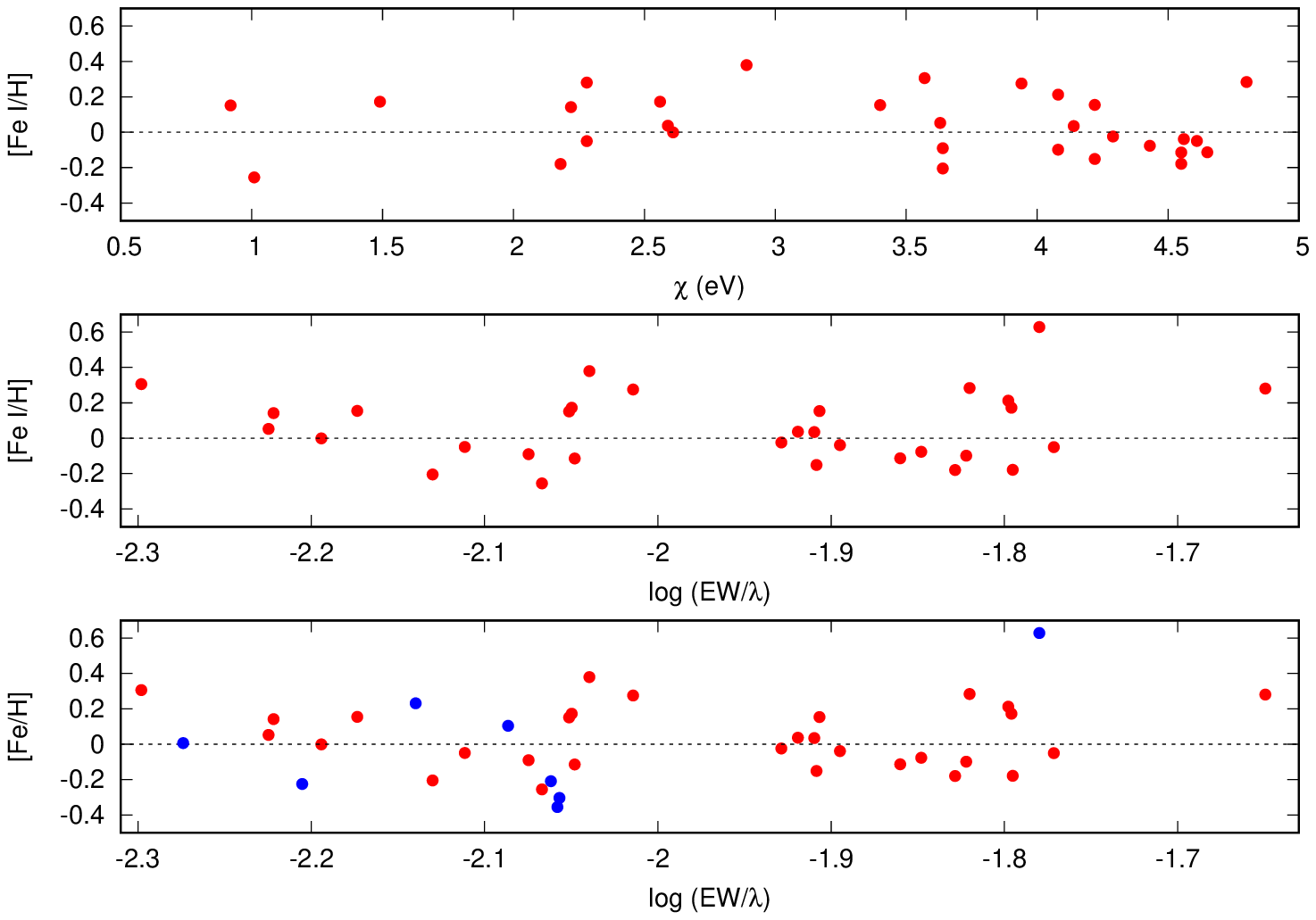}
    \caption{\textit{Top}: iron abundance $\textup{[\ion{Fe}{i}/H]}$ \textit{versus} excitation potential $\chi$ of iron lines of \ion{Fe}{i} (to determine the spectroscopic effective temperature when the trend of the graphic is null). \textit{Middle}: iron abundance \textit{versus} the reduced equivalent width $\log (EW/\lambda)$ (to determine the microturbulent velocity when the trend of the graphic is null). \textit{Bottom}: iron abundance of \ion{Fe}{i} (red dots) and \ion{Fe}{ii} (blue dots) \textit{versus} the reduced equivalent width $\log (EW/\lambda)$ (to determine the spectroscopic surface gravity, when the trend of the graphic is null) to object 6 (NGC188-3140).}
    \label{resultado-turbo}
\end{figure}

The spectroscopic effective temperature, $T_{\textrm{eff}}$, was obtained by means of the excitation equilibrium of \ion{Fe}{i} lines, i.e., removing the trend between the \ion{Fe}{i} abundance and the excitation potential of the lines. The microturbulent velocity, $\xi$, was found by requiring that there should be no trend in a plot of the \ion{Fe}{i} abundances as a function of the reduced equivalent width, $\log (EW/\lambda)$. The value of $\log g$ was obtained considering the ionisation equilibrium of Fe, that is, forcing the abundances of \ion{Fe}{i} and \ion{Fe}{ii} to be equal. The final metallicity [Fe/H] is then found, when the Fe abundance value given by the lines and the value used to build the model atmosphere are the same. These steps were repeated iteratively until all conditions were satisfied using a two sigma clipping to remove outliers. The final result obtained for an example star can be seen in Figure \ref{resultado-turbo}. Before applying the same method to our data, we used it to analyse the atmospheric parameters of the high-quality spectrum of Arcturus ($R\sim150\,000$, S/N $\sim1000$) provided by \citet{hinkle-visivel}. Arcturus is a reference star of the same spectral type as the stars in our sample and has high-quality parameters and abundances available in the literature \citep[e.g.,][]{alan-arcturus,ramirez2011,fanelli2020}. The comparison between our results and those of the literature for Arcturus is shown in Table \ref{comparacao-arcturus}. It can be seen that the values obtained here for the atmospheric parameters are similar to those in the literature. The differences are of the same order as the uncertainties.

\begin{table*}
    \caption{Comparison between atmospheric parameters and abundances for Arcturus found in this work and the literature.}
    \begin{threeparttable}
    \begin{tabular}{lccccc} \hline \hline
    & This work & 1 & 2 & 3 & 4$^{\star}$\\ \hline
    T$_{\textrm{eff}}$ (K) & 4244 $\pm$ 200 & 4280 $\pm$ 75 & 4286 $\pm$ 30 & 4290 $\pm$ 50 & 4283 $\pm$ 33\\
    $\log$ g (dex) & 1.50 $\pm$ 0.40 & 1.69 $\pm$ 0.30 & 1.66 $\pm$ 0.05 & 1.50 $\pm$ 0.10 & 1.67 $\pm$ 0.06 \\
    $\textup{[Fe/H]}$ (dex) & $-$0.58 $\pm$ 0.06 & $-$0.49 $\pm$ 0.05 & $-$0.52 $\pm$ 0.04 & $-$0.50 $\pm$ 0.07 & $-$0.57 $\pm$ 0.01 \\
    $\xi$ (km/s) & 1.66 $\pm$ 0.17 & 1.74 $\pm$ 0.20 & 1.74 & 1.7 $\pm$ 0.1 & 1.60 $\pm$ 0.05 \\ \hline
    $\textup{[C/Fe]}$ (dex) & 0.10 & --- & 0.43 $\pm$ 0.07 & 0.15 $\pm$ 0.09 & 0.14 $\pm$ 0.05\\
    $\textup{[N/Fe]}$ (dex) & 0.40 & --- & 0.05 $\pm$ 0.03 & 0.37 $\pm$ 0.13 & ---\\
    $\textup{[O/Fe]}$ (dex) & 0.34 & 0.41 & --- &0.60 $\pm$ 0.17 & 0.49 $\pm$ 0.02 \\
    $\textup{[Na/Fe]}$ (dex) & 0.21 $\pm$ 0.12 & 0.16 $\pm$ 0.12 & 0.11 $\pm$ 0.03 & --- & 0.09 $\pm$ 0.03 \\
    $\textup{[Si/Fe]}$ (dex) & 0.48 $\pm$ 0.07 & 0.26 $\pm$ 0.05 & 0.33 $\pm$ 0.04 & --- & 0.32 $\pm$ 0.02 \\
    $\textup{[Ca/Fe]}$ (dex) & 0.05 $\pm$ 0.11 & 0.04 $\pm$ 0.05 & 0.11 $\pm$ 0.04 & --- & 0.03 $\pm$ 0.02 \\
    $\textup{[Sc/Fe]}$ (dex) & 0.06 $\pm$ 0.09 & --- & 0.19 $\pm$ 0.06 & --- & 0.10 $\pm$ 0.04 \\
    $\textup{[Ti/Fe]}$ (dex) & 0.18 $\pm$ 0.16 & 0.21 $\pm$ 0.03 & 0.24 $\pm$ 0.05 & --- & 0.20 $\pm$ 0.02 \\
    $\textup{[V/Fe]}$ (dex) & 0.24 $\pm$ 0.19 & --- & 0.20 $\pm$ 0.05 & --- & --- \\
    $\textup{[Cr/Fe]}$ (dex) & 0.05 $\pm$ 0.10 & --- & $-$0.05 $\pm$ 0.04 & --- & $-$0.05 $\pm$ 0.02 \\
    $\textup{[Co/Fe]}$ (dex) & 0.20 $\pm$ 0.08 & --- & 0.09 $\pm$ 0.04 & --- & 0.02 $\pm$ 0.02 \\
    $\textup{[Ni/Fe]}$ (dex) & 0.09 $\pm$ 0.06 & --- & 0.06 $\pm$ 0.03 & --- & 0.04 $\pm$ 0.01 \\
    $\textup{[Y/Fe]}$ (dex) & $-$0.20 $\pm$ 0.12 & --- & --- & --- & $-$0.11 $\pm$ 0.08 \\ \hline
    $^{12}$C/$^{13}$C (dex) & 7 & --- & --- & 9 $\pm$ 2& --- \\
    $^{16}$O/$^{17}$O (dex) & 3500 & --- & --- & 3030 $\pm$ 530 & --- \\
    $^{16}$O/$^{18}$O (dex) & 1800 & --- & --- & 1660 $\pm$ 400 & --- \\ \hline
    \end{tabular}
    $^{\star}$ Chemical abundances from optical lines.\\
    References: (1) \cite{alan-arcturus}, (2) \cite{ramirez2011}, (3) \cite{abia2012}, and (4) \cite{fanelli2020}.
    \end{threeparttable}
    \label{comparacao-arcturus}
\end{table*}

\begin{figure}
    \includegraphics[angle=-90,width=\columnwidth]{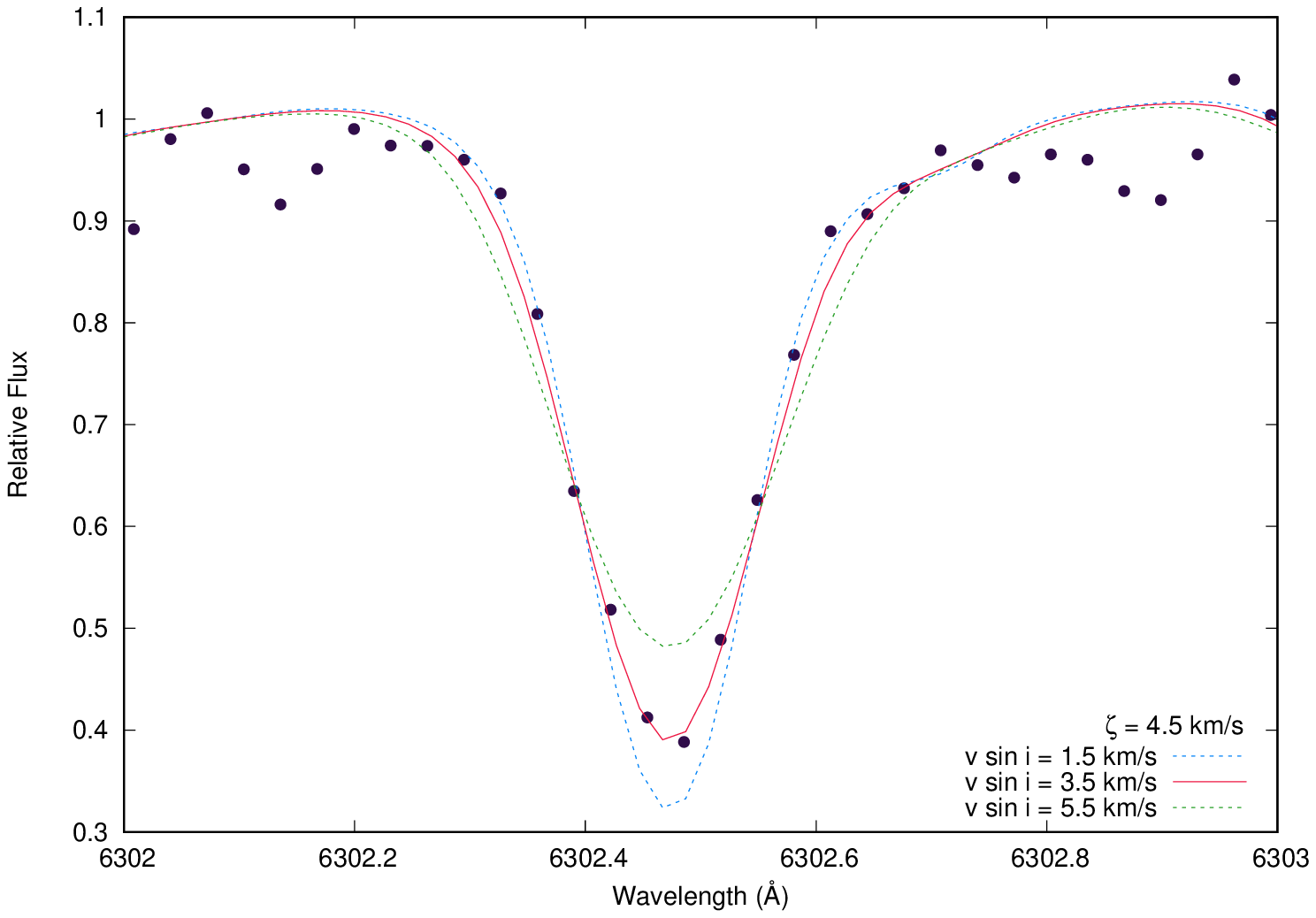}
    \includegraphics[angle=-90,width=\columnwidth]{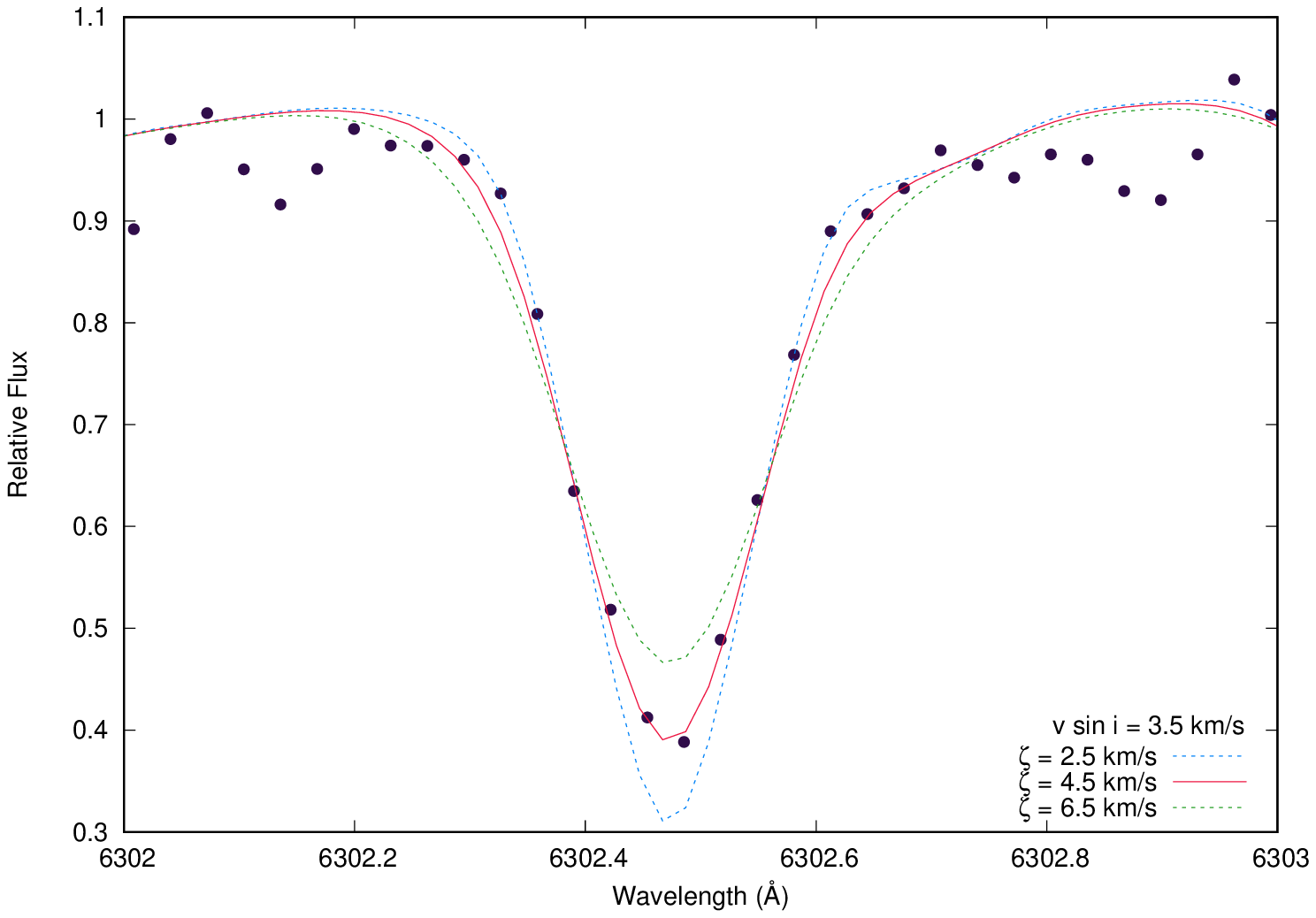}
    \caption{Observed (black dots) and synthetic (blue, red, and green dashed lines) spectra used to determine the projected rotational velocity $v\sin i$ (\textit{top}) and macroturbulence velocity $\zeta$ (\textit{bottom}) for object 8 (NGC188-2187). The best fit curve is the red one.}
    \label{parameters}
\end{figure}

In order to reproduce the spectral broadening, which is needed to determine the chemical abundances, we also determined the projected rotational velocity $v \sin i$ and the macroturbulence velocity $\zeta$. They were determined using the same spectral synthesis technique as used by \citet[][]{sales-silva,dasilveira,penhasuarez,martinez}. The method uses the \ion{Fe}{i} line at 6302.5 \AA, which is unblended, to determine the broadening parameters. For this analysis using spectral synthesis we used the \textsc{moog} software. We first analysed the spectrum of Arcturus and adopted the values derived for it ($v\sin i=4.5$ \kms, $\zeta=3.5$ \kms) as the first guess for the parameters of the remaining sample, as Arcturus is a standard K-type giant star. One parameter is determined while keeping the other constant. Afterwards, the value of the first parameter is fixed and the second is varied. The steps were repeated iteratively until a synthetic spectrum that best fits the observed spectrum was found. Figure \ref{parameters} shows an example of the determination of the projected rotational velocity $v\sin i$ and the macroturbulence velocity $\zeta$ for object 8 (NGC188-2187).

For the sample of stars with spectrum in the near IR (objects 11 to 22), the atmospheric parameters were obtained using only photometry. In the region of the observations ($\sim$ 4.6 $\mu$m), we cannot identify iron lines and therefore we cannot apply the same method used in the analysis of the visible region (i.e., the excitation and ionisation equilibrium of the Fe lines).

\subsection{Abundances}

Adopting the spectroscopic atmospheric parameters found as described above, we attempted to obtain abundances of the elements Na, Mg, Si, Ca, Sc, Ti, V, Cr, Co, Ni, Y, and Ce using the method of equivalent widths with \textsc{moog} for the NGC188 stars. As before, we used a combination of the line lists used in \citet{rodolfo2009} and \citet{arthur}, in particular because \citet{rodolfo2009} did not take into account the hyperfine structure (HFS) for the elements Sc, Co, and V. We included data on hyperfine splitting when possible from the literature \citep[such as][]{lawler89,kurucz95,lawler2014,lawler2015,lawler2019}. We also note that, as the EW of the Mg lines turned out to be too large (EW > 120 m\AA, see Table \ref{ew-tabela} in Appendix \ref{sec:ew2} for details), and thus to have line profiles that are non-Gaussian, we did not calculate its abundance.

The C, N and O abundances and the $^{12}$C/$^{13}$C isotopic ratios were obtained by spectral synthesis using \textsc{moog}. We used the values of projected rotational velocity $v\sin i$ and macroturbulence velocity $\zeta$ as fixed parameters to reproduce the spectral broadening. We adopted the solar abundances\footnote{$\log \textrm{A(X)}=12+\log\Big(\frac{n_X}{n_H}\Big)$, where A(X) is the abundance of the element X and $n_X$ is its numeric density.} from \citet{asplund}, $\log$ A(C) = 8.43, $\log$ A(N) = 7.83 and $\log$ A(O) = 8.69.

\begin{figure*}
    \includegraphics[angle=-90,width=\columnwidth]{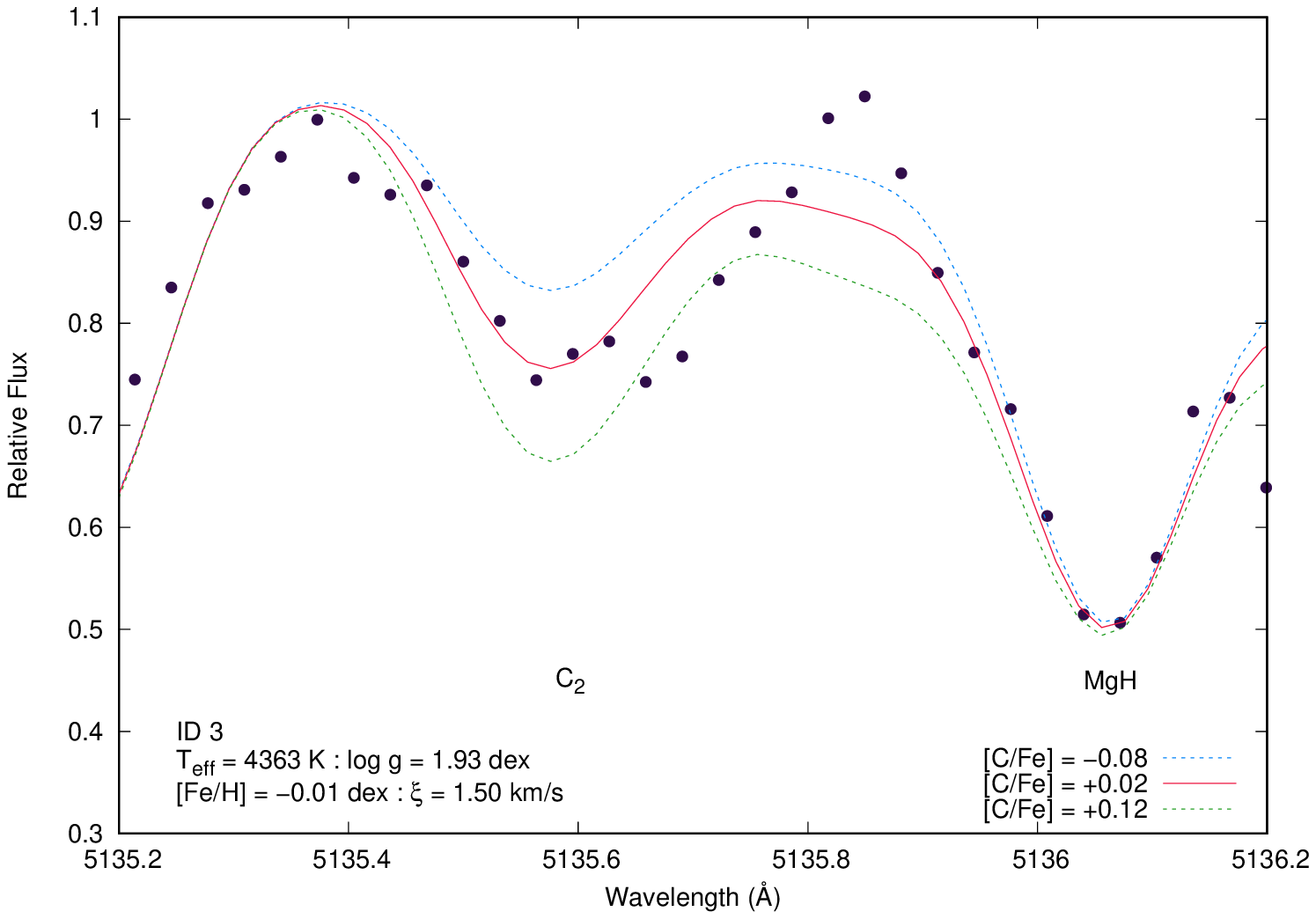}
    \includegraphics[angle=-90,width=\columnwidth]{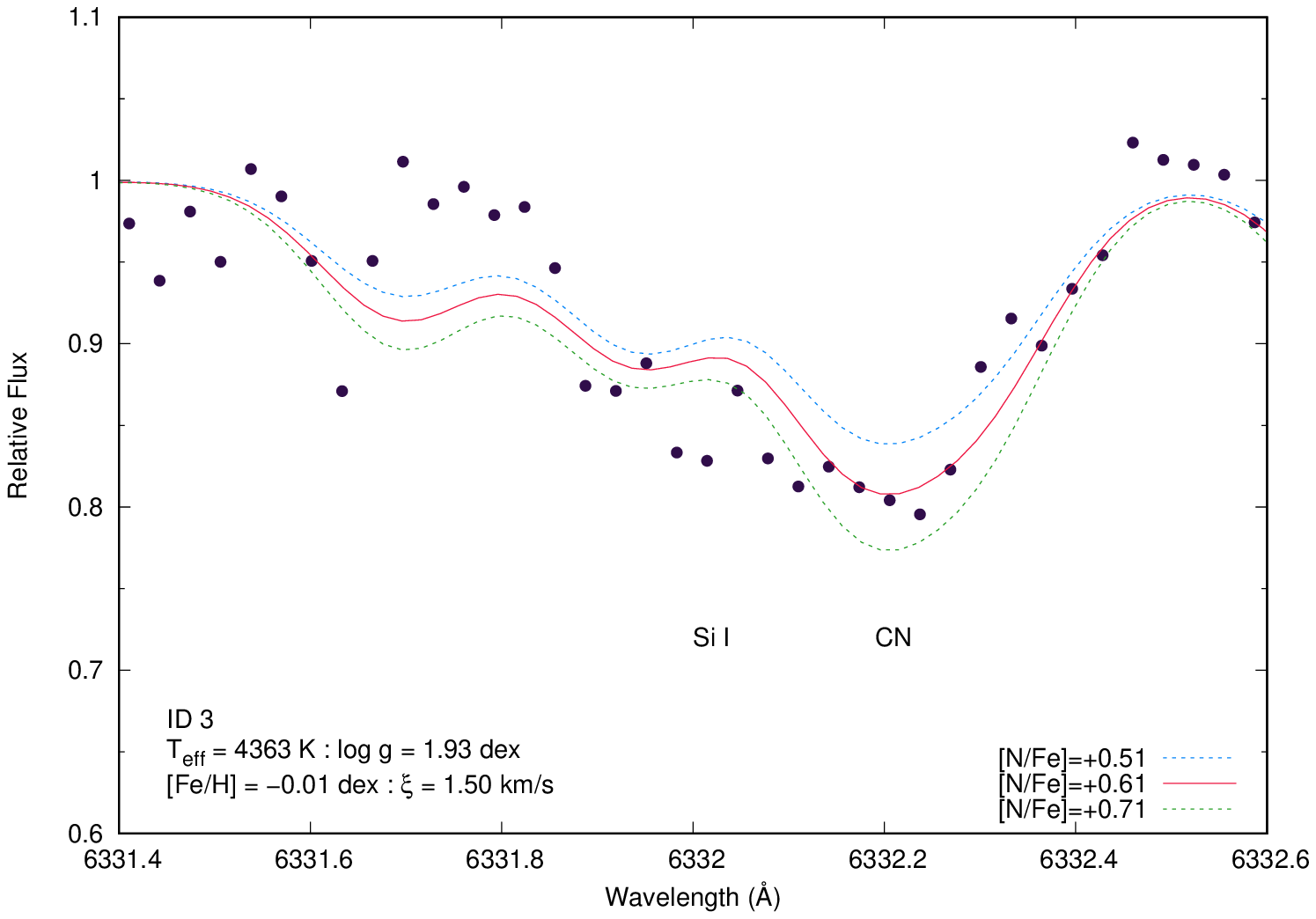}\\
    \includegraphics[angle=-90,width=\columnwidth]{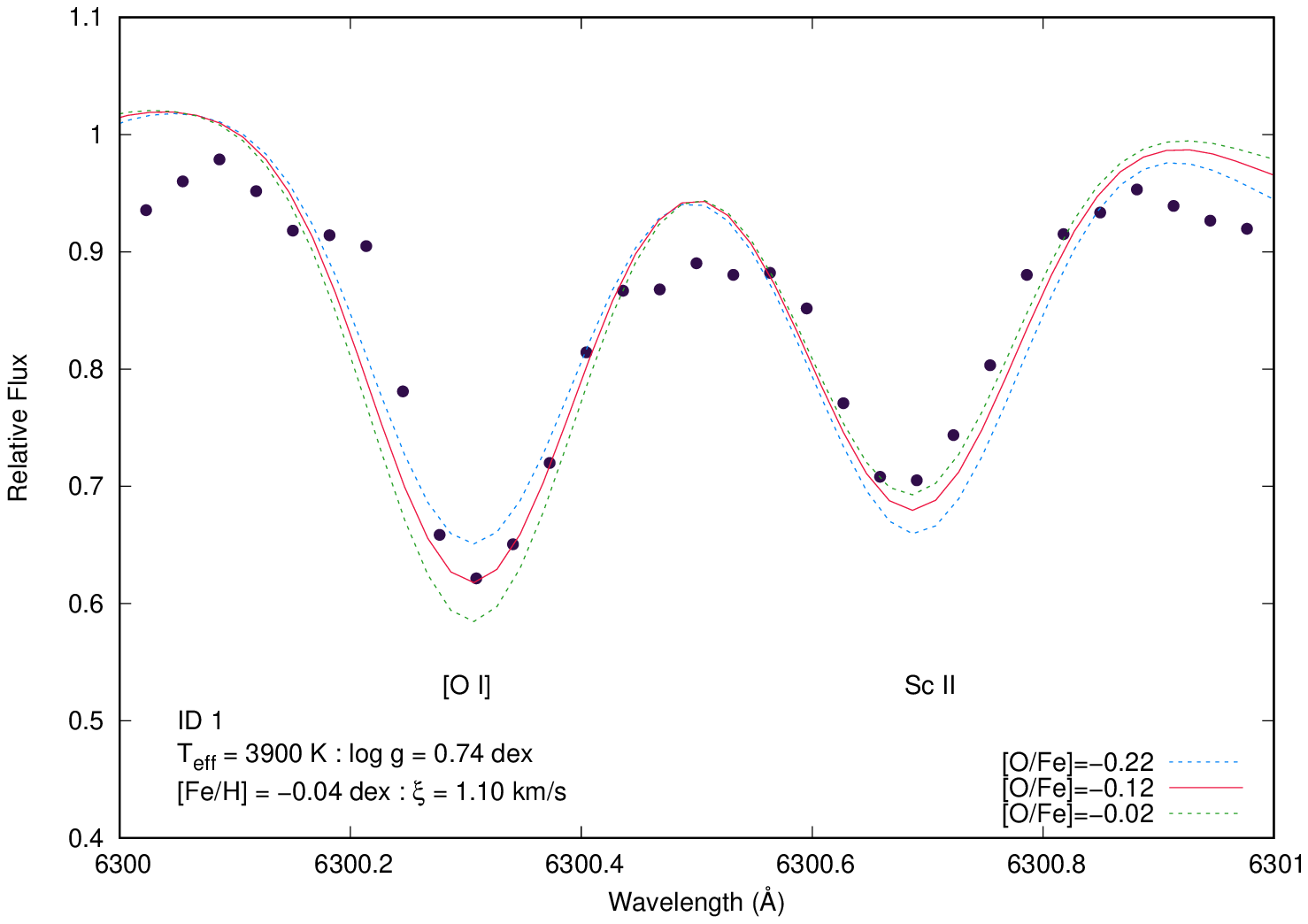}
    \includegraphics[angle=-90,width=\columnwidth]{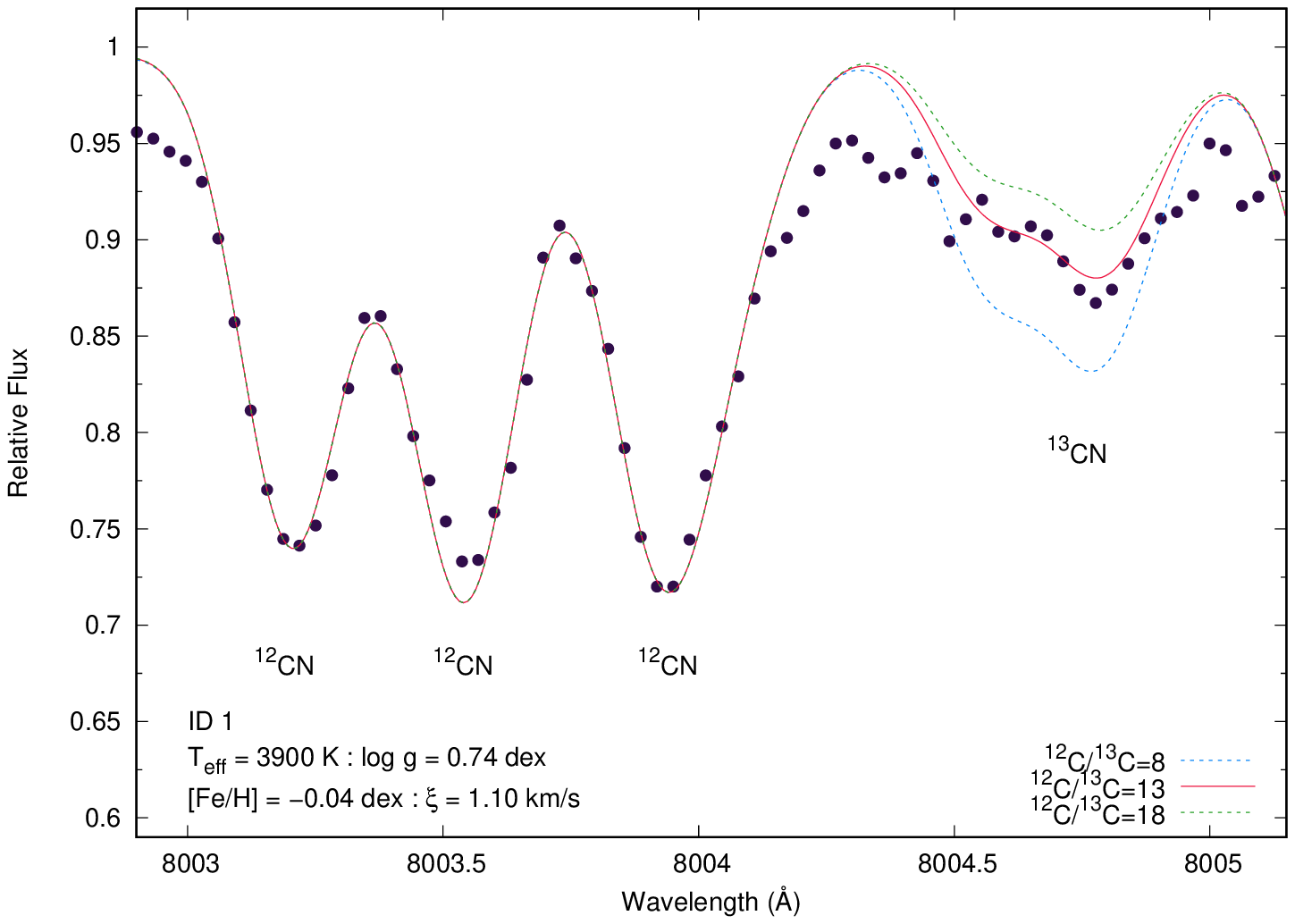}
    \caption{Comparison between the observed (black dots) and synthetic (blue, red, and green lines) spectra in the regions considered for spectral synthesis of CNO and $^{12}$C/$^{13}$C isotopic ratio. The best fit curve is the red one.}
    \label{sintese}
\end{figure*}

The carbon abundance was obtained through the C$_2$(0,1) Swan system  A$^3\Pi$-X$^3\Pi$ at 5135 \AA. The adopted dissociation potential of the C$_2$ molecule was $D_0(C_2)=6.27$ eV \citep{brooke}. The abundance found in the analysis of this feature is the abundance of all carbon isotopes ($^{12}$C +$^{13}$C). The nitrogen abundance was obtained through the red system CN(5,1) band A$^2\Pi$-X$^2\Sigma$ at 6332.18 \AA\,\citep{sneden2014}. The isotopic ratio of $^{12}$C/$^{13}$C was derived through the fitting of lines of $^{12}$CN and $^{13}$CN in the spectral region at 8005 \AA. The oxygen abundance was calculated through the forbidden line [\ion{O}{i}] at 6300.311 \AA. The forbidden [\ion{O}{i}] line is blended with a weak line of \ion{Ni}{i} at 6300.34 \AA\, and with a nearby line of \ion{Sc}{ii} at 6300.70 \AA. These lines are included in the synthesis with data from \cite{johansson} and \cite{spite}, respectively. The synthesis is exemplified in Figure \ref{sintese}.

As a test of the methodology, we first obtained the abundances of Arcturus. The results are shown in Table \ref{comparacao-arcturus}, where they are also compared with the results from the literature.
The analysis performed by \citet{alan-arcturus} used a high-resolution spectrum ($R=65\,000$) with high S/N (S/N $\sim 250$). \citet{ramirez2011} used high-resolution spectra in the visible with resolution $R\sim100\,000$ and \citet{fanelli2020} analysed high-resolution spectra in the near IR with $R=50\,000$. \citeauthor{alan-arcturus} used the same method of measurements of equivalent widths used in this work to obtain the abundances of the elements, while \citeauthor{ramirez2011} used the curve-of-growth approach to measure the abundances, and \cite{fanelli2020} used the spectral
synthesis technique. For the abundance of Eu (not included in Table \ref{comparacao-arcturus}), we obtained [Eu/Fe] = 0.23 $\pm$ 0.15 dex while in the literature a value of [Eu/Fe] $=0.30\pm0.05$ is reported \citep{overbeek}. Considering the uncertainty of the measurements, there is no important offset between our abundances for Arcturus and results from the literature.

The $^{16}$O/$^{17}$O and $^{16}$O/$^{18}$O isotopic ratios were calculated in the IR by fitting the region between 46620 and 46730 \AA\, employing the same methods adopted by \cite{harrislambert84a,harrislambert84b}, \cite{harrislambert87,harrislambert88}, \cite{abia2012} and \cite{lebzelter2015}. We used a line list that includes molecular lines of CO, C$_2$, CN, OH, SiO, MgH and SiS. The references of the line list data are given in Table \ref{moleculas}, together with the value and source of the respective molecular dissociation energies, $D_0$, which we adopted. In this IR region, CO is the most important molecule that contributes to most spectral lines \citep[see][]{hinkle-infra}. For the CO molecule, we use the line list of \citet{goorvitch}, which, according to the analysis by \citet{pavlenko}, is the one that better reproduces the abundances and isotopic ratios from the literature. The data used for the lines of C$_2$, SiO, MgH and SiS were obtained from the ExoMol database\footnote{\url{http://www.exomol.com/}}, which is a database of molecular line lists that can be used for spectral characterisation and simulation of the atmospheres of exoplanets, brown dwarfs, and cool stars.

\begin{table}
    \caption{References of the line list and dissociation potential $D_0$ considered in the infrared spectral synthesis.}
    \begin{tabular}{llll} \hline \hline
    Molec. & Ref. & $D_0$ (eV) & Ref. \\ \hline
    CO & \cite{goorvitch} & 11.09 & \cite{huber} \\
    $^{12}$C$_2$ & \cite{yurchenko} & 6.27 & \cite{sneden2014} \\
    $^{12}$C$^{13}$C & \cite{yurchenko} & 6.244 & \cite{ram} \\
    CN & \cite{sneden2014} & 7.724 & \cite{sneden2014} \\
    OH & \cite{brooke} & 4.411 & \cite{brooke} \\
    SiO & \cite{barton} & 8.26 & \cite{huber} \\
    MgH & \cite{yadin} & 1.34 & \cite{huber} \\
    SiS & \cite{upadhyay} & 6.47 & \cite{herzberg} \\ \hline
    \end{tabular}
    $D_0$: Dissociation potential
    \label{moleculas}
\end{table}

\begin{figure*}
    \includegraphics[angle=-90,width=1.93\columnwidth,trim = 2cm 0cm 1.5cm 0cm, clip]{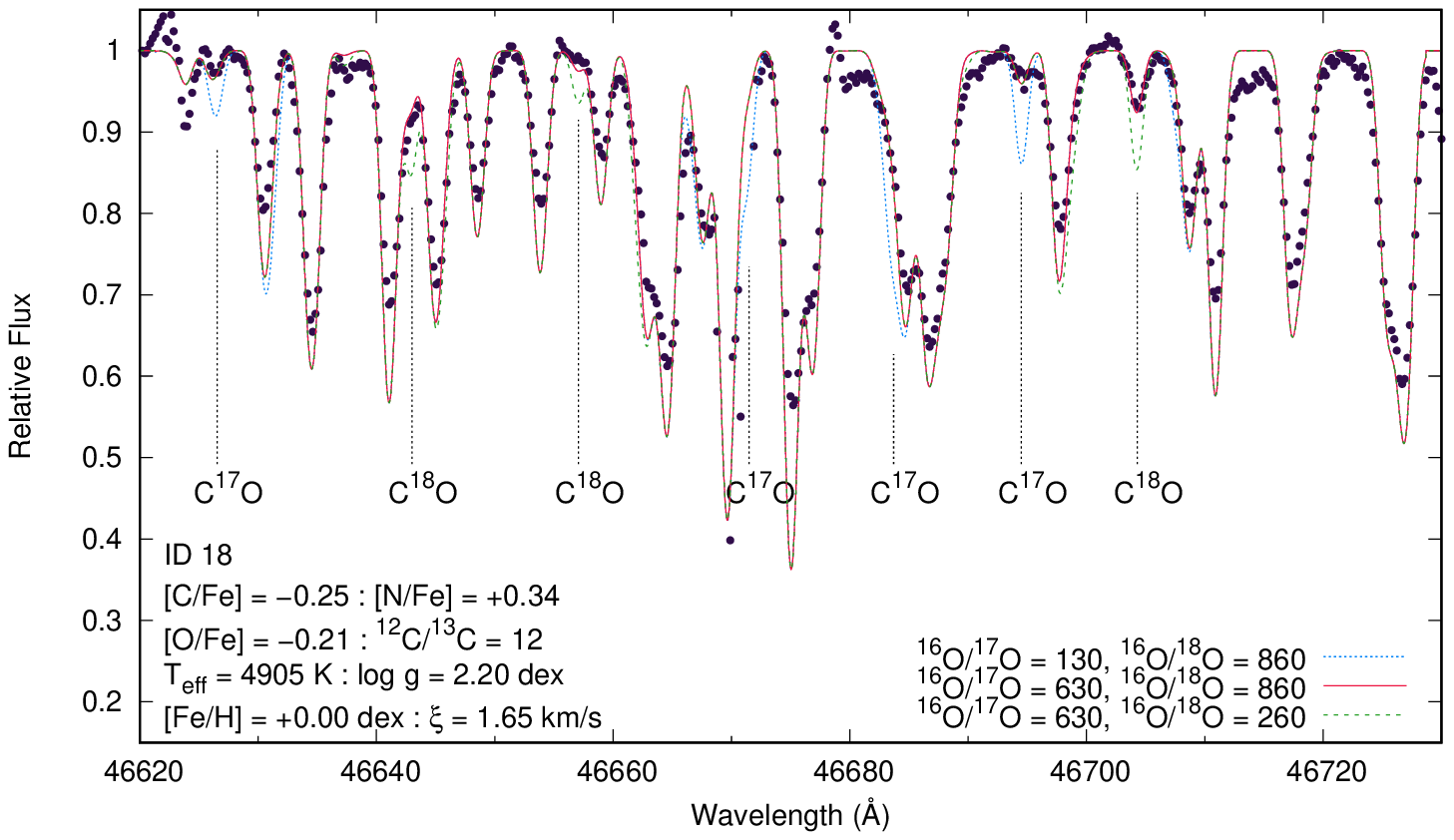}
    \caption{Spectral synthesis used to obtain $^{16}$O/$^{17}$O and $^{16}$O/$^{18}$O isotopic ratios. Black dots correspond to the observed spectrum, and coloured lines correspond to synthetic spectra. The best fit curve is the red one. Some lines of C$^{17}$O and C$^{18}$O used in the synthesis are labelled.}
    \label{sinteseO}
\end{figure*}

To calculate the oxygen isotopic ratios, we employed the photometric atmospheric parameters obtained in this work, and abundances of CNO and the $^{12}$C/$^{13}$C isotopic ratio found in the literature for each object. We adopted a Gaussian broadening of FWHM $\sim1.1$ \AA, which includes the macroturbulence velocity $\zeta$ and the projected rotational velocity $v \sin i$ to reproduce the resolution of the observed spectrum. In our procedure, we first compute the $^{16}$O/$^{17}$O isotopic ratio and then the $^{16}$O/$^{18}$O isotopic ratio, with a comparison of the lines of C$^{17}$O and C$^{18}$O between the synthetic and observed spectrum. We repeated the process iteratively until the synthetic spectrum reproduces in the best way the observed spectrum. We show in Figure \ref{sinteseO} an example of spectral synthesis for the determination of the $^{16}$O/$^{17}$O and $^{16}$O/$^{18}$O isotopic ratios. To verify the method, we also determined the isotopic ratios for Arcturus using atmospheric parameters calculated in this work, CNO abundances, and $^{12}$C/$^{13}$C isotopic ratios from \citet{abia2012}. In Table \ref{comparacao-arcturus}, our results for the oxygen isotopic ratios are compared with those obtained by \citet{abia2012}. The differences between the oxygen isotopic ratios can be related to the continuum levels of the spectra and the measurements of the weak lines of C$^{17}$O and C$^{18}$O.

As we see in Figure \ref{sinteseO}, the synthetic spectrum does not fit the centre of strong lines of CO. This result was also found in other analyses of K-type red giant stars at $\sim5\,\mu$m \citep[e.g.,][]{harrislambert84a,harrislambert84b,harrislambert88,tsuji} and is related to the limitations of the hydrostatic atmospheric model adopted to describe the external layers of the star. Strong lines of CO are produced in higher layers of the atmosphere and are more sensitive to local conditions. In comparison, weak spectral lines are formed in lower layers of the stellar atmosphere and are insensitive to high-atmosphere conditions \citep{harrislambert88,tsuji}. In K- and M-type red giant stars, stronger lines are formed in a layer of cold molecular clouds in the upper atmosphere, also known as the quasi-static molecular dissociation zone \citep{tsuji,tsuji2009}. Therefore, we considered only weak lines to calculate the oxygen isotopic ratios.

\subsection{Uncertainties}

The uncertainties in the photometric parameters were estimated by propagating the uncertainties in the magnitudes through the calibrations.
Uncertainties in the measurements of equivalent width $\sigma_{EW}$ were estimated using the expression presented by \citet{cayrel}:
\begin{align}
\sigma_{EW}=\frac{1.5}{\textrm{S/N}}\sqrt{\delta_x\cdot \textrm{FWHM}}
\end{align}
where S/N is the signal-noise ratio, $\delta_x$ is the spectral dispersion, and FWHM is the full width at half maximum of the spectral line. The FWHM was estimated as $\textrm{FWHM}\approx\Delta\lambda$, where $\Delta\lambda$ was obtained through the spectral resolution $R$ considering the typical wavelength of the spectrum $\lambda_\textrm{typical}$ ($\Delta\lambda=\frac{\lambda_\textrm{typical}}{R}$). For our data in the visible, we have $\lambda_{\textrm{typical}}\sim\lambda_{\textrm{mean}}\sim5880$ \AA, therefore,  $\textrm{FWHM}\approx\Delta\lambda\approx0.09$ \AA. Also, we have from the header of our spectra, S/N $\sim15-42$, $R=68000$ and $\delta_x\approx0.032$ \AA/pix, therefore, the $EW$ uncertainties are of about $2-6$ m\AA. Consequently, we observe that the errors are mainly influenced by the S/N and spectral resolution. The uncertainties resulting from the measurements $EW$ have only a small influence on the atmospheric parameters. Therefore, we ignored the uncertainties in the $EW$s when computing the uncertainties in the atmospheric parameters \citep[see e.g.,][]{alan,sales-silva,bagdonas}. 

To estimate the uncertainties of the spectroscopic atmospheric parameters, for $T_{\textrm{eff}}$ and $\xi$, we modified that parameter until the slope of the fit used to fix the parameter changed by its 1-$\sigma$ uncertainty. For $\log g$, the parameter was changed until the mean values of \ion{Fe}{i} and \ion{Fe}{ii} changed by the dispersion of \ion{Fe}{i}. The uncertainty in [Fe/H] is the dispersion of the abundances given by the \ion{Fe}{i} lines. We estimated the uncertainties for stars that had their abundances computed: objects 1 (NGC188-3018), 3 (NGC188-2072), 4 (NGC188-2026), 6 (NGC188-3140), and 8 (NGC188-2187). Five objects were rejected for reasons explained in Section \ref{sec:atmospheric}. The values considered for each stars can be seen in Table \ref{incerteza}. These estimates are on the high side of those found in the literature \citep[e.g.,][]{rodolfo2009,alan-arcturus,topcu2016,arthur}, but can be understood given the low S/N ratio of the spectra analysed here.

Uncertainties in the abundances were estimated by evaluating the influence of the uncertainties in the atmospheric parameters, changing one at a time. The total uncertainty was then found by the quadrature law \citep{incerteza}, considering that the errors are independent of each other:
\begin{align}
\sigma_{\textrm{total}}=\sqrt{\sigma^2_{T_{\textrm{eff}}}+\sigma^2_{\log g}+\sigma^2_{\textup{[Fe/H]}}+\sigma^2_{\xi}+\sigma^2_{\textrm{sample}}}
\end{align}
where $\sigma$ is a given uncertainty and $\sigma_{\textrm{sample}}$ is the the standard deviation of the abundances given by the individual lines of the species being evaluated. The values obtained can be seen in Table \ref{incerteza}.

\begin{table*}
    \scriptsize
    \caption{Uncertainty in abundances computed through EW measurements.}
    \begin{tabular}{lccccccclccccccc} \hhline{=======~=======}
    \multicolumn{7}{c}{Object 1 (NGC188-3018)} & & \multicolumn{7}{c}{Object 6 (NGC188-3140)} \\ \cline{1-7} \cline{9-15}
    \multirow{2}{*}{[X/Fe]} &$\Delta T_{\textrm{eff}}$ &$\Delta \log g$ &$\Delta$[Fe/H] &$\Delta\xi$ & \multirow{2}{*}{$\sigma_{\textrm{sample}}$} & \multirow{2}{*}{$\sigma_{\textrm{total}}$} & & \multirow{2}{*}{[X/Fe]} &$\Delta T_{\textrm{eff}}$ &$\Delta \log g$ &$\Delta$[Fe/H] &$\Delta\xi$ & \multirow{2}{*}{$\sigma_{\textrm{sample}}$} & \multirow{2}{*}{$\sigma_{\textrm{total}}$} \\ \cline{2-5} \cline{10-13}
    &$\pm$ 150 K &$\pm$ 0.42 dex&$\pm$ 0.15 dex&$\pm$ 0.16 \kms & & & & &$\pm$ 200 K &$\pm$ 0.40 dex&$\pm$ 0.16 dex&$\pm$ 0.17 \kms \\ \cline{1-7} \cline{9-15}
    \ion{Na}{i} &--- &--- &--- &--- &--- &--- & &\ion{Na}{i} &0.16 &0.02 &0.01 &0.04 &0.18 &0.25 \\
    \ion{Si}{i} &0.22 &0.14 &0.02 &0.04 &0.15 &0.31 & &\ion{Si}{i} &0.09 &0.07 &0.03 &0.03 &0.09 &0.15 \\
    \ion{Ca}{i} &0.11 &0.01 &0.01 &0.10 &0.10 &0.18 & &\ion{Ca}{i} &0.19 &0.03 &0.01 &0.06 &0.15 &0.25 \\
    \ion{Sc}{ii} &0.07 &0.17 &0.05 &0.11 &0.59 &0.63 & &\ion{Sc}{ii} &0.03 &0.17 &0.06 &0.03 &0.00 &0.18 \\
    \ion{Ti}{i} &0.13 &0.04 &0.02 &0.14 &0.10 &0.22 & &\ion{Ti}{i} &0.27 &0.00 &0.01 &0.07 &0.18 &0.33 \\
    \ion{V}{i} &0.12 &0.06 &0.03 &0.13 &0.20 &0.27 & &\ion{V}{i} &0.30 &0.01 &0.01 &0.09 &0.05 &0.32 \\
    \ion{V}{ii} &--- &--- &--- &--- &--- &--- & &\ion{V}{ii} &0.08 &0.16 &0.15 &0.02 &0.12 &0.27 \\
    \ion{Cr}{i} &0.07 &0.03 &0.02 &0.13 &0.08 &0.17 & &\ion{Cr}{i} &0.21 &0.02 &0.00 &0.07 &0.15 &0.27 \\
    \ion{Cr}{ii} &0.23 &0.18 &0.04 &0.05 &0.21 &0.36 & &\ion{Cr}{ii} &0.13 &0.11 &0.06 &0.04 &0.36 &0.40 \\
    \ion{Co}{i} &0.06 &0.14 &0.04 &0.12 &0.17 &0.26 & &\ion{Co}{i} &0.10 &0.05 &0.03 &0.50 &0.14 &0.53 \\
    \ion{Ni}{i} &0.11 &0.14 &0.03 &0.11 &0.11 &0.24 & &\ion{Ni}{i} &0.04 &0.07 &0.03 &0.07 &0.08 &0.14 \\
    \ion{Y}{ii} &0.02 &0.17 &0.05 &0.13 &0.00 &0.21 & &\ion{Y}{ii} &0.00 &0.16 &0.06 &0.07 &0.00 &0.19 \\
    \ion{Ce}{ii} &0.00 &0.17 &0.06 &0.04 &0.00 &0.18 & &\ion{Ce}{ii} &0.02 &0.17 &0.06 &0.07 &0.00 &0.19 \\
    \ion{Eu}{ii} &0.03 &0.17 &0.05 &0.05 &0.00 &0.19 & &\ion{Eu}{ii} &--- &--- &--- &--- &--- &--- \\ \cline{1-7} \cline{9-15}
    \multicolumn{7}{c}{Object 3 (NGC188-2072)} & & \multicolumn{7}{c}{Object 8 (NGC188-2187)} \\ \cline{1-7} \cline{9-15}
    \multirow{2}{*}{[X/Fe]} &$\Delta T_{\textrm{eff}}$ &$\Delta \log g$ &$\Delta$[Fe/H] &$\Delta\xi$ & \multirow{2}{*}{$\sigma_{\textrm{sample}}$} & \multirow{2}{*}{$\sigma_{\textrm{total}}$} & & \multirow{2}{*}{[X/Fe]} &$\Delta T_{\textrm{eff}}$ &$\Delta \log g$ &$\Delta$[Fe/H] &$\Delta\xi$ & \multirow{2}{*}{$\sigma_{\textrm{sample}}$} & \multirow{2}{*}{$\sigma_{\textrm{total}}$} \\ \cline{2-5} \cline{10-13}
    &$\pm$ 250 K &$\pm$ 0.45 dex&$\pm$ 0.17 dex&$\pm$ 0.32 \kms & & & & &$\pm$ 150 K &$\pm$ 0.25 dex&$\pm$ 0.12 dex&$\pm$ 0.40 \kms \\ \cline{1-7} \cline{9-15}
    \ion{Na}{i} &0.21 &0.01 &0.01 &0.13 &0.00 &0.25 & &\ion{Na}{i} &0.12 &0.02 &0.00 &0.10 &0.09 &0.18 \\
    \ion{Si}{i} &0.22 &0.14 &0.04 &0.06 &0.06 &0.28 & &\ion{Si}{i} &0.10 &0.05 &0.04 &0.06 &0.06 &0.14 \\
    \ion{Ca}{i} &0.25 &0.01 &0.01 &0.08 &0.03 &0.26 & &\ion{Ca}{i} &0.14 &0.03 &0.00 &0.12 &0.06 &0.19 \\
    \ion{Sc}{ii} &0.05 &0.20 &0.06 &0.05 &0.13 &0.25 & &\ion{Sc}{ii} &0.02 &0.11 &0.05 &0.06 &0.00 &0.13 \\
    \ion{Ti}{i} &0.35 &0.01 &0.00 &0.10 &0.09 &0.38 & &\ion{Ti}{i} &0.20 &0.00 &0.00 &0.12 &0.12 &0.26 \\
    \ion{V}{i} &0.34 &0.04 &0.01 &0.17 &0.10 &0.39 & &\ion{V}{i} &0.20 &0.01 &0.00 &0.17 &0.10 &0.28 \\
    \ion{V}{ii} &--- &--- &--- &--- &--- &--- & &\ion{V}{ii} &--- &--- &--- &--- &--- &--- \\
    \ion{Cr}{i} &0.23 &0.01 &0.01 &0.11 &0.08 &0.27 & &\ion{Cr}{i} &0.14 &0.00 &0.01 &0.12 &0.09 &0.20 \\
    \ion{Cr}{ii} &0.25 &0.21 &0.06 &0.07 &0.08 &0.35 & &\ion{Cr}{ii} &--- &--- &--- &--- &--- &--- \\
    \ion{Co}{i} &0.04 &0.13 &0.04 &0.14 &0.10 &0.22 & &\ion{Co}{i} &0.04 &0.05 &0.03 &0.14 &0.10 &0.19 \\
    \ion{Ni}{i} &0.04 &0.14 &0.04 &0.14 &0.08 &0.22 & &\ion{Ni}{i} &0.01 &0.05 &0.03 &0.16 &0.06 &0.18 \\
    \ion{Y}{ii} &0.01 &0.10 &0.06 &0.10 &0.00 &0.16 & &\ion{Y}{ii} &0.00 &0.10 &0.05 &0.21 &0.00 &0.24 \\
    \ion{Ce}{ii} &--- &--- &--- &--- &--- &--- & &\ion{Ce}{ii} &--- &--- &--- &--- &--- &--- \\
    \ion{Eu}{ii} &--- &--- &--- &--- &--- &--- & &\ion{Eu}{ii} &--- &--- &--- &--- &--- &--- \\  \cline{1-7} \cline{9-15}
    \multicolumn{7}{c}{Object 4 (NGC188-2026)} & & & & & & & & \\ \cline{1-7}
    \multirow{2}{*}{[X/Fe]} &$\Delta T_{\textrm{eff}}$ &$\Delta \log g$ &$\Delta$[Fe/H] &$\Delta\xi$ & \multirow{2}{*}{$\sigma_{\textrm{sample}}$} & \multirow{2}{*}{$\sigma_{\textrm{total}}$} & & & & & & & & \\ \cline{2-5}
    &$\pm$ 200 K &$\pm$ 0.34 dex&$\pm$ 0.12 dex&$\pm$ 0.17 \kms & & & & & & & & & & & \\ \cline{1-7}
    \ion{Na}{i} &0.16 &0.01 &0.01 &0.05 &0.06 &0.18 & & & & & & & & \\
    \ion{Si}{i} &0.09 &0.07 &0.02 &0.04 &0.09 &0.15 & & & & & & & & \\
    \ion{Ca}{i} &0.19 &0.03 &0.01 &0.08 &0.34 &0.40 & & & & & & & & \\
    \ion{Sc}{ii} &0.03 &0.16 &0.04 &0.04 &0.00 &0.17 & & & & & & & & \\
    \ion{Ti}{i} &0.29 &0.01 &0.01 &0.06 &0.16 &0.34 & & & & & & & & \\
    \ion{V}{i} &0.29 &0.00 &0.01 &0.06 &0.11 &0.32 & & & & & & & & \\
    \ion{V}{ii} &--- &--- &--- &--- &--- &--- & & & & & & & & \\
    \ion{Cr}{i} &0.21 &0.01 &0.00 &0.06 &0.08 &0.24 & & & & & & & & \\
    \ion{Cr}{ii} &0.15 &0.14 &0.04 &0.05 &0.13 &0.25 & & & & & & & & \\
    \ion{Co}{i} &0.11 &0.06 &0.02 &0.06 &0.12 &0.19 & & & & & & & & \\
    \ion{Ni}{i} &0.05 &0.06 &0.02 &0.09 &0.09 &0.15 & & & & & & & & \\
    \ion{Y}{ii} &0.00 &0.15 &0.04 &0.08 &0.00 &0.18 & & & & & & & & \\
    \ion{Ce}{ii} &--- &--- &--- &--- &--- &--- & & & & & & & & \\
    \ion{Eu}{ii} &--- &--- &--- &--- &--- &--- & & & & & & & & \\ \cline{1-7}
    \end{tabular}
    \label{incerteza}
\end{table*}

Several factors can affect the abundance calculation: (1) accuracy in parameters of the adopted models, (2) equivalent width measurements, (3) quality in spectra fitting, and (4) internal errors related to the adopted methods (e.g., thermodynamic approximation of models, LTE, and 1D plane-parallel atmospheric model).

To estimate the uncertainties in our spectral synthesis, we chose a representative star with atmospheric parameter values similar to the mean of the sample. For this calculation, we chose object 6 (NGC188-3140) for the visible region and object 12 (NGC3860-44) for the infrared region. The results obtained were applied to the rest of their respective sample.

Uncertainties in the oxygen isotopic ratios were derived taking into account the mean uncertainties of the photometric atmospheric parameters, of CNO abundances and of the isotopic ratio of $^{12}$C/$^{13}$C. The values used are $\Delta T_{\textrm{eff}}\approx\pm$ 189 K, $\Delta \log g\approx\pm$ 0.08 dex, $\Delta\xi\approx\pm$ 0.50 \kms, $\Delta$[C/Fe] $\approx\pm$ 0.10 dex, $\Delta$[N/Fe] $\approx\pm$ 0.13 dex, $\Delta$[O/Fe] $\approx\pm$ 0.16 dex and $\Delta^{12}$C/$^{13}$C $\approx\pm$ 2. The total uncertainty was calculated by adding the uncertainties in the parameters and input abundances quadratically. The values can be seen in Table \ref{incerteza-sintese2}.

\begin{table*}
    \caption{Uncertainty in abundances computed through spectral synthesis in the visible (object 6, NGC188-3140) and infrared regions (object 12, NGC3860-44).}
    \begin{tabular}{ccccccccc} \hline\hline
    \multicolumn{6}{c}{Object 6 (NGC188-3140)} & & & \\ \hline
    \multirow{2}{*}{[X/Fe]} & $\Delta T_{\textrm{eff}}$ & $\Delta \log g$ & $\Delta$[Fe/H] & $\Delta\xi$ & \multirow{2}{*}{$\sigma_{\textrm{total}}$} & & \\ \cline{2-5}
    &$\pm$ 200 K &$\pm$ 0.40 dex&$\pm$ 0.16 dex&$\pm$ 0.17 \kms & & & \\ \hline
    C & 0.01 & 0.05 & 0.09 & 0.10 & 0.14 & & \\
    N & 0.12 & 0.17 & 0.04 & 0.09 & 0.23 & & \\
    O & 0.05 & 0.20 & 0.05 & 0.01 & 0.21 & & \\
    $^{12}$C/$^{13}$C & 3 & 1 & 0 & 0 & 3 & & \\ \hline
    \multicolumn{9}{c}{Object 12 (NGC3860-44)} \\ \hline
    \multirow{2}{*}{ } & $\Delta T_{\textrm{eff}}$ & $\Delta \log g$ & $\Delta\xi$ & $\Delta$[C/Fe] & $\Delta$[N/Fe] & $\Delta$[O/Fe] & $\Delta^{12}$C/$^{13}$C & \multirow{2}{*}{ $\sigma_{\textrm{total}}$ } \\ \cline{2-8}
    & $\pm$ 189 K & $\pm$ 0.08 dex & $\pm$ 0.50 \kms & $\pm$ 0.10 dex & $\pm$ 0.13 dex & $\pm$ 0.16 dex & $\pm$ 2 & \\ \hline
    $^{16}$O/$^{17}$O & 420 & 50 & 50 & 100 & --- & --- & --- & 437 \\
    $^{16}$O/$^{18}$O & 400 & 70 & 100 & 215 & --- & --- & --- & 470 \\ \hline
    \end{tabular}
    \label{incerteza-sintese2}
\end{table*}

From Table \ref{incerteza-sintese2}, we see that abundances of CNO in the visible region suggests are weakly dependent on uncertainties in the microturbulent velocity ($\lesssim$ 0.10 dex) since only weak lines were used. In addition, uncertainties in the abundances of C, N, and O influence each other, considering the connection between them due to the molecular equilibrium. For the infrared sample, we see that the oxygen isotopic ratios are weakly sensitive to variations in [N/Fe], [O/Fe] and in the isotopic ratio of $^{12}$C/$^{13}$C, so that the best fit does not change when these values are modified (see Fig. \ref{sinteseO}). Only for variations larger than one standard deviation can we detect modifications in the wings of the spectral lines of CO (for $\Delta$[O/Fe] $\gtrsim0.5$ dex) and in the line depth (for $\Delta$[C/Fe] $\gtrsim0.2$ dex). When comparing the uncertainty values obtained in this work with those obtained by \citet{lebzelter2015}, their results are 4 or 5 times smaller than ours, which can be related to the different wavelength regions considered for the analysis. \citet{lebzelter2015} used the $K$-band region ($\sim$ 2.3 $\mu$m), while we used the $M$-band region ($\sim$ 4.6 $\mu$m). However, lower flux levels, line blending, and issues in defining the continuum level made them choose to analyse data only in the 2.3 $\mu$m spectrum region. Another contrast is shown in the sample of stars with a lower effective temperature compared to ours. This selection in the sample makes their molecular lines stronger, making them easier to measure. In addition, we used literature values for atmospheric parameters and chemical composition of the objects analysed, while \citet{lebzelter2015} chose a fixed value of C/O in 0.3 and calculated the oxygen abundance using the OH vibrational-rotation lines in the $H$-band ($\sim$ 1.6 $\mu$m). In addition, their results do not provide detailed information on the S/N ratios, exposure times, and error determination. Therefore, considering these differences in the methods used in both articles regarding the description of the observational spectra and the computation of the errors, we consider our results to be consistent.

\section{Results}

\subsection{Radial velocities}\label{sec:radialvelocities-result}

The kinematic results of the heliocentric radial velocity $V_h$ compared to the values of the literature $V_{h,lit}$ are shown in Figure \ref{comparacao-kinematics} and the values can be seen in Table \ref{resultado1} in Appendix \ref{sec:ew2}. For NGC188 stars, we detail their heliocentric radial velocities in a plot. The radial velocities obtained for individual objects and the mean velocity of the cluster agree with values found in the literature, considering the data uncertainty, and they agree with the current values of \citet{cantat-gaudin2020} regarding the membership of the stars.

\citet{cantat-gaudin2020} classified the object 7 (NGC188-1061) as a non-member of the cluster. It presents a radial velocity similar to the mean velocity of the cluster; however, after considering also its proper motion, they classified this star as a non-member of the cluster. For that reason, this object was removed from the abundance analysis performed in this work.

For object 5 (NGC188-1116), we measure a heliocentric radial velocity that departs significantly ($\sim$ 16 \kms) from that of the literature. This difference can be explained according to \citet{geller2008,geller2009} and \citet{jacobson2011} because it is an SB2 spectroscopic binary star (see Section \ref{sec:binaries}). According to \citet{geller2008}, NGC188 has a large dispersion in radial velocity due to its population of binaries. In our sample for NGC188, the stars with more dispersion of radial velocity are binaries, i.e. objects 3 (NGC188-2072), 5 (NGC188-1116), and 10 (NGC188-2194). Therefore, we do not use objects 3 (NGC188-2072), 5 (NGC188-1116), 7 (NGC188-1061), and 10 (NGC188-2194) in the calculation of the mean radial velocity of the cluster.

In the computation of the radial velocities of star 11 (NGC2682-MMU6495), we consider only the spectrum found in run B of observations, although the star is observed in both dates of observations (see Section \ref{sec:observacao} and Table \ref{observacao}). The spectrum observed in run A presents a low S/N (S/N $\sim$ 5), which affects the identification of spectral lines. This happened because of technical difficulties with closing the loop of the adaptive optics system for that observation. This was solved for the other objects observed during that night. In agreement with \citet{mmu}, using the mean radial velocity of NGC3532 ($\langle V_h\rangle=4.33\pm0.34$ \kms), the object 19 (NGC3532-MMU649) is not a member of the cluster. Therefore, we consider only object 18 (NGC3532-MMU19) in the calculation of the radial velocity.

\begin{figure}
    \includegraphics[height=\columnwidth, angle=-90]{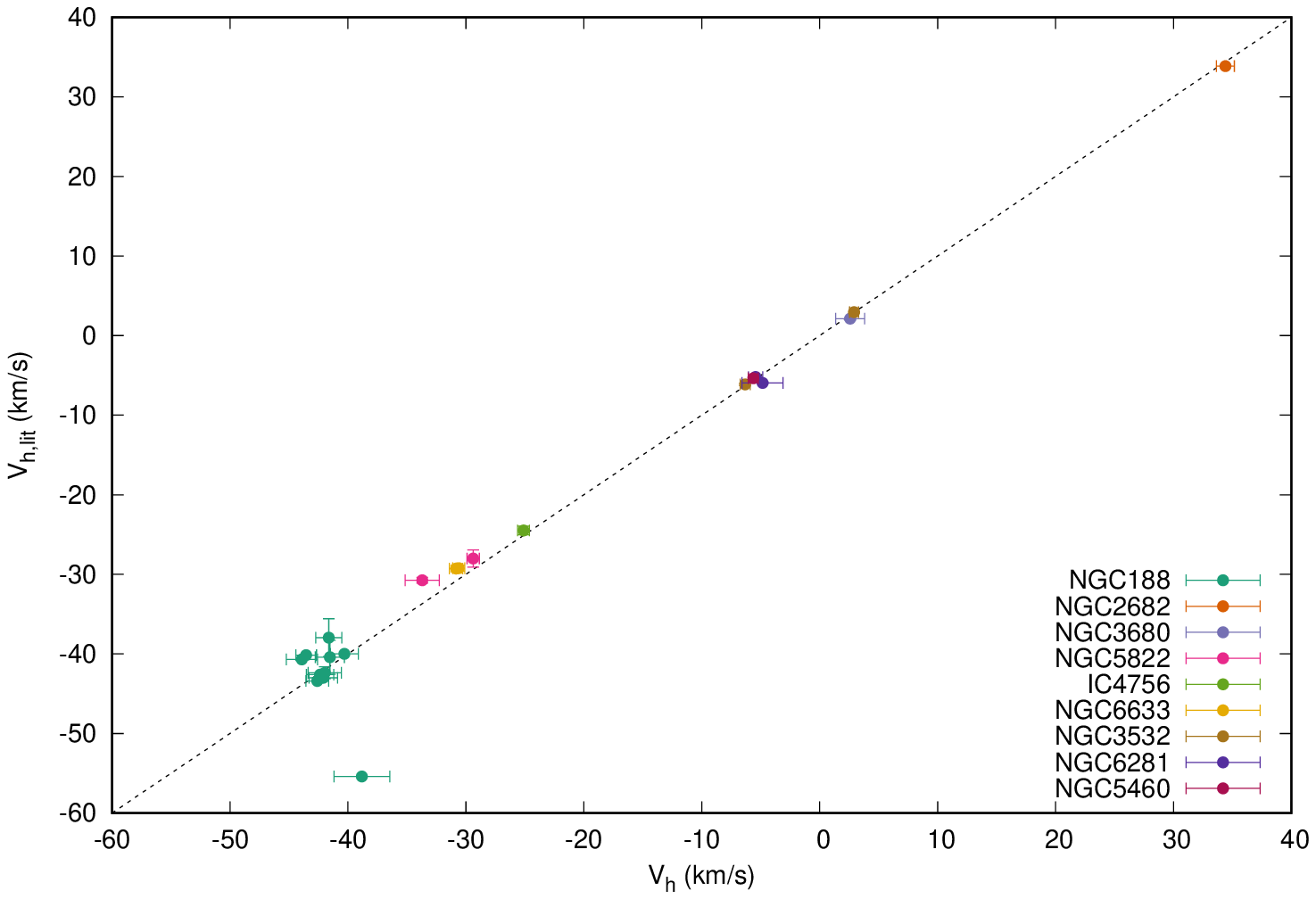}
    \includegraphics[height=\columnwidth, angle=-90]{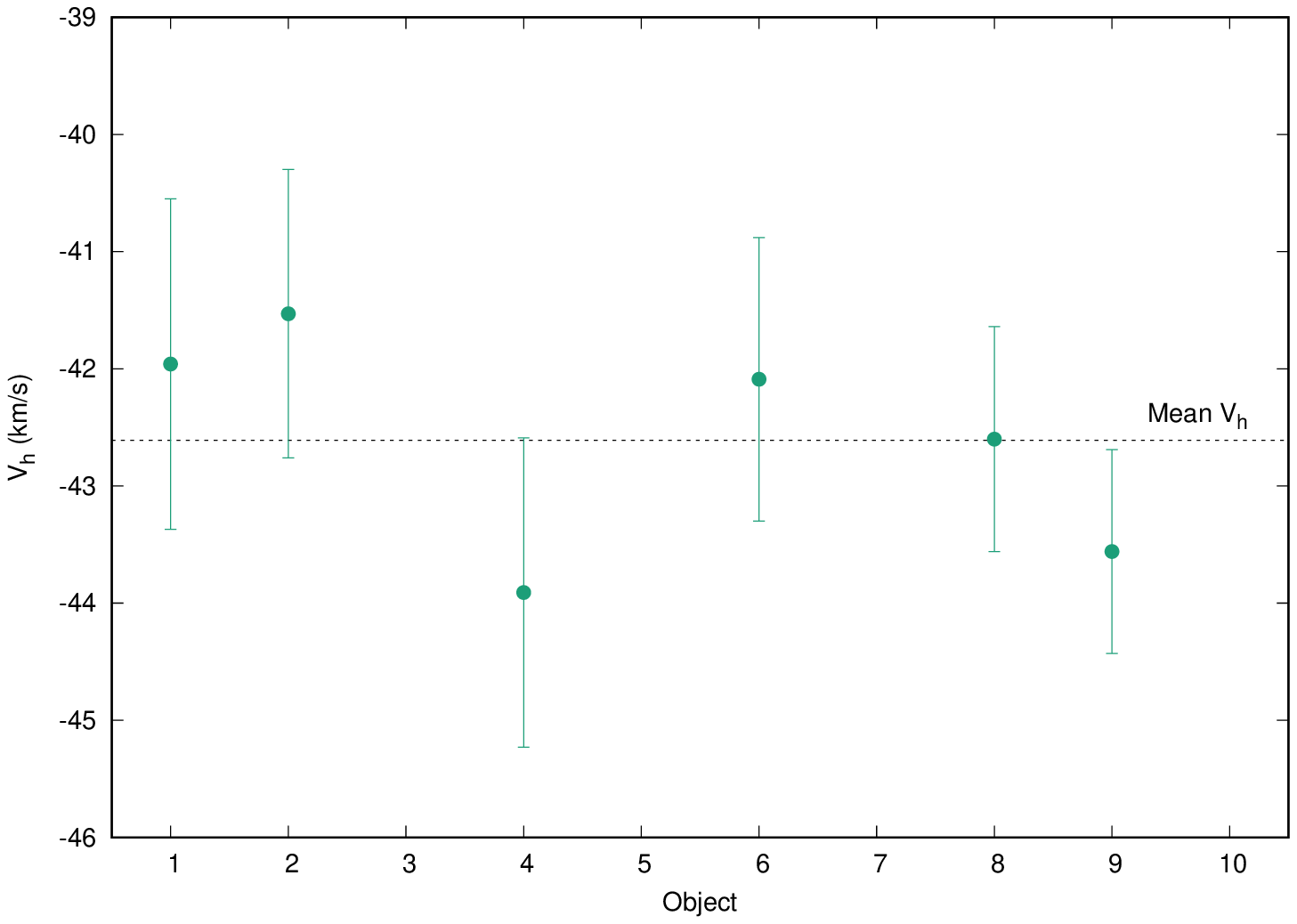}
    \caption{\textit{Top:} Comparison between the results of heliocentric radial velocity $V_h$ obtained in this work with the literature $V_{h,lit}$ for all the objects (see Table \ref{resultado1} in Appendix \ref{sec:ew2} for references). \textit{Bottom:} Results of the heliocentric radial velocity for the NGC188 cluster. The mean velocity of the NGC188 cluster is $-$42.61 $\pm$ 1.23 \kms, ignoring in the calculation the radial velocity of binary stars (objects 3 (NGC188-2072), 5 (NGC188-1116), and 10 (NGC188-2194)) and the object 7 (NGC188-1061), which is not a member of the cluster.}
    \label{comparacao-kinematics}
\end{figure}

\subsection{Atmospheric parameters}

\subsubsection{Effective temperature, surface gravity, and microtubulence velocity}\label{sec:atmospheric}

The atmospheric parameters obtained from photometric calibrations were calculated for all objects, except objects 2 (NGC188-1001), 5 (NGC188-1116), 7 (NGC188-1061), 9 (NGC188-1006), and 10 (NGC188-2194). These stars were removed from the chemical analysis for the following reasons: objects 2 (NGC188-1001) and 5 (NGC188-1116) are binary stars with high rotational speed (see Section \ref{sec:binaries}), object 7 (NGC188-1061) is a non-member of the cluster, and objects 9 (NGC188-1006) and 10 (NGC188-2194) have low S/N (S/N $\sim15$, see Table \ref{observacao}). Removing stars with low S/N improves the precision of our values for the mean atmospheric parameters compared to the literature. The atmospheric parameters of the objects are detailed in Table \ref{final2}. The atmospheric parameters are similar in photometry and spectroscopy, considering the uncertainties.

\begin{table*}
    \caption{Comparison between photometric and spectroscopic atmospheric parameters.}
    \begin{threeparttable}
    \begin{tabular}{cccccccccccc} \hline \hline
    \multirow{3}{*}{ ID } & \multicolumn{4}{c}{ Photometry } & & \multicolumn{6}{c}{ Spectroscopy } \\ \cline{2-5} \cline{7-12}
    & $T_{\textrm{eff}}$ & $\log g$ & [Fe/H] & $\xi$ & & $T_{\textrm{eff}}$ & $\log g$ & [Fe/H] & $\xi$ & $\zeta$ & $v \sin$i \\
    & (K) & (dex) & (dex) & (\kms) & & (K) & (dex) & (dex) & (\kms) & (\kms) & (\kms) \\ \hline
    1 & 4018 $\pm$ 110 & 1.53 $\pm$ 0.07 & +0.17 & 1.79 $\pm$ 0.49 & & 3900 $\pm$ 150 & 0.74 $\pm$ 0.42 & $-$0.04 $\pm$ 0.15 & 1.10 $\pm$ 0.16 & 5.5 & 4.0 \\
    3 & 4328 $\pm$ 169 & 2.21 $\pm$ 0.09 & +0.17 & 1.57 $\pm$ 0.50 & & 4363 $\pm$ 250 & 1.93 $\pm$ 0.45 & $-$0.01 $\pm$ 0.17 & 1.50 $\pm$ 0.32 & 5.0 & 4.0 \\
    4 & 4986 $\pm$ 179 & 3.05 $\pm$ 0.07 & +0.17 & 1.33 $\pm$ 0.50 & & 4788 $\pm$ 200 & 2.38 $\pm$ 0.34 & +0.02 $\pm$ 0.12 & 0.90 $\pm$ 0.17 & 6.0 & 3.0 \\
    6 & 4807 $\pm$ 165 & 3.09 $\pm$ 0.07 & +0.17 & 1.27 $\pm$ 0.50 & & 4800 $\pm$ 200 & 2.59 $\pm$ 0.40 & +0.03 $\pm$ 0.16 & 1.10 $\pm$ 0.17 & 4.5 & 2.5 \\
    8 & 4710 $\pm$ 168 & 3.09 $\pm$ 0.07 & +0.17 & 1.25 $\pm$ 0.50 & & 4780 $\pm$ 150 & 3.15 $\pm$ 0.25 & +0.19 $\pm$ 0.12 & 1.00 $\pm$ 0.40 & 4.5 & 3.5 \\
    11 & 4131 $\pm$ 162 & 1.47 $\pm$ 0.10 & +0.03 & 1.80 $\pm$ 0.49 & & --- & --- & --- & --- & --- & --- \\
    12 & 4518 $\pm$ 162 & 1.99 $\pm$ 0.08 & +0.04 & 1.66 $\pm$ 0.50 & & --- & --- & --- & --- & --- & --- \\
    13 & 4339 $\pm$ 166 & 1.61 $\pm$ 0.08 & +0.02 & 1.77 $\pm$ 0.50 & & --- & --- & --- & --- & --- & --- \\ 
    14 & 4307 $\pm$ 161 & 1.72 $\pm$ 0.08 & +0.02 & 1.73 $\pm$ 0.50 & & --- & --- & --- & --- & --- & --- \\ 
    15 & 5049 $\pm$ 167 & 2.37 $\pm$ 0.06 & +0.07 & 1.63 $\pm$ 0.50 & & --- & --- & --- & --- & --- & --- \\ 
    16$^{\star}$ & 4549 $\pm$ 433 & 1.51 $\pm$ 0.20 & +0.10 & 1.85 $\pm$ 0.51 & & --- & --- & --- & --- & --- & --- \\ 
    17 & 5080 $\pm$ 107 & 2.17 $\pm$ 0.04 & +0.10 & 1.73 $\pm$ 0.50 & & --- & --- & --- & --- & --- & --- \\ 
    18 & 4905 $\pm$ 172 & 2.20 $\pm$ 0.07 & +0.00 & 1.65 $\pm$ 0.50 & & --- & --- & --- & --- & --- & --- \\ 
    20 & 5012 $\pm$ 168 & 1.95 $\pm$ 0.06 & +0.00 & 1.78 $\pm$ 0.50 & & --- & --- & --- & --- & --- & --- \\ 
    21 & 4929 $\pm$ 167 & 1.98 $\pm$ 0.06 & +0.00 & 1.75 $\pm$ 0.50 & & --- & --- & --- & --- & --- & --- \\ 
    22 & 4824 $\pm$ 166 & 1.71 $\pm$ 0.07 & $-$0.15 & 1.80 $\pm$ 0.50 & & --- & --- & --- & --- & --- & --- \\ \hline
    \end{tabular}
    $^{\star}$ Object with very large uncertainty in magnitude due to low quality in its photometry (see Section \ref{sec:photometry}).
    \end{threeparttable}
    \label{final2}
\end{table*}

The values obtained in photometry are also similar to those obtained previously in the literature (see Table \ref{resultado-foto} in Appendix \ref{sec:ew2} for more details). The comparison cannot be made for objects 2 (NGC188-1001), 5 (NGC188-1116), and 22 (NGC5460-MMU17) because there are no equivalent data for them in the literature. The object 7 (NGC188-1061) was excluded from the analysis because it is a non-member of the cluster (see Section \ref{sec:radialvelocities-result}), and objects 9 (NGC188-1006) and 10 (NGC188-2194) were removed from the analysis because their spectra have low S/N (S/N $\sim15$, see Table \ref{observacao}). The mean differences between photometric values and spectroscopic results available from the literature are $\sim40$ K in effective temperature, $\sim0.4$ dex in surface gravity and $\sim0.3$ \kms\, in microturbulent velocity.

Comparing spectroscopic and photometric atmospheric parameters, we see agreement within the uncertainties. Considering the surface gravity, the spectroscopic values tend to be smaller (mean difference $\sim$ 0.43 dex), but is still mostly within the uncertainties ($\sigma\sim$ 0.4 dex). The uncertainties in spectroscopic surface gravity are about $\sim5$ times larger than photometric ones, while the spectroscopic microturbulence uncertainties are about $\sim3$ times smaller than photometric estimates. These values are related to the dependence in the adopted photometric calibrations, which is smaller in surface gravity than in microturbulent velocity, since to determine the photometric microturbulent velocity we used photometric $T_{\textrm{eff}}$ and $\log g$, while in the surface gravity determination, we used the mean photometric $T_{\textrm{eff}}$.

Stars 1 (NGC188-3018) and 4 (NGC188-2026) present significant differences between photometric and spectroscopic estimates of surface gravity ($\sim$ 0.7$-$0.8 dex). We believe these results are a consequence of the low S/N ratio of their spectra. The surface gravity is obtained through the ionisation equilibrium of iron lines, and the measurements of the equivalent widths of the \ion{Fe}{i} and \ion{Fe}{ii} lines is affected by a low S/N value.

\begin{figure}
    \includegraphics[angle=-90,width=\columnwidth]{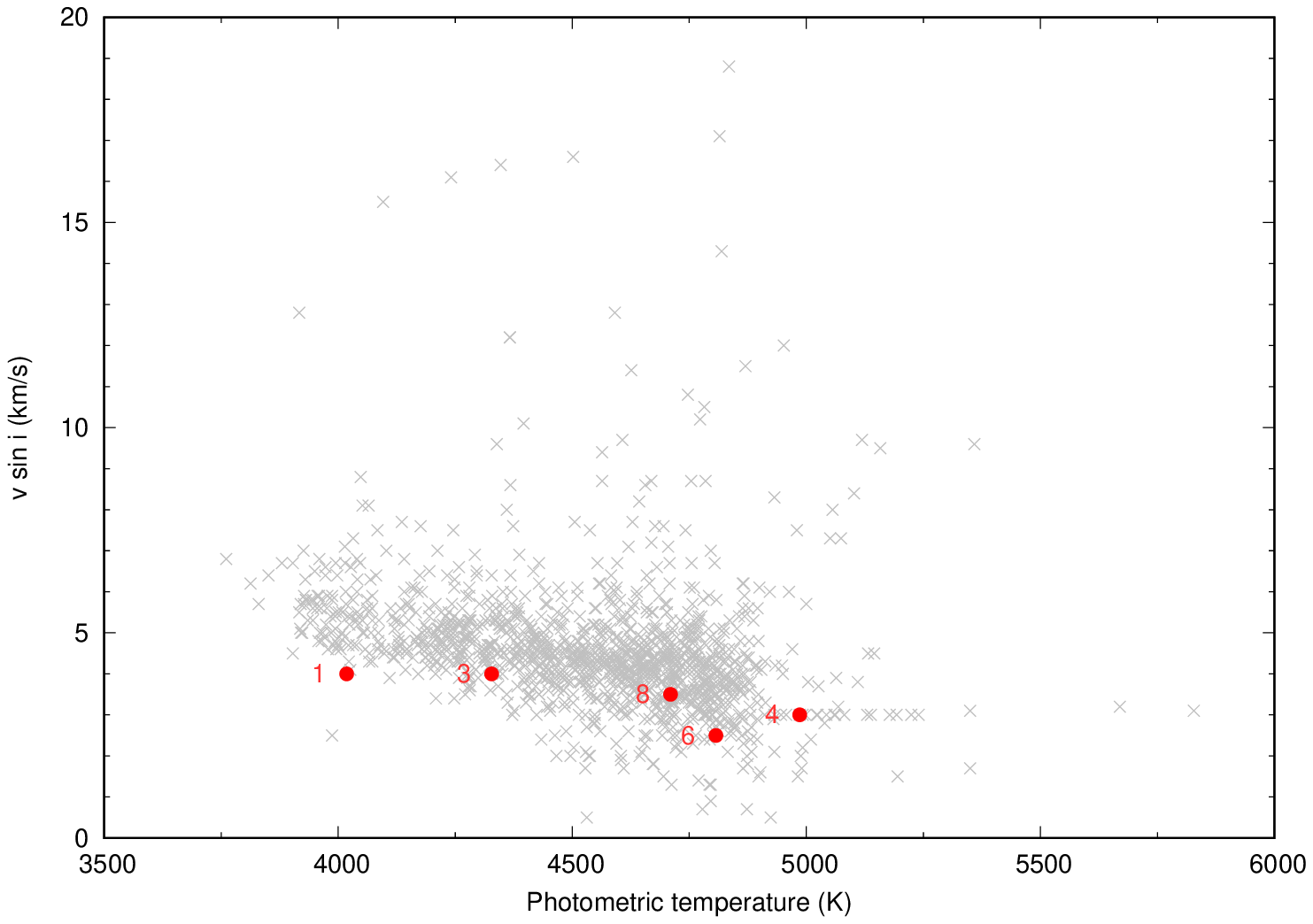}
    \caption{Projected rotational velocity $v \sin i$ of RGB stars as a function of photometric temperature. Red symbols indicate stars analysed in this work, and gray symbols indicate stars from the sample of \citet{carlberg}. Numeric values can be seen in Table \ref{final2}.}
    \label{carlberg}
\end{figure}

The values of the macroturbulence velocity $\zeta$ and $v \sin i$ calculated for the NGC188 sample can be seen in Table \ref{final2}. The mean macroturbulence velocity and $v\sin i$ obtained are $\langle\zeta\rangle=4.9$ \kms and $\langle v\sin i\rangle=3.5$ \kms, respectively. The mean values obtained are similar to those found for K-type giant stars in the literature \citep[see Table B.2 in][]{Gray2005}. As a test of the methodology considered in this work, we calculate the macroturbulence velocity $\zeta$ and $v \sin i$ of Arcturus, a standard K-type giant star. The results obtained are $\zeta=5.0$ \kms and $v \sin i=1.5$ \kms. These values are similar to those found in the literature \citep[$\zeta=5.6\pm0.2$ \kms, $v\sin i=1.5$ \kms;][]{sheminova2015}. Therefore, our results for macroturbulence velocity and $v\sin i$ are consistent with the literature on K-type giant stars. The results obtained for projected rotational velocity $v \sin i$ are consistent with those found for RGB stars \citep[$2.0\leqslant v\sin i\leqslant6.0$ \kms;][]{penhasuarez}, as determined by \citet[][]{carlberg} and can be seen in Figure \ref{carlberg}.

\subsubsection{Metallicity}\label{sec:metallicity}

\begin{table*}
    \centering
    \caption{Comparison between metallicity values calculated in this work and in the literature.}
    \begin{threeparttable}
    \begin{tabular}{cclcccclc}\hhline{====~====}
    ID & [Fe/H] (dex) & Ref. & Method & & ID & [Fe/H] (dex) & Ref. & Method \\ \cline{1-4}\cline{6-9}
    1 & +0.06 $\pm$ 0.08 & \cite{jacobson2011} & S & & 14 & +0.02 $\pm$ 0.11 & \cite{rodolfo2009} & S \\
    & +0.17 & This work & P & & & +0.02 & This work & P \\ \cline{6-9}
    & $-$0.04 $\pm$ 0.15 & This work & S & & 15 & +0.08 $\pm$ 0.08 & \cite{rodolfo2009} & S \\ \cline{1-4}
    3 & +0.13 $\pm$ 0.03 & \cite{friel2010} & S & & & +0.07 & This work & P \\ \cline{6-9}
    & +0.06 $\pm$ 0.06 & \cite{jacobson2011} & S & & 16 & +0.04 $\pm$ 0.10 & \cite{rodolfo2009} & S \\
    & +0.14 $\pm$ 0.02 & \cite{jacobson2013} & S & & & $-$0.14 $\pm$ 0.13 & \cite{morel2014} & S \\
    & +0.17 & This work & P & & & $-$0.03 $\pm$ 0.12 & \cite{morel2014} & A \\
    & $-$0.01 $\pm$ 0.17 & This work & S & & & +0.04 & \cite{luck2014} & U \\ \cline{1-4}
    4 & +0.03 $\pm$ 0.08 & \cite{jacobson2011} & S & & & $-$0.20 $\pm$ 0.19 & \cite{boeche2016} & F \\
    & +0.17 & This work & P & & & +0.10 & This work & P \\ \cline{6-9}
    & +0.02 $\pm$ 0.12 & This work & S & & 17 & +0.35 $\pm$ 0.03 & \cite{valenti2005} & F \\ \cline{1-4}
    6 & +0.01 $\pm$ 0.06 & \cite{jacobson2011} & S & & & +0.04 $\pm$ 0.10 & \cite{santos2009} & S \\
    & +0.17 & This work & P & & & +0.00 $\pm$ 0.10 & \cite{santos2009} & S \\
    & +0.03 $\pm$ 0.16 & This work & S & & & +0.11 $\pm$ 0.11 & \cite{rodolfo2009} & S \\ \cline{1-4}
    8 & $-$0.07 $\pm$ 0.07 & \cite{jacobson2011} & S & & & $-$0.01 $\pm$ 0.11 & \cite{morel2014} & S \\
    & +0.17 & This work & P & & & $-$0.07 $\pm$ 0.10 & \cite{morel2014} & A \\
    & +0.19 $\pm$ 0.12 & This work & S & & & +0.00 $\pm$ 0.10 & \cite{jacobson2016} & W \\ \cline{1-4}
    11 & +0.03 $\pm$ 0.05 & \cite{netopil} & S & & & +0.10 & This work & P \\ \cline{6-9}
    & +0.04 $\pm$ 0.06 & \cite{lit2} & C & & 18 & +0.13 $\pm$ 0.02 & \cite{luck2014} & U  \\ 
    & +0.03 & This work & P & & & +0.11 $\pm$ 0.11 & \cite{rodolfo2009} & S \\ \cline{1-4}
    12 & $-$0.28 $\pm$ 0.32 & \cite{pasquini2001} & S & & & +0.00 & This work & P \\ \cline{6-9}
    & $-$0.14 $\pm$ 0.16 & \cite{anthony2009} & S & & 20 & +0.01 $\pm$ 0.09 & \cite{rodolfo2009} & S \\
    & $-$0.13 $\pm$ 0.08 & \cite{mitschang2012} & S & & & +0.00 & This work & P \\ \cline{6-9}
    & $-$0.15 $\pm$ 0.08 & \cite{penhasuarez} & S & & 21 & +0.09 $\pm$ 0.07 & \cite{rodolfo2009} & S \\
    & +0.04 & This work & P & & & +0.00 & This work & P \\ \cline{1-4} \cline{6-9}
    13 & +0.09 $\pm$ 0.21 & \cite{luck1994} & U & & & & & \\
    & +0.03 $\pm$ 0.10 & \cite{rodolfo2009} & S & & & & & \\
    & $-$0.09 $\pm$ 0.09 & \cite{penhasuarez} & S & & & & & \\
    & +0.02 & This work & P & & & & & \\ \cline{1-4}
    \end{tabular}
    \begin{tablenotes}
      \small
      \item Labels: (A) Asteroseismology, (C) Calibration relations from the raw outputs of the pipeline, (F) Spectral fitting to a grid of synthetic spectra, (P) Photometry, (S) Spectroscopy, (U) Unweighted mean of the photometric and spectroscopic values and (W) Weighted-median value for each atmospheric parameter.
    \end{tablenotes}
    \end{threeparttable}
    \label{metallicity}
\end{table*}

In this work, we consider iron abundance [Fe/H] as the indicator of the global metallicity for the objects in the sample, considering that our objects have near-solar metallicities, and we assume a solar enhancement for $\alpha$-elements\footnote{$\alpha$-elements are produced by the successive addition of $\alpha$ particles, and some examples of them are O, Mg, Si, Ca, and Ti.} \citep[see][]{paulajofre}. The practical reason for adopting this method is that iron lines, other than hydrogen and helium lines, dominate the optical spectrum for a wide variety of main sequence stars, which makes it affordable for us to use iron lines as indicators of the metallicity \cite[e.g.,][]{paunzen, heiter, netopil}. The metallicity derived and compared to the literature star by star is given in Table \ref{metallicity} for all the objects (the isochrone fitting method implies that stars within one cluster have the same [Fe/H] value) and distinguishing the derivation methods to obtain it. We also compared our mean spectroscopic values for NGC188 with the data from the literature in Table \ref{comparacao-literatura}, differentiating the methods used to derive it.

The individual photometric results derived in this work for the metallicity of NGC188 (objects 1 to 10) are higher than the most recent values found in the literature \citep[e.g.][]{donor2020}. In the literature, it is possible to observe a higher and a lower value for the mean metallicity of the cluster. \citet{heiter} and \citet{netopil} investigated metallicities available in previous studies. They argued that the difference is related to the intermediate resolution ($R\sim18000$) of the spectra considered by \citet{jacobson2011}. Therefore, we consider our result for the photometric metallicity of the NGC188 cluster to be consistent with the mean value obtained through high-resolution spectroscopy by \citet{netopil}.

Our spectroscopic metallicity results for all stars are similar to the literature, within the uncertainties, except for object 8 (NGC188-2187). They differ from the results of \citet{jacobson2011} about 0.1 to 0.2 dex, which is smaller than the values found using high-resolution spectra. However, considering the difference in spectral resolution in the literature, our results are consistent with the mean spectroscopic values of other studies, as presented in Table \ref{comparacao-literatura}.

Similarly, for the infrared sample (objects 11 to 22), we can compare the metallicity obtained from isochrone fitting to the literature value as a validation of photometric parameters obtained, which will be used in the determination of effective temperature for those stars. We can see differences in metallicity between the values found in this work and the literature up to 0.2 dex for the objects 12 (NGC3680-44), 13 (NGC5822-1), 16 (NGC6633-78), 18 (NGC3532-MMU19), and 21 (NGC6281-4). This result is related to differences in spectral resolution in previous work in the literature, as well as methods used to obtain metallicity \citep[e.g. spectroscopy, photometry, asteroseismology;][]{netopil}.

\subsection{Abundance patterns for the metals in the NGC188 sample}

We were able to compute the abundances of Na, Si, Ca, Sc, Ti, V, Cr, Co, Ni, Y, Ce, and Eu through spectroscopic atmospheric models and equivalent width measurements for most of the NGC188 sample stars (objects 1 (NGC188-3018), 3 (NGC188-1001), 4 (NGC188-2026), 6 (NGC188-3140), and 8 (NGC188-2187)). The measurements of equivalent width obtained can be seen in Table \ref{ew-tabela} in Appendix \ref{sec:ew2} and the calculated abundances and isotopic ratios are shown in Table \ref{final}. In Table \ref{comparacao-literatura}, we compare the mean abundances found by us and in the literature for NGC188. Given the relatively low S/N values of our spectra (hence large error bars) for the stars of this cluster, we consider that the comparison is quite satisfactory.

Only object 3 (NGC188-2072) in our sample has elemental abundances previously determined in the literature \citep[see][]{friel2010}. Our results have differences between 0.1 and 0.2 dex in estimated abundances, and they are similar to the mean abundances in other works.

\begin{figure*}
    \includegraphics[angle=-90, width=1.03\columnwidth]{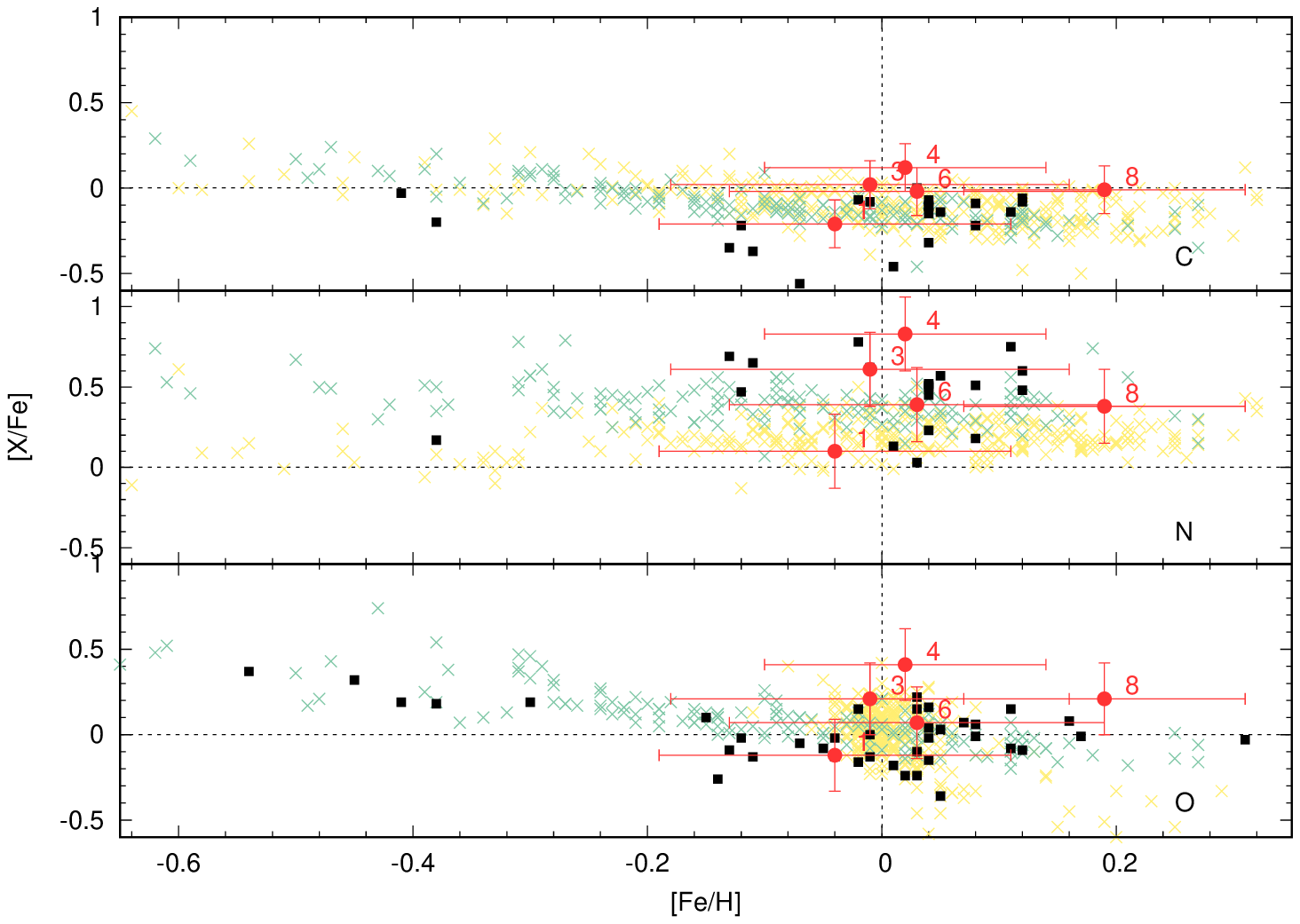}
    \includegraphics[angle=-90, width=1.03\columnwidth]{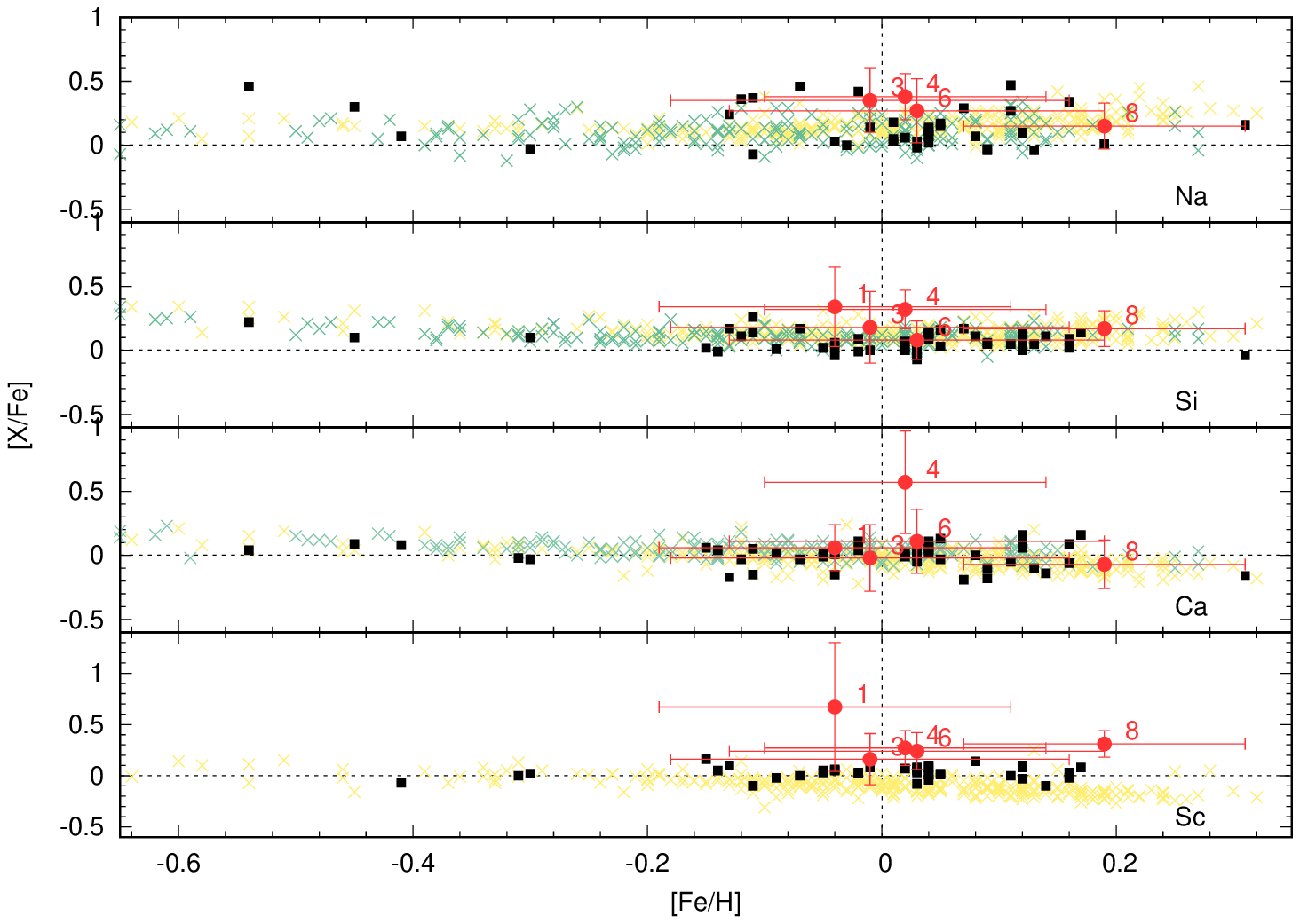}\\
    \includegraphics[angle=-90, width=1.03\columnwidth]{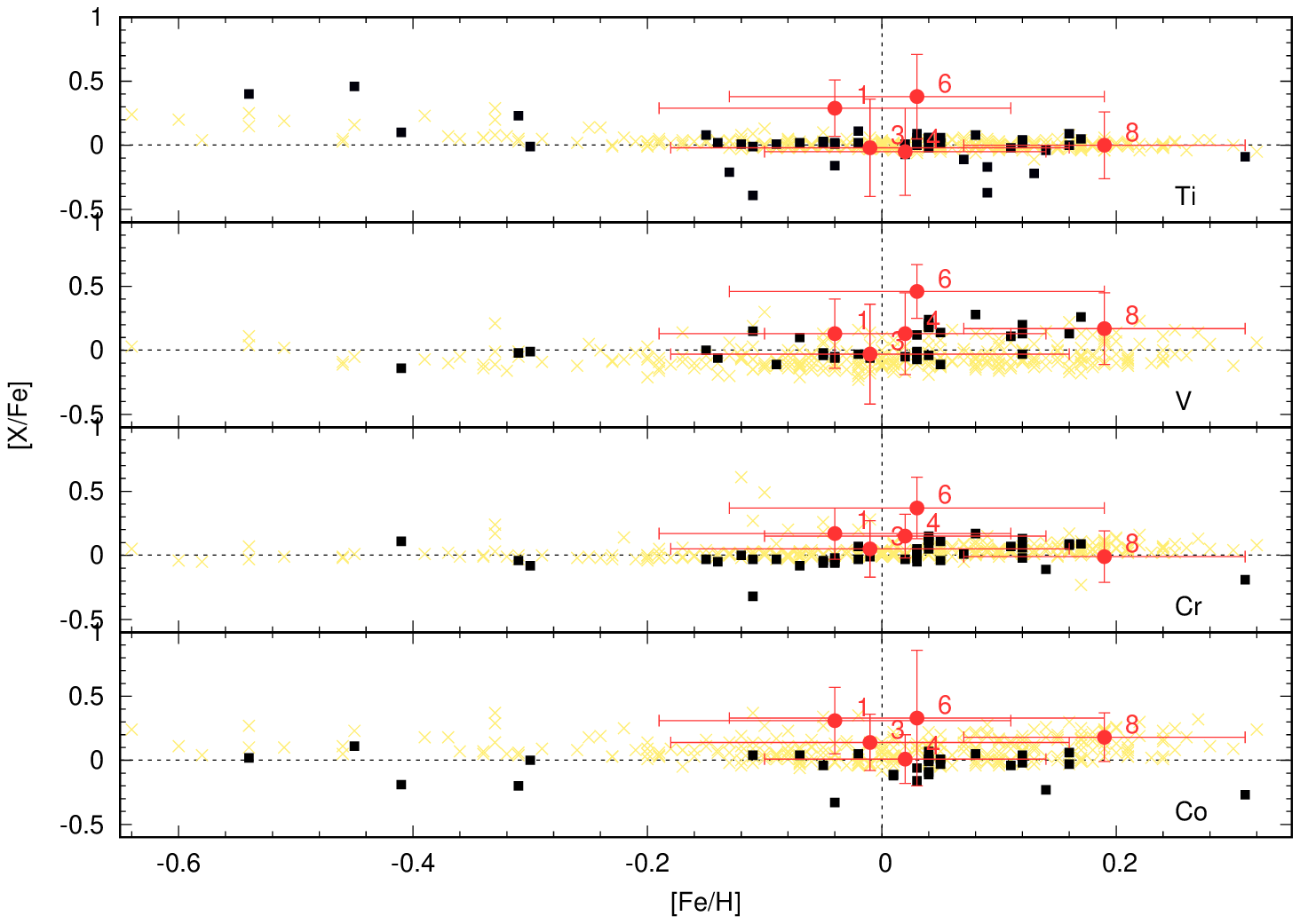}
    \includegraphics[angle=-90, width=1.03\columnwidth]{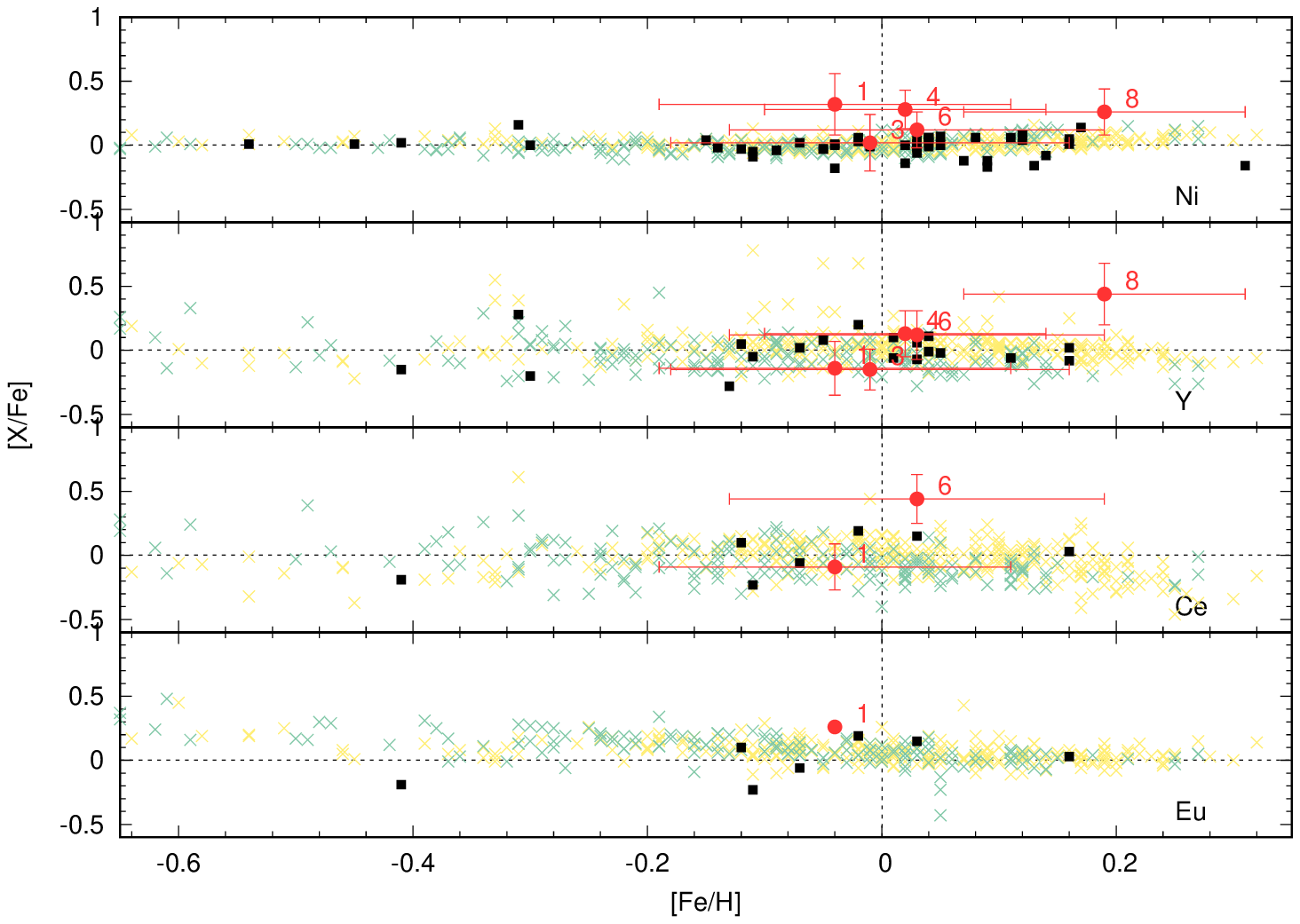}
    \caption{Abundance ratio [X/Fe] as a function of metallicity for our NGC188 sample stars (red symbols) using their individual [Fe/H] spectroscopic values. Yellow symbols represent giant field stars of \citet{luck-heiter}, green symbols represent clump giants of \citet{mishenina-2006,mishenina-2007} and black symbols represents the mean abundances of open clusters available in the literature \citep{gratton1994,brown1996,carretta2005,friel2005,yong2005,jacobson2008,rodolfo2009,villanova2009,pancino2010,friel2010,jacobson2011,carrera2011,zacs2011,santrich2013,topcu2015,topcu2016,drazultimo,drazprimeiro,occaso}.}
    \label{element}
\end{figure*}

The values of the abundance ratios from C to Eu available for individual stars are compared with data for evolved stars from the literature as a function of metallicity in Figure \ref{element}. For a better comparison with the literature value, their solar values were corrected to the same solar value adopted in this work \citep[i.e.,][]{asplund}. The dispersion in metallicity observed for NGC 188 stars could be related to the S/N of the spectra, which can affect the number of reliable lines for equivalent width measurements. For instance, while approximately 40 iron lines were selected for each star, some spectra provided more \ion{Fe}{ii} lines than others for calculating individual metallicities. The abundances of elements will be discussed in the following subsections as they appear in each sub-figure of Figure \ref{element}.

\begin{table*}
    \caption{Abundances and isotopic ratios obtained spectroscopically in this work and those adopted from the literature.}
    \begin{threeparttable}
    \begin{tabular}{cccccccccc} \hline \hline
    ID & [C/Fe] & [N/Fe] & [O/Fe] & $\log$ (C+N+O) & Ref. & $^{12}$C/$^{13}$C & Ref. & $^{16}$O/$^{17}$O & $^{16}$O/$^{18}$O \\
    & (dex) & (dex) & (dex) & (dex) & & & & & \\ \hline
    1 & $-$0.21 $\pm$ 0.14 & +0.10 $\pm$ 0.23 & $-$0.12 $\pm$ 0.21 & 8.72 $\pm$ 0.37 & 1 & 13 $\pm$ 3 & 1 & --- & --- \\
    3 & +0.02 $\pm$ 0.14 & +0.61 $\pm$ 0.23 & +0.21 $\pm$ 0.21 & 9.08 $\pm$ 0.37 & 1 & 13 $\pm$ 3 & 1 & --- & --- \\
    4 & +0.12 $\pm$ 0.14 & +0.83 $\pm$ 0.23 & +0.41 $\pm$ 0.21 & 9.30 $\pm$ 0.38 & 1 & 12 $\pm$ 3 & 1 & --- & --- \\
    6 & $-$0.02 $\pm$ 0.14 & +0.39 $\pm$ 0.23 & +0.07 $\pm$ 0.21 & 8.99 $\pm$ 0.37 & 1 & 13 $\pm$ 3 & 1 & --- & --- \\
    8 & $-$0.01 $\pm$ 0.14 & +0.38 $\pm$ 0.23 & +0.21 $\pm$ 0.21 & 9.24 $\pm$ 0.40 & 1 & 19 $\pm$ 3 & 1 & --- & --- \\
    11 & $-$0.31 $\pm$ 0.03$^{\star}$ & +0.13 $\pm$ 0.03$^{\star}$ & $-$0.05 $\pm$ 0.04$^{\star}$ & 8.81 $\pm$ 0.08$^{\star}$ & 2 & 24 $\pm$ 4$^{\star}$ & 3 & 730 $\pm$ 437 & 280 $\pm$ 470 \\
    12 & $-$0.28 $\pm$ 0.10 & +0.10 $\pm$ 0.34 & $-$0.14 $\pm$ 0.16 & 8.92 $\pm$ 0.33 & 4 & 8 & 4 & 480 $\pm$ 437 & 710 $\pm$ 470 \\
    13 & $-$0.34 $\pm$ 0.13 & +0.37 $\pm$ 0.37 & $-$0.13 $\pm$ 0.19 & 8.94 $\pm$ 0.43 & 4 & 16 & 4 & 780 $\pm$ 437 & 1070 $\pm$ 470 \\
    14 & $-$0.29 $\pm$ 0.06$^{\star}$ & +0.35 $\pm$ 0.11$^{\star}$ & $-$0.12 $\pm$ 0.08$^{\star}$ & 8.96 $\pm$ 0.16$^{\star}$ & 4 & 16 $\pm$ 5$^{\star}$ & 4 & 530 $\pm$ 437 & 960 $\pm$ 470 \\
    15 & $-$0.60 $\pm$ 0.06 & +0.55 $\pm$ 0.07 & $-$0.17 $\pm$ 0.11 & 8.96 $\pm$ 0.19 & 5 & 5 $\pm$ 1 & 5 & 180 $\pm$ 437 & 1960 $\pm$ 470 \\
    16 & $-$0.15 $\pm$ 0.09 & +0.42 $\pm$ 0.13 & $-$0.04 $\pm$ 0.14 & 9.13 $\pm$ 0.27 & 6 & 18 $\pm$ 5 & 5 & 930 $\pm$ 437 & 1160 $\pm$ 470 \\
    17 & $-$0.20 $\pm$ 0.09 & +0.46 $\pm$ 0.13 & $-$0.03 $\pm$ 0.14 & 9.13 $\pm$ 0.27 & 6 & 21 $\pm$ 4 & 5 & 330 $\pm$ 437 & 760 $\pm$ 470 \\
    18 & $-$0.25 $\pm$ 0.05 & +0.34 $\pm$ 0.07 & $-$0.21 $\pm$ 0.11 & 8.86 $\pm$ 0.18 & 5 & 12 $\pm$ 1 & 5 & 630 $\pm$ 437 & 860 $\pm$ 470 \\
    20 & $-$0.24 $\pm$ 0.05 & +0.55 $\pm$ 0.07 & $-$0.15 $\pm$ 0.11 & 8.96 $\pm$ 0.17 & 5 & 12 $\pm$ 1 & 5 & 380 $\pm$ 437 & 360 $\pm$ 470 \\
    21 & $-$0.22 $\pm$ 0.05 & +0.40 $\pm$ 0.07 & $-$0.08 $\pm$ 0.11 & 8.95 $\pm$ 0.19 & 5 & 12 $\pm$ 1 & 5 & 430 $\pm$ 437 & 720 $\pm$ 470 \\
    22 & $-$0.05 $\pm$ 0.13$^{\star\star}$ & 0.35 $\pm$ 0.12$^{\star\star}$ & $-$0.08 $\pm$ 0.14$^{\star\star}$ & 8.77 $\pm$ 0.25$^{\star\star}$ & 7 & 10 $\pm$ 2$^{\star\star}$ & 7 & 750 $\pm$ 437 & 320 $\pm$ 470 \\ \hline
    ID & [Na/Fe] & [Si/Fe] & [Ca/Fe] & [Sc/Fe] & [Ti/Fe] & [V/Fe] & [Cr/Fe] & [Co/Fe] & [Ni/Fe] \\
    & (dex) & (dex) & (dex) & (dex) & (dex) & (dex) & (dex) & (dex) & (dex) \\ \hline
    1 & --- & +0.34 $\pm$ 0.31 & +0.06 $\pm$ 0.18 & +0.67 $\pm$ 0.63 & +0.29 $\pm$ 0.22 & +0.13 $\pm$ 0.27 & +0.17 $\pm$ 0.20 & +0.31 $\pm$ 0.26 & +0.32 $\pm$ 0.24 \\
    3 & +0.35 $\pm$ 0.25 & +0.18 $\pm$ 0.28 & $-$0.02 $\pm$ 0.26 & +0.16 $\pm$ 0.25 & $-$0.02 $\pm$ 0.38 & $-$0.03 $\pm$ 0.39 & +0.05 $\pm$ 0.22 & +0.14 $\pm$ 0.22 & +0.02 $\pm$ 0.22 \\
    4 & +0.38 $\pm$ 0.18 & +0.32 $\pm$ 0.15 & +0.57 $\pm$ 0.40 & +0.27 $\pm$ 0.17 & $-$0.05 $\pm$ 0.34 & +0.13 $\pm$ 0.32 & +0.15 $\pm$ 0.17 & +0.01 $\pm$ 0.19 & +0.28 $\pm$ 0.15 \\
    6 & +0.27 $\pm$ 0.25 & +0.08 $\pm$ 0.15 & +0.11 $\pm$ 0.25 & +0.24 $\pm$ 0.18 & +0.38 $\pm$ 0.33 & +0.46 $\pm$ 0.21 & +0.37 $\pm$ 0.24 & +0.33 $\pm$ 0.53 & +0.12 $\pm$ 0.14 \\
    8 & +0.15 $\pm$ 0.18 & +0.17 $\pm$ 0.14 & $-$0.07 $\pm$ 0.19 & +0.31 $\pm$ 0.13 & +0.00 $\pm$ 0.26 & +0.17 $\pm$ 0.28 & $-$0.01 $\pm$ 0.20 & +0.18 $\pm$ 0.19 & +0.26 $\pm$ 0.18 \\ \hline
    ID & [Y/Fe] & [Ce/Fe] & [Eu/Fe] \\
    & (dex) & (dex) & (dex) \\ \hline
    1 & $-$0.14 $\pm$ 0.21 & $-$0.09 $\pm$ 0.18 & +0.26 \\
    3 & $-$0.15 $\pm$ 0.16 & --- & --- \\
    4 & +0.13 $\pm$ 0.18 & --- & --- \\
    6 & +0.12 $\pm$ 0.19 & +0.44 $\pm$ 0.19 & --- \\
    8 & +0.44 $\pm$ 0.24 & --- & --- \\ \hline
    \end{tabular}
    $^{\star}$ Mean values of cluster RGB stars.\\
    $^{\star\star}$ Values obtained for Aldebaran ($\alpha$ Tau).\\
    References regarding the CNO abundances and $^{12}$C/$^{13}$C: (1) This work, (2) \cite{souto}, (3) \cite{grazina2000}, (4) \cite{penhasuarez}, (5) \cite{rodolfo2009}, (6) \cite{morel2014} and (7) \cite{abia2012}.
    \label{final}
    \end{threeparttable}
\end{table*}

\subsubsection{Na, Si, Ca and Sc abundances}

Na abundance ratios found are similar to those of the literature for giant field stars and open clusters, although our mean $\langle$[Na/Fe]$\rangle$ is 0.21 dex higher than the mean of the literature. The values obtained are likely attributed to the large EW measurements for the Na lines in our sample (see Table \ref{ew-tabela}). The Na abundance obtained for object 3 (NGC188-2072) agrees with the value from \citet{friel2010} within the uncertainties. Consequently, we consider our results to be in agreement with the literature, despite the slightly higher values.

Si abundances have values between $-$0.08 and 0.34 dex, with a dispersion of 0.05 dex considering the mean of 0.22 dex. These values are similar to those obtained in the literature for NGC188, within their errors, for giant field stars, and for other open clusters. The results for Ca abundances are comparable to the literature values, within the uncertainties. However, object 4 (NGC188-2026) has an abundance of about 0.5 dex higher than the mean of the literature, which can be related to measurements of equivalent width in spectra with low S/N (S/N $\sim$ 27). If we exclude object 4, we obtained a mean of 0.02 dex, which is in better agreement with the literature values within their errors.

The value found for $\langle$[Sc/Fe]$\rangle$ is 0.34 dex larger than that obtained by \citet{occaso}, which is $-$0.01 dex. If we exclude object 1 (NGC188-3018), that has an extreme value, 0.67 dex, we get a mean of 0.20 dex, still 0.14 dex above \citet{occaso} value. This difference can be related to the number of spectral lines and their spectroscopic parameters used in both works. We used two spectral lines, while \citet{occaso} used another five spectral lines. One of their stars has an abundance of [Sc/Fe] $=0.14$ dex, corrected to the solar values adopted in this work, \citet{asplund}, with a difference of 0.05 dex with the rest of the sample. A more recent study performed by \citet{occaso2024} using four stars with nine Sc spectral lines indicated an abundance of [Sc/Fe]$=0.10 \pm 0.04$ dex, corrected to the solar values adopted in this work, \citet{asplund}; although our best result is 0.14 dex above their results, it still is within the error limit because of its uncertainty (0.07 dex). Therefore, we consider that our results are consistent with the literature.

\subsubsection{Ti, V, Cr and Co abundances}

The Ti abundances obtained agree with literature values for giant field stars and open clusters. The values obtained for $\langle$[Ti/Fe]$\rangle$ agree with the abundances calculated by \citet{jacobson2011} within the uncertainties.

V abundances found are $\sim$ 0.13 dex larger than the mean from literature of [\ion{V}{i}/Fe] and [\ion{V}{ii}/Fe]. \citet{occaso} used spectral synthesis of five spectral lines of \ion{V}{i} and found [\ion{V}{i}/Fe]$=0.01\pm0.05$ dex (mean for two stars), $\sim$ 0.13 dex lower than what we obtained in this work ($\langle$[\ion{V}{i}/Fe]$\rangle=0.13\pm0.07$ dex). \citet{donor2018} and recently \citet{donor2020} used measurements of equivalent widths for two spectral lines of \ion{V}{i} in the infrared and found, respectively, [\ion{V}{i}/Fe]$=0.03\pm0.08$ dex and [\ion{V}{i}/Fe]$=-0.03\pm0.14$ dex, with a large uncertainty. When we compare our results with field stars (see Figure \ref{element}), \citet{occaso}, and \citet{donor2020}, we see that our V abundance has a large deviation. In addition, according to \citet{scott}, it is expected that \ion{V}{i} exhibits NLTE effects, considering that only \ion{V}{ii} lines are expected to be formed in LTE. Therefore, \citet{scott} indicate an \textit{ad hoc} correction of $+0.1$ dex for all \ion{V}{i} lines. In conclusion, we consider that our results are consistent with the literature, which also show a wide dispersion.

Cr and Co abundances have values with the mean of 0.15 dex and 0.19 dex, respectively. Our Cr abundances are close to the solar value and similar to those obtained in the literature for the cluster, within the errors, and for other open clusters and giant stars. Co values are marginally consistent with the literature, given the errors. Cr and Co abundances obtained by \citet{friel2010} for the object 3 (NGC188-2072) are based on measurements of one spectral line, presenting a large uncertainty. Our results consider 13 spectral lines of \ion{Cr}{i} and 7 spectral lines of \ion{Co}{i}. Therefore, we conclude that our results are more precise than those of \citet{friel2010}.

\subsubsection{Ni, Y, Ce and Eu abundances}

In Figure \ref{element}, the star-to-star Ni and Y abundances agree with literature data for giant field stars and open clusters, within the errors. The mean Ni abundance is higher by $\sim$ 0.13 dex when compared the literature values. If we exclude object 1 (NGC188-3018), which has an extreme value, the mean offset becomes 0.10 dex, which remains consistent with the literature within the errors. Our mean Y abundance differs by 0.10 dex from that of \citet{slumtrup2018}, and it is consistent, within the uncertainties. This difference can be explained by the number of spectral lines used in each work. While \citet{slumtrup2018} used three spectral lines for a single star in the cluster, we used one reliable spectral line across five objects.

Ce and Eu abundances have few data available in the literature, nonetheless, the abundances we found agree with the mean of open clusters within $\sim$ 0.2 dex. Most equivalent widths for these elements are smaller than or close to the noise limit ($EW <$ 30 m\AA). Therefore, we have few abundances determined for the stars of the sample for these two species. In spite of these observation limitations, the results remain consistent with the literature within the uncertainties.

\subsection{CNO abundances and isotopic ratios}

The abundance ratios for CNO and the $^{12}$C/$^{13}$C isotopic ratio for NGC188 stars were obtained by spectral synthesis. For the infrared sample of stars (objects 11 to 22, except object 19 (NGC3532-MMU649)), we derived only the $^{16}$O/$^{17}$O and $^{16}$O/$^{18}$O isotopic ratios using spectral synthesis. In contrast, abundance ratios for C, N, O and the $^{12}$C/$^{13}$C isotopic ratios for the stars with IR data were obtained in the literature.

\subsubsection{CNO abundances}

The major difference is found for the oxygen abundance ($\sim$ 0.2 dex). \citet{friel2010} found [O/Fe] $=-$0.18 dex for the object 3 (NGC188-2072) using the spectral synthesis of the spectral line [\ion{O}{i}] at 6300.3 \AA. The authors still obtain a total mean oxygen abundance for the cluster NGC188 of [O/Fe] $=-0.04\pm0.10$ dex, using spectra of intermediate resolution ($R\sim28000$) in the optical region (5000$-$8300 \AA) for four stars in the cluster. \citet{donor2018} obtain the oxygen abundance analysing high-resolution spectra of 13 stars in the region of the $H$-band of near infrared (1.5$-$1.7 $\mu$m), with a mean of [O/Fe] $=0.02\pm0.04$ dex. Analysing 14 stars in that same region, \citet{donor2020} found a mean of [O/Fe] $=0.00\pm0.05$ dex and \citet{occaso} with five spectra at high resolution ($R\geqslant65000$) in the optical region (4000$-$9000 \AA) obtain [O/Fe] $=0.05\pm0.03$. Therefore, there is a typical difference of $\sim0.2$ dex relative to the abundances found in this work, which is of the same order of magnitude as the uncertainties in our data.

We show in Figure \ref{element} the calculated CNO abundance ratios for the NGC188 sample and compare with the literature data for other stars and open clusters. They present values similar to those found in the literature, in which there is a decrease of carbon and oxygen abundances with increasing metallicity of open clusters. Also, we see an overabundance of nitrogen in the sample ([N/Fe] > 0.0 for all-stars), as expected by stellar evolution models. The reason for this phenomenon observed in giant stars is that carbon inside the star is depleted in favour of nitrogen through the CN cycle \citep{2000A&A...354..169G,mishenina-2006,luck-heiter}. The sum of the CNO abundances for NGC188 stars varies from 8.72 to 9.30 dex (see Table \ref{final}), with an average of 9.07 $\pm$ 0.09 dex. This value is close to the solar one \citep[8.92 $\pm$ 0.10 dex,][]{asplund}, which almost agrees with the solar average metallicity of the NGC188, $\langle$[Fe/H]$\rangle=0.04\pm0.04$ dex.

Objects 3 (NGC188-2072) and 4 (NGC188-2026) present the highest nitrogen abundance values. The possible cause is related to the spectral synthesis analysis of nitrogen. In the region used for that, there is a blend with the spectral line of \ion{Si}{i} that makes it difficult to reliably fit a synthetic spectrum to the noisy observed one (see Figure \ref{sintese}). We cannot clearly identify the evolution of these elements along the RGB through our abundance of carbon and nitrogen. Therefore, we consider the carbon and nitrogen abundances derived to be consistent with the data in the literature within the uncertainties.

\subsubsection{$^{12}$C/$^{13}$C isotopic ratio}

\begin{figure*}
    \includegraphics[angle=-90,width=1.03\columnwidth]{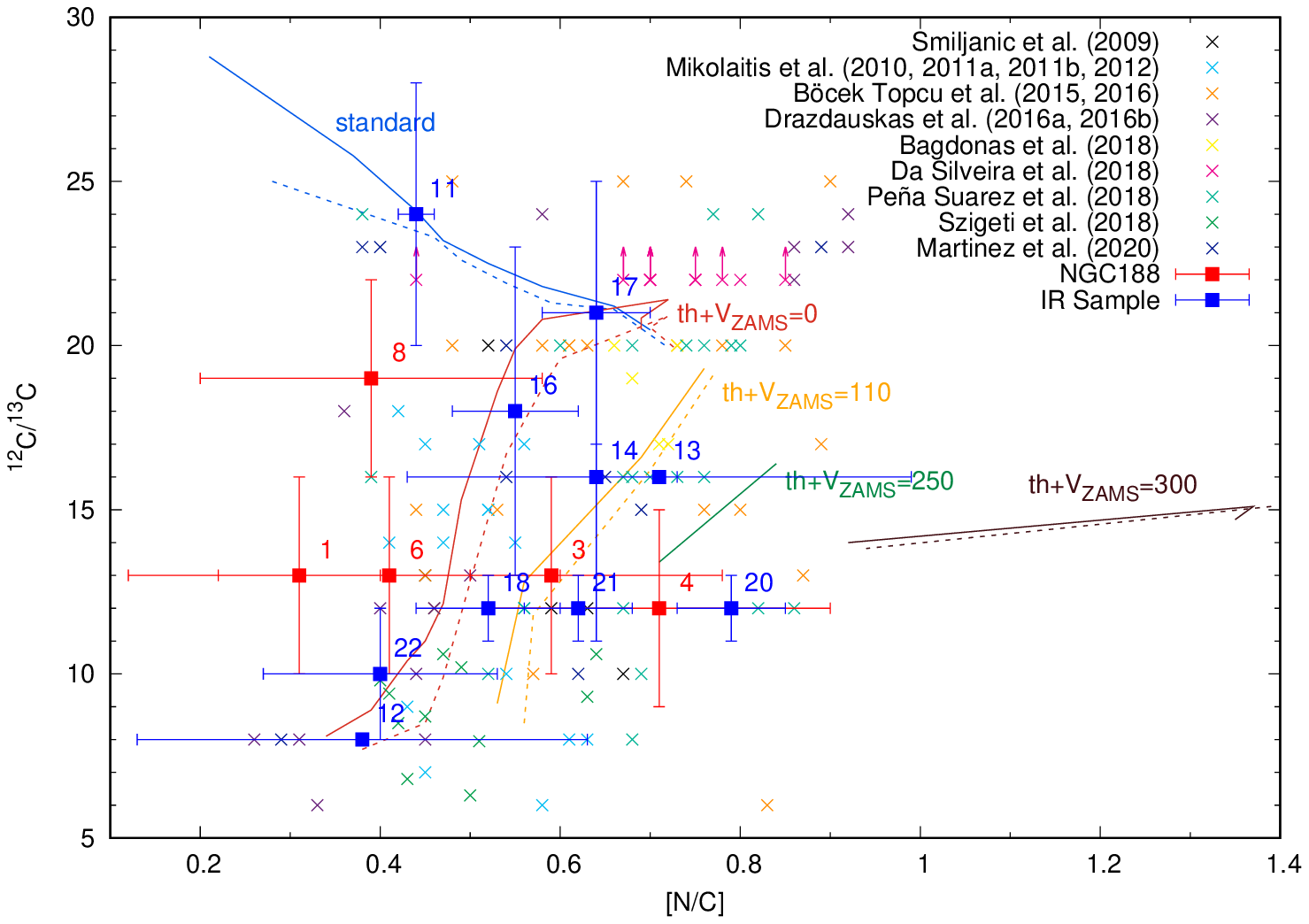}
    \includegraphics[angle=-90,width=1.0\columnwidth]{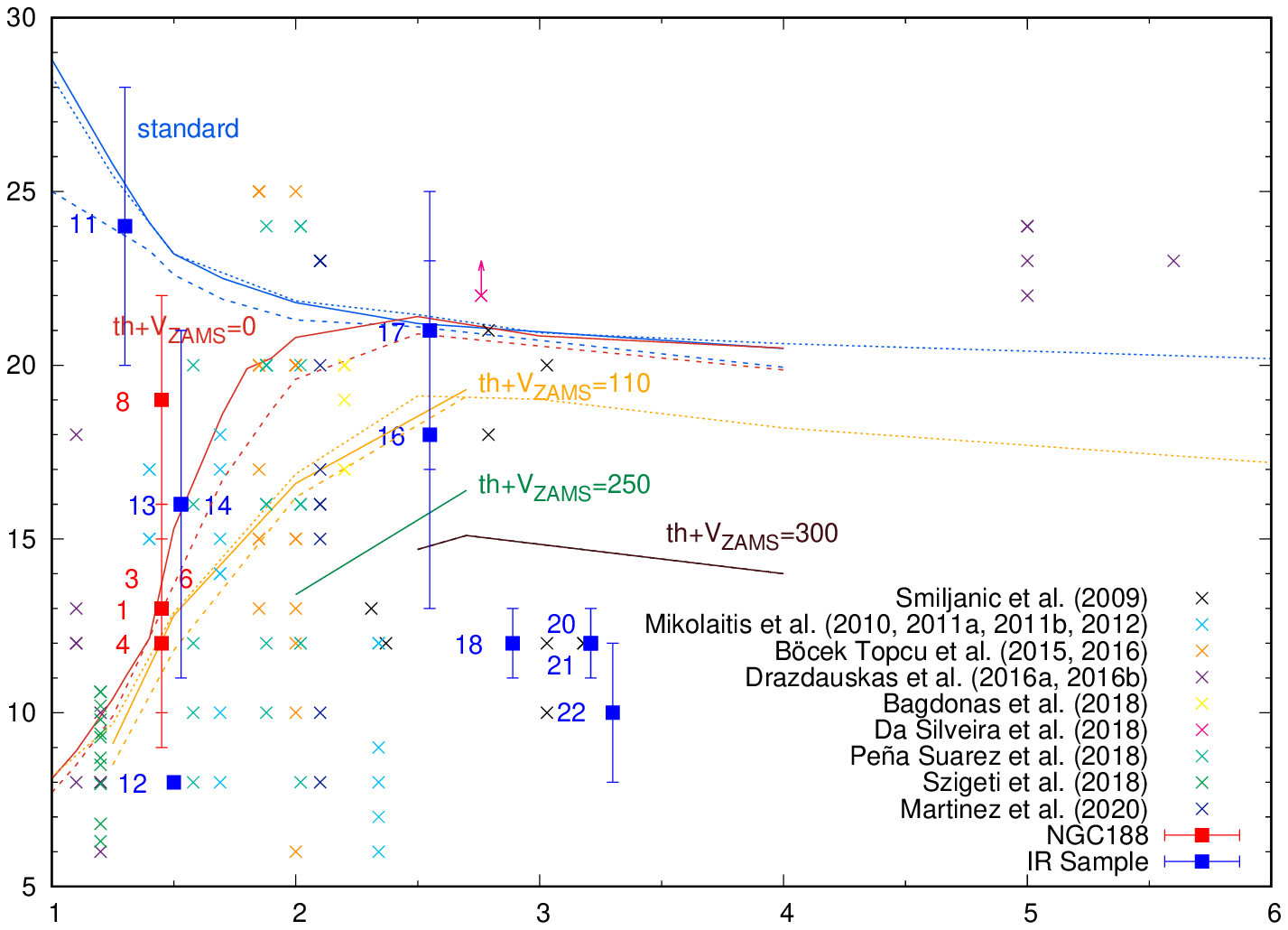}
    \caption{$^{12}$C/$^{13}$C isotopic ratio as a function of abundance ratio $\textup{[N/C]}$ (\emph{left}) and turn-off mass (\emph{right}), compared with open clusters data from the literature \citep{rodolfo2009,mikolaitis2010,mikolaitis2011a,mikolaitis2011b,mikolaitis2012,topcu2015,topcu2016,drazultimo,drazprimeiro,bagdonas,dasilveira,penhasuarez,szigeti,martinez} and models of \citet{charbonnel-mixing} at the tip of the RGB (continuous lines) and at the end of the second dredge-up on the early-AGB (dashed lines). The predictions of the standard models are in blue. The salmon lines are for models including thermohaline diffusion. The orange, green, and brown lines are for models including both thermohaline and rotation-induced mixing for different initial rotation velocities at the zero age main sequence. The standard models from \citet{lagarde} of first dredge-up (FDUP) and thermohaline diffusion with rotation-induced mixing (TH+V) are indicated with dotted lines.}
    \label{ratio-resultado1}
\end{figure*}

The results of $^{12}$C/$^{13}$C isotopic ratio as a function of both the abundance ratio [N/C] and the turn-off mass can be seen in the right and left panels of Figure \ref{ratio-resultado1}, respectively. Despite the fact that these objects are distributed along the RGB, stars 1 (NGC188-3018), 3 (NGC188-2072), and 6 (NGC188-3140) have similar $^{12}$C/$^{13}$C isotopic ratio. Our values have error bars with a large value due to the composition of uncertainties, mostly because of $\log g$ (see Table \ref{incerteza}). Also, our results are in excellent agreement with data found in the literature for other open clusters (see references in figures, as well as other papers, e.g. \citealt{sales-silva,grazina2015,grazina2016,bagdonas,penhasuarez})
and in field giant stars \citep[e.g.][not shown in figures]{1998A&A...332..204C,tsuji,abia2012}.

We also compare in Figure \ref{ratio-resultado1} the cluster data with predictions of the stellar evolution models of \citet{charbonnel-mixing} and \citet{lagarde} computed at solar metallicity ($Z = 0.014$) with the same evolution code and very similar assumptions on the treatment of internal mixing processes (see Table \ref{modelos}). In standard models, which include only convection as a mixing process, the surface carbon isotopic ratio decreases from its initial value only during the first dredge-up, when the deepening convective envelope engulfs the $^{13}$C peak that has build-up on the main sequence. This is clearly not sufficient to explain the low values of $^{12}$C/$^{13}$C and their behaviour as a function of [C/N] and stellar mass, as already noticed by many other studies.

However, model predictions change significantly when thermohaline diffusion and rotation-induced mixing are included in the computations. As explained in the original papers, the respective efficiencies of these two processes depend on the mass domain. The impact of thermohaline diffusion, which sets in at the RGB bump, dominates in the stars with masses $M\lesssim2.2\,M_{\astrosun}$ (more massive stars do not cross the RGB bump, as they ignite central He in non-degenerate conditions), and it is stronger for the lowest stellar masses. On the other hand, rotation-induced mixing, which already modifies the internal isotope abundance profiles when the stars are on the main sequence, can explain the low carbon isotopic ratios as well as the higher [N/C] values for more massive stars, as well as the dispersion over the entire mass range. 

Given the relatively large uncertainties in the derivation of both ratios, it is, however, not possible to draw an obvious sequence of their behaviour along the evolution phases we probe. We note, e.g., that object 8 (NGC188-2187) has an isotopic ratio larger than object 6 (NGC188-3140), despite both objects being at the same evolutionary stage. This can indicate that the thermohaline process has just started in object 8 (NGC188-2187). Since the values of $^{12}$C/$^{13}$C we obtained are lower than the standard model predictions for most of the stars in our sample, we conclude that probably most of the sample has already experienced the effect of thermohaline diffusion and/or rotation-induced mixing depending on their mass.

\begin{table*}
    \caption{Literature models used in the comparison of the results for $^{12}$C/$^{13}$C, $^{16}$O/$^{17}$O and $^{16}$O/$^{18}$O.}
    \begin{threeparttable}
    \begin{tabular}{llc} \hline \hline
        Label & Model & Ref.\\ \hline
         standard & Standard (no mixing process beyond convection), first and second dredge-up & 1 \\ 
         th+V$_{\textrm{ZAMS}}=0$ & Only thermohaline mixing (without rotation) & 1 \\
         th+V$_{\textrm{ZAMS}}=110$ & Thermohaline and rotation-induced mixing with initial velocity V$_{\textrm{ZAMS}}=110$ \kms & 1 \\
         th+V$_{\textrm{ZAMS}}=250$ & Thermohaline and rotation-induced mixing with initial velocity V$_{\textrm{ZAMS}}=250$ \kms & 1 \\
         th+V$_{\textrm{ZAMS}}=300$ & Thermohaline and rotation-induced mixing with initial velocity V$_{\textrm{ZAMS}}=300$ \kms & 1 \\
         FDUP & Standard model, first dredge-up & 1, 2, 3, 4$^{\star}$, 5\\
         TH+V & Thermohaline and rotation-induced mixing with initial velocity $90\leqslant V_{\textrm{ZAMS}} \leqslant137$ \kms & 3 \\ \hline
         & Initial chemical composition changed$^{\star\star}$ & \\ \hline
         C18OHH & $^{16}$O/$^{17}$O $=2696$ and $^{16}$O/$^{18}$O $=249$ & 6 \\
         C16OLL & $^{16}$O/$^{17}$O $=1348$ and $^{16}$O/$^{18}$O $=249$ & 6 \\
         C16OL & $^{16}$O/$^{17}$O $=1617$ and $^{16}$O/$^{18}$O $=299$ & 6 \\ \hline
         & Nuclear reaction rates of $^{17}$O and $^{18}$O changed$^{\star\star\star}$ & \\ \hline
         O17L & Low rate of $^{17}$O and recommended for $^{18}$O & 6 \\
         O18L & Recommended rate of $^{17}$O and low rate of $^{18}$O & 6 \\
         O17H & High rate of $^{17}$O and recommended for $^{18}$O & 6 \\
         O18H & Recommended rate of $^{17}$O and high rate of $^{18}$O & 6 \\
         O17BH & Recommended rate of $^{17}$O by \cite{bruno2016} & 5 \\
         O17BL & Recommended rate of $^{17}$O by \cite{bruno2016} and initial chemical composition $^{16}$O/$^{17}$O $=1797$ & 5 \\ \hline
    \end{tabular}
    V$_{\textrm{ZAMS}}$: Velocity of stellar rotation at Zero Age Main Sequence.\\
    $^{\star}$ Nuclear reaction rate $^{17}$O(p,$\alpha$)$^{14}$N from \cite{iliadis} and solar initial chemical composition \citep[$^{16}$O/$^{17}$O $=2696$,][]{lodders} for \cite{straniero} model.\\
    $^{\star\star}$ Solar values as references to initial chemical composition: $^{16}$O/$^{17}$O $=2696$ and $^{16}$O/$^{18}$O $=499$ \citep{lodders}.\\
    $^{\star\star\star}$ Nuclear reaction rate compared to recommend values from \cite{iliadis} to $^{17}$O(p,$\alpha$)$^{14}$N  and $^{18}$O(p,$\alpha$)$^{15}$N.\\
    References: (1) \cite{charbonnel-mixing}, (2) \citet{boothroyd-sackmann}, (3) \cite{lagarde}, (4) \citet{karakas-lattanzio}, (5) \citet{straniero} and (6) \citet{lebzelter2015}.
    \end{threeparttable}
    \label{modelos}
\end{table*}

\subsubsection{$^{16}$O/$^{17}$O and $^{16}$O/$^{18}$O isotopic ratios}

\begin{figure*}
    \includegraphics[angle=-90,width=1.03\columnwidth]{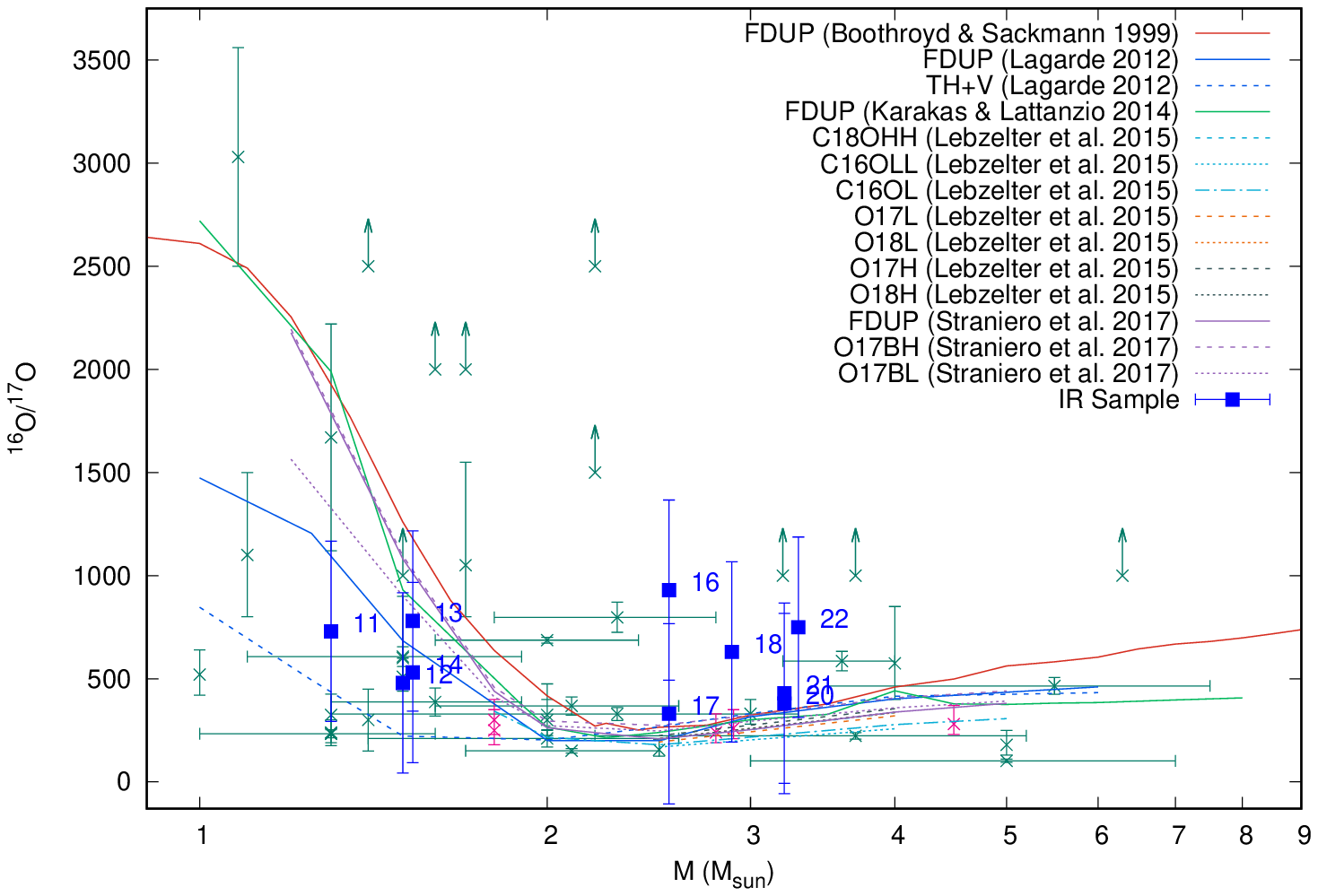}
    \includegraphics[angle=-90,width=1.03\columnwidth]{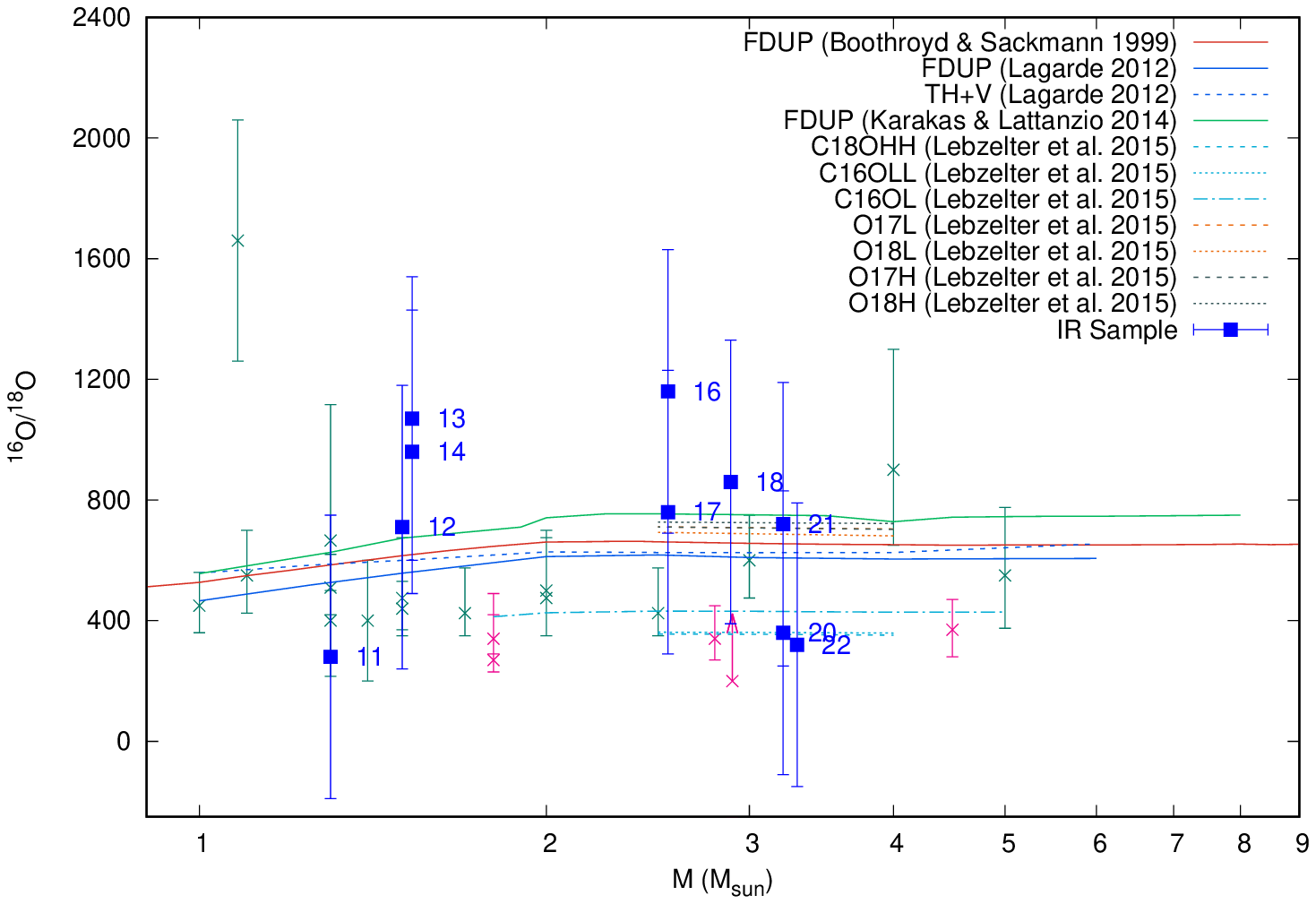}
    \caption{$^{16}$O/$^{17}$O (\textit{left}) and $^{16}$O/$^{18}$O (\textit{right}) isotopic ratios as function of turn-off mass for analysed stars (blue symbols), compared with literature data. Green symbols represent giant field starts \citep[][all the lower limits for $^{16}$O/$^{17}$O are from Tsuji 2008]{harrislambert84b,harrislambert85b,harrislambert87,harrislambert88,tsuji,abia2012} and pink symbols represent open cluster members \citep{lebzelter2015}. Literature models presented in Table \ref{modelos} are represented as lines. Continuous lines correspond to models of standard first dredge-up \citep[FDUP;][]{boothroyd-sackmann,lagarde,karakas-lattanzio,lebzelter2015,straniero}. Dashed and dotted lines correspond to models of thermohaline and rotation-induced mixing from \citet[][TH+V]{lagarde}, models obtained through variations in parameters in the stellar evolution code from \citet{lebzelter2015} and \citet{straniero}.}
    \label{oxy}
\end{figure*}

\begin{figure*}
    \includegraphics[angle=-90,width=1.03\columnwidth]{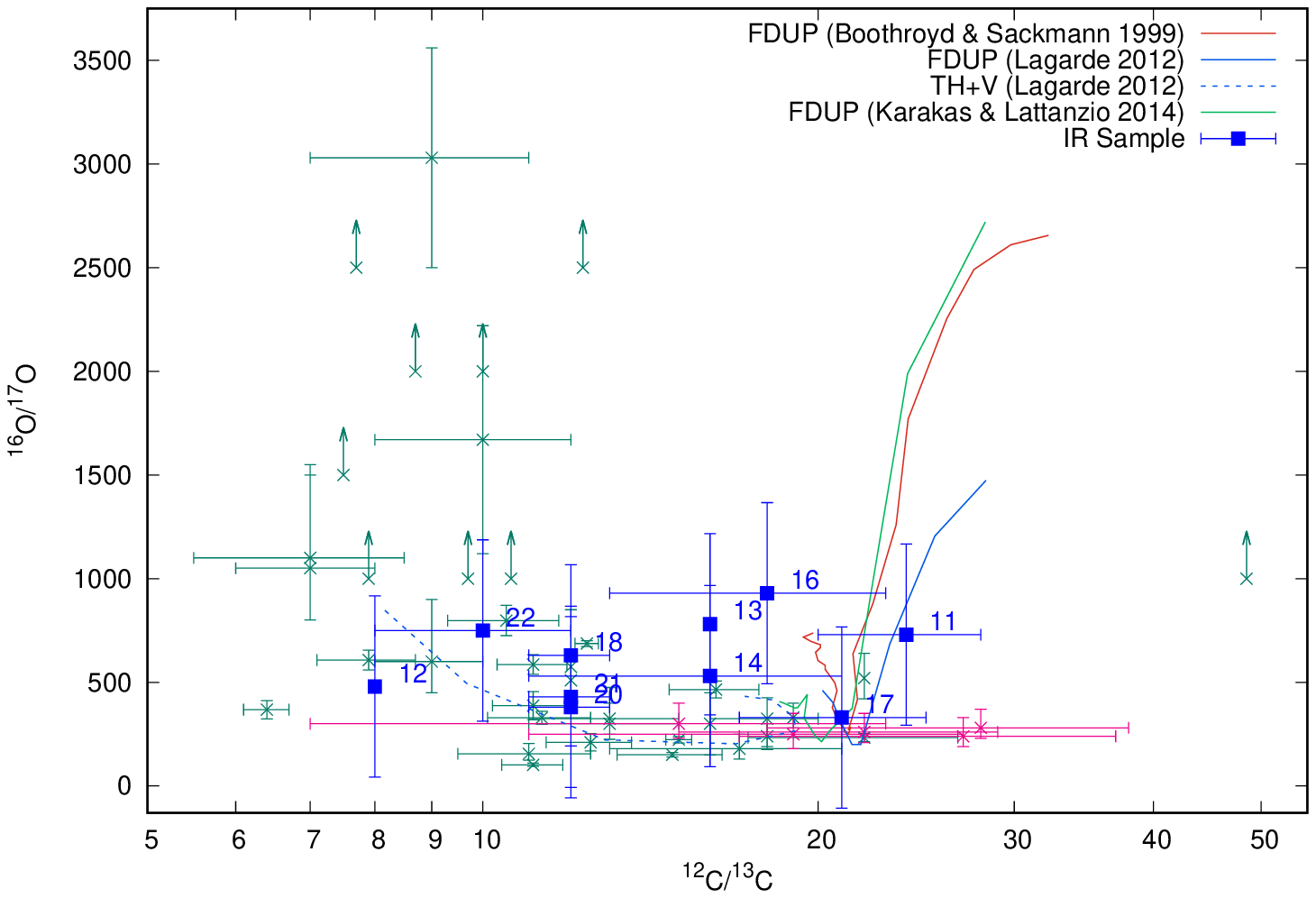}
    \includegraphics[angle=-90,width=1.03\columnwidth]{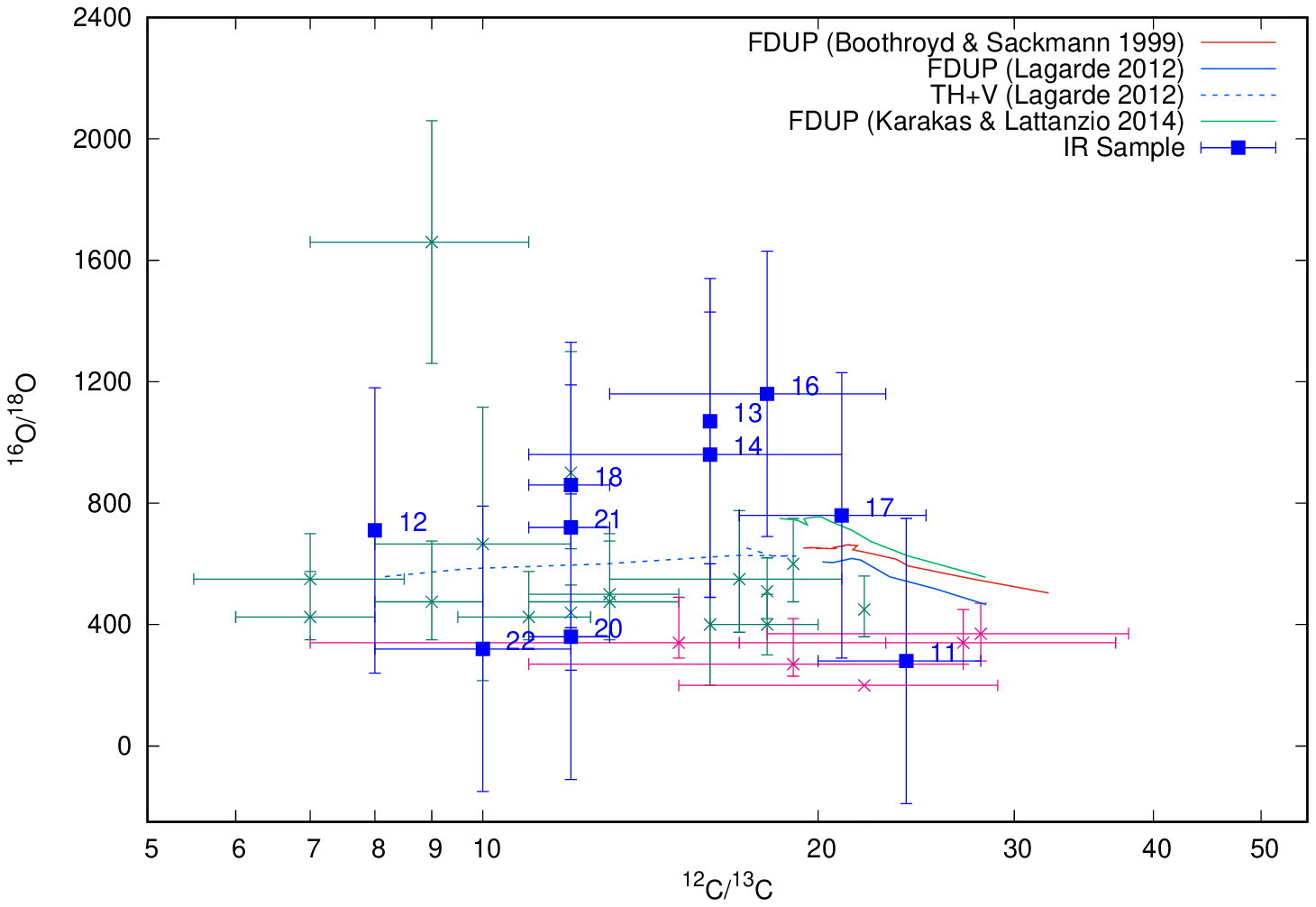}
    \caption{$^{16}$O/$^{17}$O (\textit{left}) and $^{16}$O/$^{18}$O (\textit{right}) as a function of $^{12}$C/$^{13}$C isotopic ratio for analysed stars (blue symbols), compared with literature data.  Green symbols represent giant field starts \citep{harrislambert84b,harrislambert85b,harrislambert87,harrislambert88,tsuji,abia2012} and pink symbols represent open cluster members \citep{lebzelter2015}.  Literature models presented in Table \ref{modelos} are represented as lines. Continuous lines correspond to models of standard first dredge-up \citep[FDUP;][]{boothroyd-sackmann,lagarde,karakas-lattanzio}. Dashed line corresponds to model of thermohaline and rotation-induced mixing from \citet[][TH+V]{lagarde}.}
    \label{oxy-carbon}
\end{figure*}

We show the results obtained for $^{16}$O/$^{17}$O and $^{16}$O/$^{18}$O isotopic ratios of the infrared sample (objects 11 to 22) as a function of the turn-off mass of the clusters in Figure \ref{oxy}. As in \citet{rodolfo2009}, we exclude object 15 (IC4756-69) from the following discussion, as it belongs to a spectroscopic binary system and is suspected to have undergone mass transfer from its low-mass companion \citep{vanderswaelmen}. For the rest of the IR sample, the oxygen isotopic ratios range between $380\leqslant \,^{16}\textrm{O/}^{17}\textrm{O}\leqslant930$ and $280\leqslant\,^{16}\textrm{O/}^{18}\textrm{O}\leqslant1160$. Within the error bars, however, they are consistent with a relatively flat distribution as a function of stellar mass. As can be seen in the figures, this is consistent, considering the uncertainties, with the oxygen isotopic ratios found in the literature \citep{harrislambert84a,harrislambert84b,harrislambert87,harrislambert88,tsuji,abia2012,lebzelter2015}. Some objects analysed by \citet{tsuji} present a lower estimate of $^{16}$O/$^{17}$O isotopic ratio due to unknown impurities in the weak spectral lines of C$^{17}$O.  

We compare oxygen isotopic data to the predicted post-first dredge-up values from different stellar evolution models (references in figures and in Table \ref{modelos}) in Figure \ref{oxy}. 
The standard models (full lines) account for the $^{16}$O/$^{17}$O ratios for our stars with masses higher than $\sim$2\,M$_{\odot}$, as well as for stars from the literature within the same mass domain. In this mass range, the $^{17}$O peak that formed during the main sequence is almost fully engulfed by the convective envelope when the stars become giants. However, in less massive stars, the base of the convective envelope reaches only the external part of the $^{17}$O peak during the first dredge-up, leading to a more modest decrease of the $^{16}$O/$^{17}$O at the end of this episode. The lower post dredge-up values of the standard models of \citet{lagarde} actually result simply from a lower initial value of this isotopic ratio in their calculations (1582 instead of $\sim$ 2700 in the other studies), which
leads to a fair agreement with the data within the entire mass range. Good agreement is also obtained with the model predictions of \citet{straniero} who used updated values for the $^{17}$O(p,$\alpha$)$^{14}$N reaction rate \citep{bruno2016}. Finally, while thermohaline mixing on the RGB does not affect the $^{16}$O/$^{17}$O ratio, rotation-induced mixing spreads out the internal $^{17}$O peak during the main sequence, allowing more of this isotope to be ingested in the convective envelope during the first dredge-up, hence leading to  lower post dredge-up values of the isotopic ratio compare to standard models in the low-mass domain  (\citealt{lagarde}; see also \citealt{Halabi2015} who studied the impact of convective overshoot on the internal CNO abundance profiles).

We note that our objects 16 (NGC6633-78), 18 (NGC3532-MMU19), and 22 (NGC5460-MMU17) have $^{16}$O/$^{17}$O isotopic ratio slightly higher than all the model predictions, possibly due to the location of the continuum level and the blending of the weak lines of C$^{17}$O as in the case of stars with lower limits \citep[e.g. see][]{tsuji}. Given these different theoretical uncertainties, as well as the relatively large observational error bars which are larger than the differences between the models, it is difficult to use the $^{16}$O/$^{17}$O ratios to precisely identify the best model.

On the other hand, we see in Figure \ref{oxy} that within the error bars, all the standard models reproduce fairly well the $^{16}$O/$^{18}$O ratios we derived for our IR sample stars over the entire mass range (all the theoretical studies we consider assumed the same initial value for $^{16}$O/$^{18}$O, with  \citealt{lebzelter2015} using this as a free parameter). This also applies to the models including rotation-induced mixing and thermohaline diffusion, as these two mechanisms barely affect the post dredge-up value of this isotopic ratio \citep{lagarde}. This is due to the fact that $^{18}$O is only destroyed through the CNO cycle, in layers that are less deep than the position of the $^{17}$O peak, implying that the post dredge-up $^{16}$O/$^{18}$O value is insensitive to variations of the mixing length for convection and of the proton-capture rates adopted in the stellar models \citep{lebzelter2015}. We thus confirm that this isotopic ratio is not a good tracer of non-standard mixing processes in stars. Consequently, the only possibility to fit the lowest ratios found in some of our stars (objects 11 (NGC2682-MMU6495), 20 (NGC6281-3), and 22 (NGC5460-MMU17)) and in the literature, is to combine a higher initial $^{18}$O and a lower initial $^{16}$O \citep{lebzelter2015}. Given the observational uncertainties, it is however difficult to conclude, or to exclude, that the star-to-star differences of the $^{16}$O/$^{18}$O ratio reflect initial variations at the birth of the red giant progenitors.

Finally, in Figure \ref{oxy-carbon} we plot the oxygen isotopic ratios as a function of the carbon isotopic ratio for our sample stars and for the data from the literature. We also compare the observed values to some of the theoretical models discussed previously. The figure summarises the conclusions of our analysis well: While rotation-induced mixing and thermohaline diffusion (with respective efficiencies that depend on the stellar masses) are required to explain the low $^{12}$C/$^{13}$C values exhibited by red giants, the oxygen isotopic ratios do not bring additional valuable constraints on the non-standard mixing processes.

\section{Conclusions}

This work considered a sample of nine open clusters in which 22 objects were analysed using spectroscopy in visible (objects 1 to 10, from NGC188 cluster) and infrared (objects 11 to 22). The spectra analysed had their heliocentric radial velocities obtained using \textsc{iraf} software. The results agree with literature values of the radial velocity and we found that object 19 (NGC3532-MMU649) is a non-member of NGC3532.

We determined the photometric atmospheric parameters (effective temperature $T_{\textrm{eff}}$, surface gravity $\log g$, metallicity [Fe/H] and microturbulent velocity $\xi$) for all sample using mean values of photometric calibrations from \citet[$J-K$, $V-K$, $B-V$ and $J-H$,][]{alonso}, \citet[$V-K$,][]{van-Belle} and \citet[$V-K$,][]{Huang}. The results agree with the values of the literature, considering the uncertainties.

For NGC188 stars (objects 1 to 10), we obtained spectroscopic atmospheric parameters, macroturbulence velocity $\zeta$, and projected rotational velocity $v\sin i$. The results obtained are similar to the values in the literature. We also calculated from measurements of equivalent width the abundances of Na, Si, Ca, Ti, V, Cr, Co, Ni, Y, Cu, and Eu. The values obtained are similar to those of giant stars in the field and in open clusters available in the literature. Using spectral synthesis, we calculated abundances of CNO and $^{12}$C/$^{13}$C isotopic ratios. The results are similar to those obtained in the literature for open clusters and support the importance of thermohaline diffusion in stars with masses below $\sim$ 2~M$_{\odot}$ and of rotation-induced mixing in more massive stars as predicted by \citet{lagarde}.

For objects 11 to 22, using photometric atmospheric parameters obtained before and literature values for CNO abundances and $^{12}$C/$^{13}$C isotopic ratios, we determined for the first time their $^{16}$O/$^{17}$O and $^{16}$O/$^{18}$O isotopic ratios through spectral synthesis. The derived values are similar to those of giant stars in the literature. We confirm that while the determination of $^{16}$O/$^{17}$O in red giants can help constraining the effect of rotation-induced mixing in low-mass stars when they were on the main sequence, $^{16}$O/$^{18}$O does not bring key additional information on non-standard mixing processes. Within the observational uncertainties, the data may actually reflect initial variations of the oxygen isotopes. Future determinations of oxygen isotopic ratios using the under-exploited infrared regions, especially at $\sim4.6\,\mu$m, shall bring a new light on this possibility.

\section*{Acknowledgements}

This study was funded by the Coordenação de Aperfeiçoamento de Pessoal de Nível Superior, Brasil (CAPES) - Funding Code 001 for the master scholarship linked to the Academic Excellency Program (PROEX) of CAPES, process 88882.345474/2019-01. CC acknowledges support by the Swiss National Science Foundation (Project 200020-192039). R. Smiljanic acknowledges support from the National Science Centre, Poland, research grant 2024/52/L/ST9/00220. Based on observations obtained with Brazilian time at the Canada-France-Hawaii Telescope (CFHT) under the program 09BB02, provided by an agreement between the Laboratório Nacional de Astrofísica (LNA/MCTI) and the CFHT, which is operated by the National Research Council of Canada (NRC), the Institut National des Sciences de l’Univers (INSU) of the Centre National de la Recherche Scientifique (CNRS) of France, and the University of Hawaii (UH). The observations at the CFHT were performed with care and respect from the summit of Maunakea which is a significant cultural and historic site. The ESPaDOnS data were reduced using the Upena pipeline, which uses the software Libre-ESpRIT, written by J.-F. Donati from Observatoire Midi-Pyrenees, and provided by CFHT. This research made use of the SIMBAD database, and the VizieR catalogue access tool operated at the CDS, Strasbourg, France, and of NASA Astrophysics Data System Bibliographic Services. \textsc{iraf} is distributed by the National Optical Astronomy Observatories (NOAO), which are operated by the Association of Universities for Research in Astronomy (AURA), under cooperative agreement with the National Science Foundation (NSF). Based on observations made with European Southern Observatory (ESO) telescopes at the La Silla Paranal Observatory under programmes ID 089.D-0716(A) and 089.D-0716(B). This publication makes use of data products from the Two Micron All Sky Survey (2MASS), which is a joint project of the University of Massachusetts and the Infrared Processing and Analysis Center/California Institute of Technology, funded by the National Aeronautics and Space Administration and the National Science Foundation. This research has made use of the WEBDA database, operated at the Department of Theoretical Physics and Astrophysics of the Masaryk University.

\section*{Data Availability}

The data underlying this article will be shared on reasonable request to the corresponding author.



\bibliographystyle{mnras}
\bibliography{example} 



\appendix

\section{Tables}\label{sec:ew2}

\begin{table*}
    \centering
    \caption{Heliocentric radial velocities $V_h$ for the stellar sample compared with the literature $V_{h,lit}$ and with the \emph{Gaia} DR3 radial velocities $V_{h,DR3}$. The normalised unit weight error (RUWE) \emph{Gaia} parameter is also given: it is close to $1.0$ if the fit to the \emph{Gaia} raw data is satisfactory, while a value larger than $\sim 1.4$ betrays some problems, often related to unrecognised multiplicity \citep{lindegren}.}
    \begin{threeparttable}
    \begin{tabular}{ccccccc}  \hline \hline
    ID & $V_h$ (km s$^{-1}$) & $V_{h,lit}$ (km s$^{-1}$) & Ref. & \emph{Gaia} DR3 ID & $V_{h,DR3}$ (km s$^{-1}$) & RUWE\\ \hline
    1 & $-41.96\pm 1.41$ & $-42.37\pm 0.74$ & 1 & 573934972233977344 & $-41.64\pm 0.17$ & $0.944$ \\ 
    2 & $-41.53\pm 1.23$ & $-40.42\pm2.13$ & 1 & 573941393208434688 & $-41.81\pm 0.37$ & $0.890$ \\
    3 & $-41.62\pm 1.10$ & $-37.96\pm 2.38$ & 1 & 573943252930299392 & $-39.31\pm 1.47$ & $1.148$ \\ 
    4 & $-43.91\pm 1.32$ & $-40.7$ & 2 & 573936174824230912 & $-39.46\pm 1.66$ & $0.977$ \\
    5 & $-38.82\pm 2.36$ & $-55.4$ & 2 & 573941053907094144 & $-38.09\pm 9.15$ & $0.986$ \\
    6 & $-42.09\pm 1.21$ & $-43.0$ & 2 & 573956069112687488 & $-43.41\pm 2.08$ & $0.958$ \\
    7 & $-40.30\pm 1.18$ & $-40.0$ & 2 & 573941122626558208 & $-38.14\pm 1.75$ & $1.522$ \\
    8 & $-42.60\pm 0.96$ & $-43.4$ & 2 & 573940813388946176 & $-42.75\pm 1.90$ & $1.019$ \\
    9 & $-43.54\pm 0.87$ & $-40.16$ & 3 & 573941397504463616 & $-44.08\pm 4.95$ & $0.955$ \\
    10 & $-42.36\pm 1.16$ & $-42.6$ & 2 & 573935075312637696 & $-44.55\pm 5.58$ & $1.216$ \\
    11 & $34.40\pm 0.76$ & $33.86\pm 0.19$ & 1 & 605015309794935552 & $34.26\pm 0.13$ & $1.033$ \\
    12 & $2.59\pm 1.23$ & $1.49\pm 0.15$ & 4 & 5382185151122208768 & $1.86\pm 0.13$ & $3.724$ \\ 
    13 & $-33.96\pm 1.45$ & $-30.97\pm 0.14$ & 5 & 5887670572485812608 & $-30.30\pm 0.16$ & $3.586$ \\ 
    14 & $-29.38\pm 0.52$ & $-28.01\pm 1.08$ & 1 & 5887641435423836544 & $-29.07\pm 0.15$ & $1.027$ \\ 
    15 & $-25.11\pm 0.50$ & $-24.49\pm 0.53$ & 5 & 4284005306836919936 & $-27.12\pm 0.75$ & $4.199$ \\ 
    16 & $-30.61\pm 0.50$ & $-29.22\pm 0.38$ & 1 & 4477460391998886144 & $-29.27\pm 0.13$ & $0.962$ \\
    17 & $-30.83\pm 0.56$ & $-28.66\pm 0.17$ & 1 & 4477223378527434880 & $-28.71\pm 0.13$ & $0.958$ \\
    18 & $2.92\pm 0.38$ & $4.03\pm 0.13$ & 1 & 5338660227476877056 & $3.82\pm 0.14$ & $2.736$ \\
    19 & $-6.32\pm 0.43$ & $-6.49\pm 0.02$ & 6 & 5338694690297345152 & $-6.44\pm 0.12$ & $0.887$ \\
    20 & $-4.84\pm 1.74$ & $-5.37\pm 0.15$ & 1 & 5976333437787480448 & $-5.55\pm 0.12$ & $0.862$ \\
    21 & $-5.43\pm 0.60$ & $-4.78\pm 0.13$ & 1 & 5976330517209580288 & $-4.81\pm 0.12$ & $0.953$ \\
    22 & $-5.63\pm 0.42$ & $-5.36\pm 0.14$ & 7 & 6091100701979396864 & $-6.10\pm 0.12$ & $1.100$ \\ \hline
    \end{tabular}
    \flushleft
    References regarding the heliocentric radial velocity found in the literature: (1) \cite{cantat-gaudin2020}, (2) \cite{jacobson2011}, (3) \cite{geller2008}, (4) \cite{mermilliod90}, and (5) \cite{mermilliod95}, (6) \cite{soubiran}, (7) \cite{mmu}, and \emph{Gaia} DR3 data from \cite{gaiadr3}. 
    \label{resultado1}
    \end{threeparttable}
\end{table*}
\begin{table*}
\caption{Equivalent widths of lines used in the analysis of abundances in stars of NGC188 cluster.}
\label{ew-tabela}
\begin{tabular}{ccccccccc} \hline \hline
    $\lambda$ & \multirow{2}{*}{ Elem. } & $\chi$ & \multirow{2}{*}{ $\log$ gf } & 1 & 3 & 4 & 6 & 8 \\
    (\AA) & & (eV) & & (m\AA) & (m\AA) & (m\AA) & (m\AA) & (m\AA) \\ \hline
    6154.22 &\ion{Na}{i} &2.10 &$-$1.547 &126.2 &111.9 &81.7 &70.0 &89.8 \\
    6160.75 &\ion{Na}{i} &2.10 &$-$1.246 &147.3 &128.9 &105.7 &109.4 &97.1 \\
    5528.42 &\ion{Mg}{i} &4.35 &$-$0.498 &440.3 &244.5 &162.6 &257.9 &428.0 \\
    5711.09 &\ion{Mg}{i} &4.35 &$-$1.724 &149.8 &168.6 &148.8 &153.1 &138.2 \\
    5772.15 &\ion{Si}{i} &5.08 &$-$1.750 &61.9 &77.5 &81.2 &64.4 &61.8 \\
    6125.03 &\ion{Si}{i} &5.61 &$-$1.465 &29.4 &40.6 &65.1 &48.1 &48.4 \\
    6131.58 &\ion{Si}{i} &5.62 &$-$1.557 &25.1 &38.8 &41.3 &38.7 &45.0 \\
    6131.86 &\ion{Si}{i} &5.62 &$-$1.617 &39.1 &42.5 &29.4 &32.4 &35.4 \\
    6142.53 &\ion{Si}{i} &5.62 &$-$1.296 &30.5 &44.2 &55.6 &36.4 &41.4 \\
    6145.08 &\ion{Si}{i} &5.62 &$-$1.311 &0.0 &49.0 &57.4 &46.5 &51.0 \\
    6155.14 &\ion{Si}{i} &5.62 &$-$0.755 &81.8 &89.7 &107.4 &106.7 &102.4 \\
    5867.57 &\ion{Ca}{i} &2.93 &$-$1.570 &93.4 &76.6 &99.0 &71.8 &57.8 \\
    6122.23 &\ion{Ca}{i} &1.88 &$-$0.180 &289.3 &259.8 &254.4 &172.4 &231.5 \\
    6156.03 &\ion{Ca}{i} &2.52 &$-$2.506 &72.4 &57.7 &69.1 &51.7 &38.7 \\
    6161.29 &\ion{Ca}{i} &2.52 &$-$1.266 &157.8 &139.8 &148.1 &85.6 &113.0 \\
    6166.44 &\ion{Ca}{i} &2.52 &$-$1.142 &157.1 &135.4 &93.9 &97.6 &104.2 \\
    6169.04 &\ion{Ca}{i} &2.52 &$-$0.800 &185.6 &155.1 &132.4 &140.9 &146.2 \\
    6169.56 &\ion{Ca}{i} &2.51 &$-$0.480 &196.9 &170.0 &148.7 &159.2 &159.8 \\
    6493.78 &\ion{Ca}{i} &2.52 &$-$0.109 &192.0 &180.5 &168.4 &180.6 &150.7 \\
    6499.65 &\ion{Ca}{i} &2.52 &$-$0.818 &165.6 &143.7 &162.0 &150.7 &143.4 \\
    5318.34 &\ion{Sc}{ii} &1.36 &$-$2.015 &60.3 &45.8 &47.6 &43.5 &37.9 \\
    5334.22 &\ion{Sc}{ii} &1.50 &$-$2.203 &91.8 &40.2 &20.5 &15.4 &13.0 \\
    5145.47 &\ion{Ti}{i} &1.46 &$-$0.540 &159.7 &127.5 &77.5 &83.8 &84.2 \\
    5295.78 &\ion{Ti}{i} &1.07 &$-$1.590 &128.3 &85.9 &46.5 &70.2 &58.1 \\
    5299.98 &\ion{Ti}{i} &1.05 &$-$2.300 &154.7 &88.3 &40.2 &37.5 &62.7 \\
    5338.33 &\ion{Ti}{i} &0.83 &$-$2.730 &97.2 &73.5 &29.0 &98.9 &59.6 \\
    5351.07 &\ion{Ti}{i} &2.78 &$-$0.067 &0.0 &54.7 &28.9 &0.0 &0.0 \\
    5766.33 &\ion{Ti}{i} &3.29 &0.289 &72.5 &56.4 &24.3 &44.1 &31.1 \\
    6121.01 &\ion{Ti}{i} &1.88 &$-$1.420 &110.3 &63.0 &51.7 &50.2 &47.1 \\
    6126.22 &\ion{Ti}{i} &1.07 &$-$1.206 &164.2 &122.9 &73.9 &109.3 &83.2 \\
    6497.68 &\ion{Ti}{i} &1.44 &$-$2.020 &110.4 &70.9 &5.6 &37.3 &31.0 \\
    5846.27 &\ion{V}{i} &3.13 &0.788 &44.5 &49.0 &30.3 &22.0 &39.3 \\
    6002.65 &\ion{V}{i} &1.05 &$-$1.570 &105.6 &76.3 &45.2 &23.4 &44.8 \\
    6039.69 &\ion{V}{i} &1.06 &$-$0.650 &152.5 &108.0 &64.5 &94.1 &73.4 \\
    6111.65 &\ion{V}{i} &1.04 &$-$0.740 &185.9 &141.0 &75.1 &80.9 &96.7 \\
    6119.53 &\ion{V}{i} &1.06 &$-$0.360 &162.5 &116.1 &72.0 &93.6 &78.7 \\
    6135.37 &\ion{V}{i} &1.05 &$-$0.760 &157.0 &113.9 &55.2 &77.5 &80.6 \\
    6150.15 &\ion{V}{i} &0.30 &$-$1.780 &203.1 &146.7 &96.1 &85.2 &99.0 \\
    6504.19 &\ion{V}{i} &1.18 &$-$1.280 &119.3 &91.8 &54.7 &51.3 &63.6 \\
    5303.22 &\ion{V}{ii} &2.28 &$-$2.047 &0.0 &25.9 &4.7 &32.9 &16.8 \\
    6028.28 &\ion{V}{ii} &2.49 &$-$2.122 &18.5 &21.6 &21.9 &30.4 &18.9 \\
    5122.12 &\ion{Cr}{i} &1.03 &$-$3.033 &104.1 &137.8 &56.7 &65.0 &56.9 \\
    5296.69 &\ion{Cr}{i} &0.98 &$-$1.992 &273.0 &200.5 &146.2 &168.6 &146.4 \\
    5300.75 &\ion{Cr}{i} &0.98 &$-$2.000 &182.5 &142.8 &103.6 &115.8 &117.3 \\
    5304.18 &\ion{Cr}{i} &3.46 &$-$0.670 &62.8 &54.1 &32.3 &72.2 &45.6 \\
    5312.88 &\ion{Cr}{i} &3.45 &$-$0.550 &64.1 &36.5 &41.1 &35.4 &43.8 \\
    5318.78 &\ion{Cr}{i} &3.44 &$-$0.670 &80.2 &65.2 &49.1 &68.9 &21.6 \\
    5329.12 &\ion{Cr}{i} &2.91 &$-$0.064 &163.2 &123.8 &96.4 &77.3 &246.4 \\
    5340.44 &\ion{Cr}{i} &3.44 &$-$0.730 &76.2 &70.3 &58.6 &113.9 &56.1 \\
    5348.32 &\ion{Cr}{i} &1.00 &$-$1.210 &0.0 &202.0 &154.8 &0.0 &0.0 \\
    5783.07 &\ion{Cr}{i} &3.32 &$-$0.500 &103.4 &79.3 &65.0 &73.4 &59.9 \\
    5783.87 &\ion{Cr}{i} &3.32 &$-$0.320 &130.3 &106.7 &77.6 &92.8 &88.3 \\
    5787.99 &\ion{Cr}{i} &3.32 &$-$0.083 &113.2 &106.0 &82.4 &73.8 &78.8 \\
    5788.39 &\ion{Cr}{i} &3.01 &$-$1.490 &73.4 &53.3 &16.6 &32.2 &35.3 \\
    5844.61 &\ion{Cr}{i} &3.01 &$-$1.870 &61.6 &43.7 &20.4 &31.2 &39.8 \\
    5863.96 &\ion{Cr}{i} &3.12 &$-$2.140 &0.0 &34.8 &31.6 &13.0 &36.6 \\
    6135.78 &\ion{Cr}{i} &4.80 &0.550 &0.0 &42.1 &44.5 &47.7 &44.9 \\
    6501.21 &\ion{Cr}{i} &0.98 &$-$3.924 &101.8 &72.7 &30.1 &54.0 &47.3 \\
    6630.02 &\ion{Cr}{i} &1.03 &$-$3.560 &135.8 &98.0 &48.9 &53.8 &56.4 \\
    5305.87 &\ion{Cr}{ii} &3.83 &$-$2.160 &34.7 &34.0 &38.3 &31.6 &28.7 \\
    5310.70 &\ion{Cr}{ii} &4.07 &$-$2.280 &26.9 &21.4 &28.7 &30.4 &17.3 \\
    5313.59 &\ion{Cr}{ii} &4.07 &$-$1.650 &34.6 &47.3 &54.4 &54.5 &25.9 \\ \hline
\end{tabular}
\end{table*}

\begin{table*}
\contcaption{}
\begin{tabular}{ccccccccc} \hline \hline
    $\lambda$ & \multirow{2}{*}{ Elem. } & $\chi$ & \multirow{2}{*}{ $\log$ gf } & 1 & 3 & 4 & 6 & 8 \\
    (\AA) & & (eV) & & (m\AA) & (m\AA) & (m\AA) & (m\AA) & (m\AA) \\ \hline
    5334.88 &\ion{Cr}{ii} &4.07 &$-$1.826 &53.6 &48.0 &59.7 &119.2 &0.0 \\
    5133.69 &\ion{Fe}{i} &4.18 &0.360 &225.3 &220.7 &178.4 &218.5 &186.4 \\
    5141.75 &\ion{Fe}{i} &2.42 &$-$2.240 &191.8 &179.3 &129.3 &148.1 &150.7 \\
    5143.73 &\ion{Fe}{i} &2.20 &$-$3.790 &103.0 &0.0 &68.2 &0.0 &76.0 \\
    5293.97 &\ion{Fe}{i} &4.14 &$-$1.870 &0.0 &68.9 &49.3 &0.0 &57.9 \\
    5294.55 &\ion{Fe}{i} &3.64 &$-$2.860 &56.3 &47.8 &0.0 &0.0 &37.9 \\
    5295.32 &\ion{Fe}{i} &4.41 &$-$1.690 &56.5 &51.9 &29.1 &0.0 &43.0 \\
    5307.36 &\ion{Fe}{i} &1.61 &$-$2.910 &196.9 &170.8 &135.3 &176.6 &142.4 \\
    5315.07 &\ion{Fe}{i} &4.37 &$-$1.550 &74.8 &87.1 &48.3 & 0.0 &61.1 \\
    5320.05 &\ion{Fe}{i} &3.64 &$-$2.540 &70.4 &58.9 &40.3 &44.8 &44.9 \\
    5321.11 &\ion{Fe}{i} &4.43 &$-$0.951 &81.6 &0.0 &68.1 &75.5 &77.1 \\
    5322.05 &\ion{Fe}{i} &2.28 &$-$2.803 &142.0 &129.1 &104.3 &119.3 &97.4 \\
    5326.14 &\ion{Fe}{i} &3.57 &$-$2.071 &96.0 &95.9 &73.7 &0.0 &0.0 \\
    5339.94 &\ion{Fe}{i} &3.27 &$-$0.630 &183.2 &184.7 &164.7 &179.8 &180.7 \\
    5358.10 &\ion{Fe}{i} &3.29 &$-$3.400 &0.0 &0.0 &27.3 &0.0 &0.0 \\
    5367.47 &\ion{Fe}{i} &4.42 &0.443 &134.9 &153.7 &153.5 &144.8 &142.3 \\
    5369.97 &\ion{Fe}{i} &4.37 &0.536 &135.6 &171.2 &171.3 &135.6 &154.3 \\
    5568.81 &\ion{Fe}{i} &3.63 &$-$2.950 &18.6 &47.3 &36.9 &33.2 &25.7 \\
    5759.27 &\ion{Fe}{i} &4.65 &$-$2.070 &24.2 &23.9 &27.5 &18.8 &24.6 \\
    5760.35 &\ion{Fe}{i} &3.64 &$-$2.490 &73.4 &66.9 &46.6 &42.7 &54.7 \\
    5775.09 &\ion{Fe}{i} &4.22 &$-$1.298 &95.1 &95.5 &76.8 &71.3 &89.0 \\
    5778.47 &\ion{Fe}{i} &2.59 &$-$3.430 &95.1 &93.8 &63.6 &69.6 &69.3 \\
    5784.69 &\ion{Fe}{i} &3.40 &$-$2.532 &91.1 &83.9 &72.4 &71.7 &58.3 \\
    5838.42 &\ion{Fe}{i} &3.94 &$-$2.340 &62.6 &62.2 &26.5 &56.5 &39.7 \\
    5849.70 &\ion{Fe}{i} &3.69 &$-$2.990 &47.0 &38.8 &36.3 &21.1 &34.4 \\
    5852.19 &\ion{Fe}{i} &4.55 &$-$1.300 &63.7 &0.0 &0.0 &52.4 &0.0 \\
    5853.18 &\ion{Fe}{i} &1.49 &$-$5.280 &90.1 &74.2 &28.7 &52.2 &51.6 \\
    5855.09 &\ion{Fe}{i} &4.61 &$-$1.478 &57.1 &49.9 &49.5 &45.3 &57.6 \\
    5856.08 &\ion{Fe}{i} &4.29 &$-$1.328 &70.1 &0.0 &62.5 &69.0 &70.9 \\
    5858.77 &\ion{Fe}{i} &4.22 &$-$2.260 &60.9 &52.1 &0.0 &39.3 &45.9 \\
    5859.61 &\ion{Fe}{i} &4.55 &$-$0.418 &107.0 &105.7 &93.3 &93.9 &106.9 \\
    5862.36 &\ion{Fe}{i} &4.55 &$-$0.125 &113.4 &125.1 &117.4 &0.0 &0.0 \\
    6003.03 &\ion{Fe}{i} &3.88 &$-$1.120 &126.7 &130.6 &131.1 &126.6 &118.4 \\
    6007.96 &\ion{Fe}{i} &4.65 &$-$0.597 &85.8 &132.8 &88.7 &82.9 &0.0 \\
    6008.58 &\ion{Fe}{i} &3.88 &$-$0.982 &155.2 &177.2 &114.7 &125.7 &119.5 \\
    6015.25 &\ion{Fe}{i} &2.22 &$-$4.680 &72.4 &0.0 &10.8 &36.1 &34.9 \\
    6019.36 &\ion{Fe}{i} &3.57 &$-$3.360 &52.0 &43.2 &0.0 &30.3 &0.0 \\
    6024.07 &\ion{Fe}{i} &4.55 &$-$0.120 &130.4 &134.1 &108.7 &234.4 &154.1 \\
    6027.06 &\ion{Fe}{i} &4.08 &$-$1.089 &105.6 &0.0 &88.5 &90.8 &101.3 \\
    6034.04 &\ion{Fe}{i} &4.31 &$-$2.420 &37.9 &0.0 &0.0 &26.5 &31.7 \\
    6035.34 &\ion{Fe}{i} &4.29 &$-$2.590 &24.2 &0.0 &29.9 &25.4 &0.0 \\
    6054.10 &\ion{Fe}{i} &4.37 &$-$2.310 &37.4 &0.0 &0.0 &16.7 &34.1 \\
    6120.25 &\ion{Fe}{i} &0.92 &$-$5.950 &96.9 &96.7 &50.0 &54.4 &47.2 \\
    6151.62 &\ion{Fe}{i} &2.18 &$-$3.299 &140.8 &140.3 &102.5 &91.3 &106.2 \\
    6157.73 &\ion{Fe}{i} &4.08 &$-$1.260 &144.8 &0.0 &151.1 &98.1 &89.4 \\
    6165.37 &\ion{Fe}{i} &4.14 &$-$1.474 &87.2 &95.6 &125.0 &75.9 &74.8 \\
    6173.34 &\ion{Fe}{i} &2.22 &$-$2.880 &172.6 &144.9 &100.0 &97.0 &126.9 \\
    6475.63 &\ion{Fe}{i} &2.56 &$-$2.942 &138.5 &130.8 &0.0 &103.6 &120.1 \\
    6481.87 &\ion{Fe}{i} &2.28 &$-$2.984 &161.7 &137.0 &122.1 &109.7 &121.3 \\
    6495.74 &\ion{Fe}{i} &4.84 &$-$0.940 &57.9 &90.4 &0.0 &0.0 &0.0 \\
    6496.47 &\ion{Fe}{i} &4.80 &$-$0.570 &102.4 &92.1 &0.0 &98.3 &93.5 \\
    6498.95 &\ion{Fe}{i} &0.96 &$-$4.699 &194.4 &145.8 &0.0 &128.9 &0.0 \\
    6627.56 &\ion{Fe}{i} &4.55 &$-$1.680 &55.3 &63.1 &47.5 &27.3 &54.1 \\
    6633.42 &\ion{Fe}{i} &4.84 &$-$1.490 &0.0 &60.9 &46.6 &70.5 &55.2 \\
    6633.76 &\ion{Fe}{i} &4.56 &$-$0.799 &83.7 &98.4 &84.5 &84.5 &89.6 \\
    6646.98 &\ion{Fe}{i} &2.61 &$-$3.990 &82.8 &65.6 &36.6 &42.5 &43.6 \\
    6648.08 &\ion{Fe}{i} &1.01 &$-$5.428 &112.4 &93.1 &0.0 &57.0 &61.5 \\
    5132.67 &\ion{Fe}{ii} &2.81 &$-$3.980 &43.3 &56.2 &0.0 &25.4 &38.3 \\
    5256.94 &\ion{Fe}{ii} &2.89 &$-$4.182 &0.0 &0.0 &40.3 &48.0 &0.0 \\
    5264.81 &\ion{Fe}{ii} &3.23 &$-$3.120 &52.7 &59.1 &54.7 &87.4 &53.3 \\
    5325.56 &\ion{Fe}{ii} &3.22 &$-$3.120 &51.3 &51.1 &54.9 &28.3 &0.0 \\ 
    5414.08 &\ion{Fe}{ii} &3.22 &$-$3.540 &32.9 &44.4 &42.8 &25.9 &28.2 \\ \hline
\end{tabular}
\end{table*}

\begin{table*}
\contcaption{}
\begin{tabular}{ccccccccc} \hline \hline
    $\lambda$ & \multirow{2}{*}{ Elem. } & $\chi$ & \multirow{2}{*}{ $\log$ gf } & 1 & 3 & 4 & 6 & 8 \\
    (\AA) & & (eV) & & (m\AA) & (m\AA) & (m\AA) & (m\AA) & (m\AA) \\ \hline
    5425.26 &\ion{Fe}{ii} &3.20 &$-$3.160 &51.4 &58.0 &59.8 &47.6 &50.6 \\
    6084.10 &\ion{Fe}{ii} &3.20 &$-$3.780 &29.0 &0.0 &0.0 &44.1 &0.0 \\
    6113.33 &\ion{Fe}{ii} &3.22 &$-$4.110 &21.1 &17.7 &27.1 &0.0 &24.7 \\
    6129.70 &\ion{Fe}{ii} &3.20 &$-$4.700 &9.6 &16.1 &12.3 &29.3 &11.8 \\
    6149.24 &\ion{Fe}{ii} &3.89 &$-$2.720 &0.0 &42.6 &47.2 &50.4 &38.7 \\
    6247.56 &\ion{Fe}{ii} &3.89 &$-$2.310 &23.6 &54.5 &68.5 &54.2 &49.6 \\
    6369.46 &\ion{Fe}{ii} &2.89 &$-$4.160 &20.8 &16.0 &43.9 &33.9 &0.0 \\
    6416.93 &\ion{Fe}{ii} &3.89 &$-$2.650 &0.0 &51.3 &49.7 &40.0 &41.9 \\
    6456.39 &\ion{Fe}{ii} &3.90 &$-$2.100 &49.9 &62.5 &0.0 &56.5 &60.0 \\
    5301.04 &\ion{Co}{i} &1.71 &$-$2.000 &117.1 &115.3 &64.0 &74.5 &79.7 \\
    5325.28 &\ion{Co}{i} &4.02 &0.091 &41.1 &49.8 &36.9 &70.5 &39.8 \\
    5342.70 &\ion{Co}{i} &4.02 &0.741 &60.6 &62.6 &43.4 &97.1 &51.3 \\
    5352.05 &\ion{Co}{i} &3.58 &0.060 &0.0 &71.7 &46.8 &0.0 &0.0 \\
    5359.20 &\ion{Co}{i} &4.15 &0.244 &27.2 &27.5 &26.1 &40.9 &0.0 \\
    5369.59 &\ion{Co}{i} &1.74 &$-$1.765 &160.5 &130.4 &103.9 &86.0 &101.2 \\
    6117.00 &\ion{Co}{i} &1.78 &$-$2.490 &93.9 &78.4 &53.0 &40.0 &44.4 \\
    6490.34 &\ion{Co}{i} &2.04 &$-$2.520 &95.2 &65.4 &31.6 &17.7 &35.6 \\
    6632.47 &\ion{Co}{i} &2.28 &$-$2.000 &86.7 &73.6 &35.9 &49.7 &47.9 \\
    5137.08 &\ion{Ni}{i} &1.68 &$-$1.940 &155.4 &159.3 &127.6 &163.1 &134.9 \\
    5593.74 &\ion{Ni}{i} &3.90 &$-$0.840 &76.1 &64.5 &79.7 &79.4 &72.9 \\
    5760.83 &\ion{Ni}{i} &4.11 &$-$0.800 &95.0 &94.7 &65.6 &38.7 &66.9 \\
    5847.01 &\ion{Ni}{i} &1.68 &$-$3.460 &124.2 &99.0 &68.5 &74.2 &75.5 \\
    6007.31 &\ion{Ni}{i} &1.68 &$-$3.400 &106.0 &95.8 &73.1 &78.5 &66.1 \\
    6053.68 &\ion{Ni}{i} &4.24 &$-$1.070 &59.2 &34.5 &54.9 &53.7 &61.0 \\
    6111.06 &\ion{Ni}{i} &4.09 &$-$0.870 &61.4 &61.6 &88.0 &57.8 &68.5 \\
    6128.99 &\ion{Ni}{i} &1.68 &$-$3.430 &114.0 &93.5 &90.2 &78.3 &77.4 \\
    6130.13 &\ion{Ni}{i} &4.27 &$-$0.960 &37.6 &42.9 &37.4 &37.2 &41.5 \\
    6635.15 &\ion{Ni}{i} &4.42 &$-$0.820 &55.8 &49.0 &43.8 &50.5 &51.8 \\
    6643.64 &\ion{Ni}{i} &1.68 &$-$2.220 &209.4 &181.6 &136.1 &147.4 &151.9 \\
    5119.12 &\ion{Y}{ii} &0.99 &$-$1.360 &62.1 &51.6 &47.9 &46.6 &50.8 \\
    5289.82 &\ion{Y}{ii} &1.03 &$-$1.850 &21.3 &19.1 &8.1 &25.5 &14.2 \\
    5330.58 &\ion{Ce}{ii} &0.87 &$-$0.400 &26.2 &27.3 &11.4 &40.9 &9.0 \\
    6043.39 &\ion{Ce}{ii} &1.21 &$-$0.480 &31.3 &17.0 &22.7 &25.3 &18.2 \\
    6645.11 &\ion{Eu}{ii} &1.38 &$-$0.200 &36.4 &29.0 &19.6 &22.6 &17.5 \\ \hline
\end{tabular}
\end{table*}
\begin{table*}
    \caption{Comparison between photometric results obtained in this work with the literature, except for objects 2, 5, 7, 9, 10, 19 and 22.}
    \begin{threeparttable}
    \begin{tabular}{ccccclc} \hline \hline
    ID & T$_{\textrm{eff}}$ (K) & log g (dex) & [Fe/H] (dex) & $\xi$ (km/s) & Ref. & Method \\ \hline
    1 & 4100 $\pm$ 200 & 1.70 $\pm$ 0.20 & +0.06 $\pm$ 0.08 & 1.50 $\pm$ 0.20 & \citet{jacobson2011} & S \\
    & 4018 $\pm$ 160 & 1.54 $\pm$ 0.09 & +0.17 & 1.79 $\pm$ 0.54 & This work & P \\ \hline
    3 & 4400 $\pm$ 100 & 2.10 $\pm$ 0.20 & +0.13 $\pm$ 0.03 & 1.50 $\pm$ 0.20 & \citet{friel2010} & S \\
    & 4400 $\pm$ 200 & 2.30 $\pm$ 0.20 & +0.06 $\pm$ 0.06 & 1.50 $\pm$ 0.20 & \citet{jacobson2011} & S \\
    & 4400 $\pm$ 50 & 2.20 $\pm$ 0.10 & +0.14 $\pm$ 0.02 & 1.50 $\pm$ 0.10 & \citet{jacobson2013} & S \\
    & 4328 $\pm$ 210 & 2.21 $\pm$ 0.10 & +0.17 & 1.57 $\pm$ 0.57 & This work & P \\ \hline
    4 & 5100 $\pm$ 200 & 3.20 $\pm$ 0.20 & +0.03 $\pm$ 0.08 & 1.50 $\pm$ 0.20 & \citet{jacobson2011} & S \\
    & 4986 $\pm$ 226 & 3.05 $\pm$ 0.09 & +0.17 & 1.33 $\pm$ 0.61 & This work & P \\ \hline
    6 & 5000 $\pm$ 200 & 3.30 $\pm$ 0.20 & +0.01 $\pm$ 0.06 & 1.50 $\pm$ 0.20 & \citet{jacobson2011} & S \\
    & 4807 $\pm$ 202 & 3.09 $\pm$ 0.08 & +0.17 & 1.27 $\pm$ 0.58 & This work & P \\ \hline
    8 & 4800 $\pm$ 200 & 3.20 $\pm$ 0.20 & $-$0.07 $\pm$ 0.07 & 1.50 $\pm$ 0.20 & \citet{jacobson2011} & S \\
    & 4710 $\pm$ 215 & 3.09 $\pm$ 0.09 & +0.17 & 1.25 $\pm$ 0.59 & This work & P \\ \hline
    11 & 4250 $\pm$ 62 & 1.00 $\pm$ 0.14 & +0.03 $\pm$ 0.05 & --- & \citet{netopil} & S \\
    & 4160 $\pm$ 71 & 1.55 $\pm$ 0.13 & +0.04 $\pm$ 0.06 & --- & \citet{lit2} & C \\
    & 4131 $\pm$ 214 & 1.47 $\pm$ 0.11 & +0.03  & 1.80 $\pm$ 0.57 & This work & P \\ \hline
    12 & 4508 $\pm$ 200 & 2.00 $\pm$ 0.30 & $-$0.28 $\pm$ 0.32 & 1.70 $\pm$ 0.30 & \citet{pasquini2001} & S \\
    & 4475  & --- & $-$0.14 $\pm$ 0.16 & --- & \citet{anthony2009} & S \\
    & 4800 $\pm$ 50 & 2.60 $\pm$ 0.10 & $-$0.13 $\pm$ 0.08 & 2.10 $\pm$ 0.10 & \citet{mitschang2012} & S \\
    & 4500 $\pm$ 200 & 1.70 $\pm$ 0.30 & $-$0.15 $\pm$ 0.08 & 1.50 $\pm$ 0.30 & \citet{penhasuarez} & S \\
    & 4518 $\pm$ 196 & 1.99 $\pm$ 0.09 & +0.04  & 1.66 $\pm$ 0.57 & This work & P \\ \hline
    13 & 4800 $\pm$ 200 & 2.50 $\pm$ 0.25 & +0.09 $\pm$ 0.21 & 2.50 $\pm$ 0.50 & \citet{luck1994} & U \\
    & 4470 $\pm$ 60 & 2.00 $\pm$ 0.26 & +0.03 $\pm$ 0.10 & 1.38 $\pm$ 0.08 & \citet{rodolfo2009} & S \\
    & 4400 $\pm$ 200 & 1.90 $\pm$ 0.30 & $-$0.09 $\pm$ 0.09 & 1.40 $\pm$ 0.30 & \citet{penhasuarez} & S \\
    & 4339 $\pm$ 212 & 1.61 $\pm$ 0.10 & +0.02  & 1.77 $\pm$ 0.57 & This work & P \\ \hline
    14 & 4425 $\pm$ 60 & 1.95 $\pm$ 0.26 & +0.02 $\pm$ 0.11 & 1.34 $\pm$ 0.08 & \citet{rodolfo2009} & S \\
    & 4307 $\pm$ 192 & 1.72 $\pm$ 0.09 & +0.02  & 1.73 $\pm$ 0.56 & This work & P \\ \hline
    15 & 5130 $\pm$ 60 & 3.00 $\pm$ 0.26 & +0.08 $\pm$ 0.08 & 1.31 $\pm$ 0.08 & \citet{rodolfo2009} & S \\
    & 5049 $\pm$ 204 & 2.37 $\pm$ 0.08 & +0.07  & 1.63 $\pm$ 0.59 & This work & P \\ \hline
    16 & 4370 $\pm$ 60 & 1.80 $\pm$ 0.26 & +0.04 $\pm$ 0.10 & 1.51 $\pm$ 0.08 & \citet{rodolfo2009} & S \\
    & 4315 $\pm$ 90 & 1.72 $\pm$ 0.23 & $-$0.14 $\pm$ 0.13 & 1.69 $\pm$ 0.08 & \citet{morel2014} & S \\
    & 4290 $\pm$ 65 & 1.85 $\pm$ 0.16 & $-$0.03 $\pm$ 0.12 & 1.68 $\pm$ 0.07 & \citet{morel2014} & A \\
    & 4369  & 1.46  & +0.04  & 1.62  & \citet{luck2014} & U \\
    & 4370 $\pm$ 30 & 1.83 $\pm$ 0.12 & $-$0.20 $\pm$ 0.19 & --- & \citet{boeche2016} & F \\
    & 4549 $\pm$ 822 & 1.51 $\pm$ 0.36 & +0.10  & 1.85 $\pm$ 1.28 & This work & P \\ \hline
    17 & 5245 $\pm$ 44 & 3.11 $\pm$ 0.06 & +0.35 $\pm$ 0.03 & 0.85  & \citet{valenti2005} & F \\
    & 5118 $\pm$ 29 & 2.83 $\pm$ 0.25 & +0.04 $\pm$ 0.10 & 1.65 $\pm$ 0.03 & \citet{santos2009} & S \\
    & 4979 $\pm$ 72 & 2.75 $\pm$ 0.12 & +0.00 $\pm$ 0.10 & 1.58 $\pm$ 0.10 & \citet{santos2009} & S \\
    & 5015 $\pm$ 60 & 2.85 $\pm$ 0.26 & +0.11 $\pm$ 0.11 & 1.44 $\pm$ 0.08 & \citet{rodolfo2009} & S \\
    & 5035 $\pm$ 80 & 2.74 $\pm$ 0.19 & $-$0.01 $\pm$ 0.11 & 1.55 $\pm$ 0.07 & \citet{morel2014} & S \\
    & 5055 $\pm$ 55 & 2.56 $\pm$ 0.05 & $-$0.07 $\pm$ 0.10 & 1.58 $\pm$ 0.06 & \citet{morel2014} & A \\
    & 4991 $\pm$ 114 & 2.61 $\pm$ 0.22 & +0.00 $\pm$ 0.10 & --- & \citet{jacobson2016} & W \\
    & 5080 $\pm$ 169 & 2.17 $\pm$ 0.06 & +0.10  & 1.73 $\pm$ 0.56 & This work & P \\ \hline
    18 & 5000 $\pm$ 200 & 2.36 $\pm$ 0.25 & +0.13 $\pm$ 0.02 & 2.00 $\pm$ 0.50 & \citet{luck1994} & U \\
    & 4995 $\pm$ 60 & 2.65 $\pm$ 0.26 & +0.11 $\pm$ 0.11 & 1.52 $\pm$ 0.08 & \citet{rodolfo2009} & S \\
    & 4905 $\pm$ 219 & 2.20 $\pm$ 0.08 & +0.00  & 1.65 $\pm$ 0.60 & This work & P \\ \hline
    20 & 4915 $\pm$ 60 & 2.30 $\pm$ 0.26 & +0.01 $\pm$ 0.09 & 1.64 $\pm$ 0.08 & \citet{rodolfo2009} & S \\
    & 5012 $\pm$ 212 & 1.95 $\pm$ 0.08 & +0.00  & 1.78 $\pm$ 0.60 & This work & P \\ \hline
    21 & 5015 $\pm$ 60 & 2.50 $\pm$ 0.26 & +0.09 $\pm$ 0.07 & 1.70 $\pm$ 0.08 & \citet{rodolfo2009} & S \\
    & 4929 $\pm$ 210 & 1.98 $\pm$ 0.08 & +0.00  & 1.75 $\pm$ 0.59 & This work & P \\ \hline
    \end{tabular}
    Labels: (A) Asteroseismology, (C) Calibration relations from the raw outputs of the pipeline, (F) Spectral fitting to a grid of synthetic spectra, (P) Photometry, (S) Spectroscopy, (U) Unweighted mean of the photometric and spectroscopic values and (W) Weighted-median value for each atmospheric parameter.
    \label{resultado-foto}
    \end{threeparttable}
\end{table*}
\begin{table*}
    \centering
    \caption{Comparison between mean abundances of NGC188 compared with the literature.}
    \begin{threeparttable}
    \begin{tabular}{cclcccclc} \hhline{====~====}
    & $\langle\textup{[X/Fe]}\rangle$ & Ref. & Method & & & $\langle\textup{[X/Fe]}\rangle$ & Ref. & Method \\ \cline{1-4}\cline{6-9}
    $\textup{[Fe/H]}$ & 0.12 $\pm$ 0.02 & \cite{friel2010} & S & & Sc & $-$0.01 $\pm$ 0.04 & \cite{occaso} & S \\
    & $-$0.03 $\pm$ 0.04 & \cite{jacobson2011} & S & & & 0.33 $\pm$ 0.07 & This work & S \\ \cline{6-9}
    & 0.12 $\pm$ 0.04 & \cite{jacobson2013} & S & & Ti & 0.05 $\pm$ 0.12 & \cite{friel2010} & S \\
    & 0.11 $\pm$ 0.04 & \cite{netopil} & S & & & 0.14 $\pm$ 0.05 & \cite{jacobson2011} & S \\
    & 0.12 $\pm$ 0.04 & \cite{overbeek} & S & & & 0.06 $\pm$ 0.05 & \cite{occaso} & S \\ 
    & 0.14 $\pm$ 0.03 & \cite{donor2018} & F & & & $-$0.12 $\pm$ 0.04 & \cite{slumtrup2018} & S \\
    & 0.13 $\pm$ 0.01 & \cite{carrera2019} & F & & & 0.03 $\pm$ 0.03 & \cite{donor2020} & F \\
    & 0.03 $\pm$ 0.03 & \cite{occaso} & S & & & 0.12 $\pm$ 0.07 & This work & S \\ \cline{6-9}
    & 0.04 $\pm$ 0.01 & \cite{slumtrup2018} & S & & V & 0.03 $\pm$ 0.08 & \cite{donor2018} & F \\ 
    & 0.09 $\pm$ 0.02 & \cite{donor2020} & F & & & 0.01 $\pm$ 0.05 & \cite{occaso} & S \\
    & 0.04 $\pm$ 0.04 & This work & S & & & $-$0.03 $\pm$ 0.14 & \cite{donor2020} & F \\ \cline{1-4}
    C & $-$0.02 $\pm$ 0.04 & This work & S & & & 0.17 $\pm$ 0.07 & This work & S \\ \cline{1-4}\cline{6-9}
    N & 0.46 $\pm$ 0.07 & This work & S & & Cr & 0.11 $\pm$ 0.08 & \cite{friel2010} & S \\ \cline{1-4}
    O & 0.04 $\pm$ 0.10 & \cite{friel2010} & S & & & $-$0.01 $\pm$ 0.06 & \cite{donor2018} & F \\ 
    & 0.02 $\pm$ 0.04 & \cite{donor2018} & F & & & $-$0.01 $\pm$ 0.01 & \cite{carrera2019} & F \\
    & 0.05 $\pm$ 0.03 & \cite{occaso} & S & & & 0.04 $\pm$ 0.02 & \cite{occaso} & S \\
    & 0.00 $\pm$ 0.05 & \cite{donor2020} & F & & & $-$0.03 $\pm$ 0.05 & \cite{slumtrup2018} & S \\
    & 0.15 $\pm$ 0.05 & This work & S & & & 0.04 $\pm$ 0.04 & \cite{donor2020} & F \\ \cline{1-4}
    Na & 0.15 $\pm$ 0.03 & \cite{friel2010} & S & & & 0.15 $\pm$ 0.05 & This work & S \\ \cline{6-9}
    & 0.10 $\pm$ 0.05 & \cite{jacobson2011} & S & & Co & 0.09 $\pm$ 0.07 & \cite{friel2010} & S \\
    & $-$0.01 $\pm$ 0.05 & \cite{carrera2019} & F & & & 0.13 $\pm$ 0.11 & \cite{donor2018} & F \\
    & 0.15 $\pm$ 0.03 & \cite{slumtrup2018} & S & & & 0.08 $\pm$ 0.07 & \cite{donor2020} & F \\
    & $-$0.01 $\pm$ 0.18 & \cite{donor2020} & F & & & 0.19 $\pm$ 0.07 & This work & S \\ \cline{6-9}
    & 0.29 $\pm$ 0.06 & This work & S & & Ni & 0.08 $\pm$ 0.05 & \cite{jacobson2011} & S \\ \cline{1-4}
    Si & 0.17 $\pm$ 0.08 & \cite{friel2010} & S & & & 0.04 $\pm$ 0.02 & \cite{donor2018} & F \\
    & 0.25 $\pm$ 0.05 & \cite{jacobson2011} & S & & & 0.03 $\pm$ 0.01 & \cite{carrera2019} & F \\
    & 0.01 $\pm$ 0.02 & \cite{donor2018} & F & & & 0.09 $\pm$ 0.04 & \cite{occaso} & S \\
    & 0.00 $\pm$ 0.01 & \cite{carrera2019} & F & & & 0.12 $\pm$ 0.03 & \cite{slumtrup2018} & S \\
    & 0.09 $\pm$ 0.03 & \cite{occaso} & S & & & 0.04 $\pm$ 0.03 & \cite{donor2020} & F \\
    & 0.14 $\pm$ 0.04 & \cite{slumtrup2018} & S & & & 0.20 $\pm$ 0.04 & This work & S \\ \cline{6-9}
    & 0.03 $\pm$ 0.01 & \cite{donor2020} & F & & Y & $-$0.02 $\pm$ 0.06 & \cite{slumtrup2018} & S \\
    & 0.22 $\pm$ 0.05 & This work & S & & & 0.08 $\pm$ 0.06 & This work & S \\ \cline{1-4}\cline{6-9}
    Ca & $-$0.03 $\pm$ 0.06 & \cite{friel2010} & S & & Ce & 0.17 $\pm$ 0.16 & This work & S \\ \cline{6-9}
    & $-$0.04 $\pm$ 0.06 & \cite{jacobson2011} & S & & Eu & $-$0.18 $\pm$ 0.12 & \cite{jacobson2013} & S \\
    & $-$0.02 $\pm$ 0.02 & \cite{donor2018} & F & & & $-$0.12 $\pm$ 0.07 & \cite{overbeek} & S \\
    & $-$0.02 $\pm$ 0.01 & \cite{carrera2019} & F & & & 0.26 & This work & S \\ \cline{6-9}
    & 0.04 $\pm$ 0.06 & \cite{occaso} & S & & & & \\
    & $-$0.09 $\pm$ 0.08 & \cite{slumtrup2018} & S & & & & \\
    & $-$0.02 $\pm$ 0.04 & \cite{donor2020} & F & & & & \\
    & 0.13 $\pm$ 0.07 & This work & S & & & \\ \cline{1-4}
    \end{tabular}
    Labels: (F) Spectral fitting to a grid of synthetic spectra and (S) Spectroscopy.
    \end{threeparttable}
    \label{comparacao-literatura}
\end{table*}


\bsp	
\label{lastpage}
\end{document}